%% file: EC_forecasts.tex
\documentclass[longauth]{aa}

\usepackage{graphicx}
\usepackage{natbib}
\usepackage{scalerel}
\usepackage{comment}%added from new version
\usepackage{array}%added by me

\usepackage[table]{xcolor}

\defcitealias{Gabarra-EP31}{G23}
\defcitealias{EP-Quai}{Q26}
\usepackage{txfonts}
\usepackage[pdfencoding=auto,psdextra]{hyperref}
\hypersetup{
    colorlinks=true,
    linkcolor=blue,
    filecolor=magenta,      
    urlcolor=blue,
    citecolor=blue
}
\makeatletter
\renewcommand*\aa@pageof{, page \thepage{} of \pageref*{LastPage}}
\makeatother

\usepackage[utf8]{inputenc}
\renewcommand{\linenumbers}[0]{} %added from new version --> to deactivate line numbering
\usepackage{euclid}

\begin{document}
%
% Put the title and authors of your (Standard Project) paper here
%
 \title{\Euclid}
\subtitle{Predicting scaling relations for
emission-line galaxies using simulated data}
%\title{\Euclid\/ Collaboration. TBD. 
%Predicting scaling relations for emission-line galaxies using simulated data 
%\thanks{This paper is published on behalf of the Euclid Consortium}
%}

%Forecasts on Star Formation, Dust-Extinction, and Metallcity for emission-line galaxies from mock catalogs and simulations}

%% please do not edit the author list once you copy it from the
%% Publication Portal -- contact ECEB Bureau for changes
\newcommand{\fssp}{{\small\texttt{FastSpec}}\xspace}
\newcommand{\egg}{{\small\texttt{EGG}}\xspace}

\input{author_list}

\abstract{
The \Euclid\ spectroscopic surveys are expected to increase the number of observed emission-line galaxies substantially compared to previous near-infrared surveys. In this work, we assess the ability of the Euclid Wide Survey (EWS) and Euclid Deep Survey (EDS), at its final (DR3) depth, to recover key scaling relations from emission-line measurements. We investigate the star-forming main sequence (MS), dust–mass relation (DMR), and mass–metallicity relation (MZR), linking stellar mass to SFR, dust attenuation, and gas-phase metallicity, respectively, using MAMBO (mocks with abundance matching in Bologna) spectroscopic simulations processed with the official \Euclid\ pipelines.
The MS can be traced over $z\in[0,4]$ using Pa\,$\beta$, \ha, and [\ion{O}{ii}]$\lambda\lambda3727,3729$ as SFR tracers. The shallow EWS flux limit biases the recovered MS towards higher normalisation and a flatter slope by selecting the brightest, most strongly star-forming galaxies. The deeper EDS enables almost complete sampling with Pa\,$\beta$ and \ha at $\logten(M_*/M_\odot)\geq10.3$ over $z\in[0,1.88]$, with only mild biases at the lower and upper ends of this redshift range. The [\ion{O}{ii}]-based MS becomes increasingly biased towards higher SFRs at the highest redshifts. The DMR can be traced through the F(\ha)/F(\hb) Balmer decrement over $z\in[1.45,1.88]$ and $z\in[0.9,1.88]$ in the EWS and EDS, respectively. The nebular colour excess agrees well with the true values, although the requirement of detecting the faint \hb line restricts the sample to less obscured galaxies at fixed $M_*$. This effect is stronger in the shallower EWS. In the EDS, the DMR is better sampled, with further improvement expected if real galaxies experience milder attenuation than predicted by MAMBO, as suggested by DMR measurements from previous surveys.  Finally, a ‘golden sample’ of several hundred galaxies per deg$^2$ at $z\in[1.49, 1.88]$, detected in multiple optical rest-frame emission lines, recovers gas-phase metallicities with typical accuracy of 0.1 dex at $\logten(M_*/M_\odot)\geq10$, while lower-mass galaxies show biases up to 0.15 dex. Relative to the MAMBO \HE-band-limited sample, the golden sample is biased towards high-SFR and low-metallicity galaxies, preferentially tracing the upper MS and lower MZR. Despite the simplified slitless spectroscopic simulations lacking effects such as mutual spectral contamination and detector persistence, these forecasts provide a quantitative benchmark for future \Euclid\ spectroscopic analyses.}

%
% Provide up to five key words:
%
    \keywords{Galaxies: evolution -- Galaxies: star formation -- Galaxies: fundamental parameters}
    %from the list in
     %https://www.aanda.org/for-authors/latex-issues/information-files#pop}
%
% Add short versions of title and author list for page headings
%
   \titlerunning{Predicting scaling relations for emission-line galaxies using simulated data}
   \authorrunning{Euclid Collaboration: Mancini, C. et al. }
   
\maketitle
\section{\label{sc:Intro}Introduction}

The \Euclid\ mission \citep{Laureijs11, Racca2016, Scaramella-EP1, EuclidSkyOverview} is designed to investigate the nature of dark energy and dark matter through its two main cosmological probes: weak gravitational lensing (WL) and galaxy clustering (GC). Over its planned six-year duration, \Euclid\ will map approximately one third of the extragalactic sky, covering about\,\num{14000}\,deg$^{2}$ away from the ecliptic and Galactic planes as part of the Euclid Wide Survey \citep[EWS;][]{Scaramella-EP1}. The EWS will be complemented by the Euclid Deep Survey (EDS; Scaramella et al., in prep.), covering 53\,deg$^{2}$ and reaching two magnitudes deeper in the \HE band by the end of the mission, with a signal-to-noise ratio (S/N) roughly four times higher in emission-line flux than the EWS.

The mission has already begun delivering data products, starting with an initial public Quick Data Release \citep[Q1;][]{Q1-TP001}. The first major public data release (DR1, at the end of 2026)  will be followed by DR2 and DR3 approximately 2.5 and 5 years later, respectively, progressively reaching the final survey depth and sky coverage.

\Euclid\ has a 0.5\,deg$^2$ field of view, observed simultaneously by two instruments. The Visual Imager (VIS) operates with a broadband filter \IE $\simeq$ ($r$+$i$+$z$) with a $5\,\sigma$ detection limit of $m_{\rm AB}\leq 26.5$ for extended sources \citep{EuclidSkyVIS}. The Near-Infrared Spectrometer and Photometer (NISP) has both a photometric channel (NISP-P), with three filters \YE, \JE, and \HE\ reaching $m_{\rm AB}\leq 24$ at $5\,\sigma$ for point-like sources \citep{EuclidSkyNISP}, and a spectroscopic channel (NISP-S). NISP-S is a slitless spectrometer with two red grisms (RGS, %covering the \ensuremath{\rm RG_\sfont{E}} band 
1.21--1.89\,$\micron$) and a blue grism (BGS, %covering the {\ensuremath{\rm BG_\sfont{E}}} band 
0.93--1.36\,$\micron$), the latter 
operating exclusively in the EDS and calibration fields. 
NISP-S was designed to meet the requirements for: i) a resolving power of $\mathcal{R}\geq380$ and $\mathcal{R}\geq260$ for \ang{;;0.5} diameter sources in the RGS and BGS, respectively; and ii) a $3.5\,\sigma$ line-flux sensitivity limit of $2\times10^{-16}\,\mathrm{erg\,s^{-1}\,cm^{-2}}$ in the EWS at $1.6\,\micron$ (also for \ang{;;0.5} diameter sources). These requirements were established to enable accurate redshift measurements, with $\sigma(z)\leq0.001(1+z)$, for more than 1000 galaxies deg$^{-2}$ in the range $z\in [1, 2]$ via \ha detection \citep{Laureijs11,Pozzetti_2016}. On-sky measurements from Q1 confirmed that the NISP-S sensitivity meets these expectations \citep{Q1-TP007}, while the spectral resolution exceeded the requirements, reaching $\mathcal{R}\simeq500$ for the RGS \citep{Q1-TP006}.
At the end of the mission, the EDS is expected to reach sensitivities of $5 \times 10^{-17}\,\mathrm{erg\, s^{-1}\, cm^{-2}}$ and $6 \times 10^{-17}\,\mathrm{erg\, s^{-1}\, cm^{-2}}$ with the RGS and BGS, respectively \citep[for more details see][]{EuclidSkyOverview}.

In the coming years, \Euclid\ will deliver the largest near-infrared spectroscopic datasets to date. Beyond redshift measurements, the tens of millions of spectra obtained will enable the study of fundamental physical and chemical properties of galaxies through rest-frame optical and near-infrared emission lines across a wide range of redshifts. The main targets of the mission, namely \ha-line emitters over $z\in [0.84, 1.88]$, are observed close to ‘cosmic noon’, a crucial epoch in galaxy evolution when star formation and black hole growth reached their peak \citep{Madau_2014}. This represents a step forward of several orders of magnitude in sample size compared to previous near-infrared spectroscopic surveys, which were limited to a few hundred to a few thousand galaxies at $z \geq 1.3$, such as 3D-HST \citep{vanDokkum_2011}, FMOS-COSMOS \citep{Silverman_2015}, MOSDEF \citep{Kriek_2015}, KMOS$^{3\rm D}$ \citep{Wisnioski_2019}, and recent NIRSpec/JWST campaigns \citep{Neufeld_2024}.
Combined with stellar masses ($M_*$) and other photometrically-derived physical properties \citep{Q1-TP005, EP-Enia}, these measurements will enable the study of three key scaling relations for emission-line galaxies, as described below.

The luminosity of recombination lines, such as \ha and Pa\,$\beta$, and of forbidden lines, such as the [\ion{O}{ii}]$\lambda\lambda$3727,3729 doublet (hereafter, [\ion{O}{ii}]), are direct tracers of the instantaneous star formation rate (SFR), re-emitting the ionising radiation from young, massive stellar populations with lifetimes shorter than 20\,Myr \citep{Kennicutt_1998}. The correlation between SFR and $M_*$ defines the so-called star-forming main sequence \citep[MS;][]{Noeske_2007, Daddi_2007a, Elbaz_2007, Speagle_2014, Popesso_2023}, which exhibits a nearly constant slope and small intrinsic scatter across cosmic time, suggesting that most galaxies build up their stellar mass through smooth and continuous gas accretion rather than through stochastic merger-driven starbursts \citep{Rodighiero_2011}. Its normalisation increases with redshift, reflecting the higher star formation efficiency at earlier epochs. Whilst the redshift evolution of the MS has been extensively studied up to $z = 5$ using photometric SFR indicators, including rest-frame UV \citep{Wuyts_2011, Nordon_2013, Pannella_2015} and far-infrared luminosity \citep{Daddi_2007a, Elbaz_2007, Schreiber_2015}, its evolution as traced through nebular emission lines remains comparatively less well explored.

The second scaling relation accessible with \Euclid\ spectroscopy is the dust–mass relation (DMR), linking $M_*$ with dust attenuation in nebular regions, as traced by the Balmer decrement, that is the deviation of the observed F(\ha)/F(\hb) flux ratio from its theoretical case~B recombination value \citep{Osterbrock_Ferland_2006}. Dust absorbs ultraviolet and optical radiation from young stars and re-emits it in the infrared \citep{Draine_2003}, thereby dimming both the stellar continuum and nebular emission lines. In the local Universe, emission lines suffer roughly twice the attenuation experienced by the stellar continuum \citep{Fanelli_1988ApJ, Calzetti_1994, Calzetti_2000}, a difference interpreted via a two-component dust model comprising both a diffuse interstellar medium and dense, optically thick stellar birth clouds \citep{Charlot_2000}. Whether the empirical nebular–to–stellar colour excess ratio \citep[$f = 0.44$;][]{Calzetti_1994, Calzetti_2000} holds at high redshift remains debated \citep{ForsterSchreiber_2009, Reddy_2010, Mancini_2011, Kashino_2013, RodriguezMunoz2022}. The DMR has been established in the local Universe \citep{Garn_Best_2010}, recently also on resolved scales \citep{Mailvaganam_2026}. It has been shown to hold to $z \sim 2$ \citep{Dominguez_2013, Whitaker_2017, Shapley_2022, Maheson_2024},  and possibly up to $z = 6.5$ \citep{Shapley_2023}, but previous studies at $z > 1$ have been limited to samples of a few hundred galaxies, owing to the difficulty of simultaneously detecting both \ha and \hb at high redshift. 

The third relation is the mass–metallicity relation (MZR), linking $M_*$ with the gas-phase metallicity $Z$, which traces the abundance of elements heavier than helium, relative to hydrogen, in the interstellar medium. Metallicity is most commonly inferred from strong rest-frame optical emission-line ratios calibrated empirically or theoretically \citep{Kewley_Dopita_2002, Pettini_Pagel_2004, Maiolino_2008, Curti_2017, Shapley_2015}, since the more accurate direct-temperature ($T_{\rm e}$) method is generally restricted to bright, low-redshift targets. The gas-phase metallicity is governed by the baryon cycle, in which enrichment from stellar evolution is counteracted by inflows of pristine gas and outflows driven by stellar and AGN feedback \citep{Fabian_2012, Heckman_2014, King_2015, Heckman_2015}. The MZR has been thoroughly characterised in large local samples \citep{Tremonti_2004} and traced up to $z = 3$ \citep{Erb_2006, Maiolino_2008, Mannucci_2010,  Zahid_2014, Sanders_2018}. %, Cresci_2012, Wuyts_2014, Kashino_2017, Ly_2016, 
A secondary dependence on SFR has been identified \citep{Ellison_2008, Mannucci_2010, Curti_2020}, motivating the fundamental metallicity relation \citep[FMR;][]{Mannucci_2010}, which links $M_*$, $Z$, and SFR, and accounts for much of the scatter in the MZR at low stellar mass as well as its strong redshift evolution. 

In this work, we assess the contribution of \Euclid\ to studying the cosmic evolution of these three scaling relations, using a set of realistic simulations to produce forecasts based exclusively on individual emission-line detections for both the EWS and EDS. The simulations reach the final BGS and RGS depths expected for DR3, at the end of the mission, assuming 15 RGS and 25 BGS visits in the EDS. 
The NISP-S wavelength coverage enables the MS to be studied over a broad redshift range using different emission lines as SFR indicators, while the DMR and MZR are probed over narrower redshift windows, mainly around cosmic noon. However, the relatively shallow line-flux sensitivity limits of the surveys introduce substantial observational biases. One of the main aims of this paper is therefore to quantify their impact on the recovery of the three scaling relations.
This analysis is part of a series of studies, based on the same set of simulations, aimed at illustrating the capabilities and limitations of the \Euclid\ spectroscopic sample for galaxy evolution science. Companion papers address related topics, including forecasts of line-emitter number counts (Euclid Collaboration: Cassata et al., in prep., hereafter EC:C) and investigations of stacking techniques to enable the detection of fainter emission lines beyond the nominal survey flux limits \citep[][hereafter \citetalias{EP-Quai}]{EP-Quai}.

The structure of the paper is as follows. In Sect.~\ref{sc:Data}, we describe  the mock catalogue, and the \Euclid\ spectroscopic simulations. 
Section~\ref{sc:Analysis} presents the redshift and flux measurements derived using the official \Euclid\ pipeline applied to the simulated data. 
Forecasts for the MS, DMR, and the MZR are presented in Sect.~\ref{sc:ScalingRel}. A summary and the main conclusions are provided in Sect.~\ref{sc:Conclusions}.
Throughout the paper we adopt the $\Lambda$CDM cosmology with  $\Omega$\textsubscript{m}$\,=\,$ 0.315, $\Omega$\textsubscript{$\Lambda$}$\,=\,$0.685, and $H_0$\,=\,$ 67.3$\,km\,s$^{-1}$\,Mpc$^{-1}$. All magnitudes are expressed in the AB system \citep{Oke_1983}.

\section{\label{sc:Data}  Mock data and spectroscopic simulations} 

\subsection{\label{sc:MAMBO} The MAMBO catalogue}
The forecasts presented in this work are based on a mock catalogue and the corresponding galaxy spectra, constructed using the mocks with abundance matching in Bologna \citep[MAMBO;][Bolzonella et al., in prep.]{Girelli_2021,Lopez-Lopez_2024}. Below, we briefly summarise the main steps followed to build the MAMBO catalogue and spectra. For further details on the generation of the simulated \Euclid dataset we refer the reader to the companion paper by \citetalias{EP-Quai}.

The MAMBO mock catalogue was built from a 3.14 deg\textsuperscript{2} lightcone extracted from \cite{Henriques_2015}, based on the Millennium cosmological dark matter (DM)  $N$-body simulation \citep{Springel_2005}. 
The lightcone spans the redshift range $z\in [0,10]$ and contains DM haloes with $M_{\rm Halo} \geq 10^{10.24}\ M_\odot\ h^{-1}$, where $h$ is the dimensionless Hubble parameter, defined as $h \equiv H_0 / (100 \text{ km s}^{-1} \text{ Mpc}^{-1})$. 
The evolution of the DM structures was rescaled to the \citet{Planck_2014} cosmology following the method of \citet{Angulo_2010}.
The DM haloes and sub-haloes are then associated with the stellar mass of galaxies using the stellar–to–halo mass relation (SHMR) by \cite{Girelli_2021} 
calibrated on observations from the Sloan Digital Sky Survey \citep[SDSS;][]{York_2000}, and the Cosmological Evolution Survey \citep[COSMOS;][]{Scoville2007} field.
The stellar mass function (SMF) in the local Universe \citep{Baldry_2008} combined with that derived by \citet{Ilbert_2013} at higher redshifts was then used to build the galaxy populations associated with the DM haloes. A star-forming (SF) or quiescent (Q) classification was assigned to each galaxy, based on the relative fractions of the SF and Q populations at a given stellar mass and redshift, as derived by \citet{Peng_2010} at $z=0$ and \citet{Ilbert_2013}  at $z\in[0,4]$. The SF/Q fraction at $z>4$ is instead extrapolated from the results at lower redshifts, since the SMF divided by SF/Q type at these redshifts still suffer by large uncertainties.

MAMBO employs a modified version of the Empirical Galaxy Generator code \citep[\egg;][]{Schreiber_2017} to generate key galaxy properties, including SFRs, dust attenuation, metallicities, photometry from the UV to the submillimetre, and emission-line strengths. The \egg framework makes use of empirical scaling relations, such as the MS of star-forming galaxies from \cite{Schreiber_2015} and the FMR from \cite{Curti_2020}, to assign SFRs and metallicities, respectively. Galaxies are then randomly distributed on the two-colour $UVJ$ diagram \citep{Williams_2009} according to their redshift, stellar mass ($M_\ast$), and SF/Q classification. A spectral energy distribution (SED) from the \cite{Bruzual_2003} stellar library is preassigned to the corresponding location on the $UVJ$ diagram, from which the synthetic photometry in the different bands are subsequently derived. The UV luminosity ($L_{\rm UV}$), reddening ($A_V$) and colour excess $E(B-V)_{\rm star}=A_V/R_V$ for stellar continuum, are likewise inherited from the assigned SED, assuming the \citet{Calzetti_2000} attenuation law (with $R_V = 4.05$). Although \egg\ does not explicitly impose a relation between stellar mass and colour excess, such as the one derived from the Balmer decrement \citep{Garn_Best_2010}, a correlation between these quantities naturally emerges from the adopted SEDs. This trend is consistent with that inferred from the mass-dependent infrared excess, $\mathrm{IRX}=\logten(L_{\rm IR}/L_{\rm UV})$, used to derive the total infrared luminosity, $L_{\rm IR}$, despite the two quantities being based on independent empirical prescriptions. We refer the reader to \citet{Schreiber_2017} for further details.

The \ha line luminosity is assigned to each galaxy from its SFR using the \citet{Kennicutt_1998} calibration, while other hydrogen lines are derived from standard line ratios assuming Case B recombination at $T$=$10^4\, \rm K$, following \citet{Osterbrock_Ferland_2006}.
Additional emission lines are assigned based on $M_*$, SFR, and Z, calibrated using SDSS observations and a redshift evolution model\footnote{Details of the MAMBO procedure and the prescriptions adopted to assign fluxes to the various emission lines are available at \url{https://github.com/xalolo/MAMBO}.} \citep{Shapley_2015}. In the mock spectra, Gaussian emission line profiles are simulated using a mass-dependent gas velocity dispersion $\sigma_{\mathrm{gas}}$, as derived by \citet{Bezanson_2018}.  
The number counts of \ha emitters in the MAMBO catalogue have been calibrated using the empirical models of \citet{Pozzetti_2016}, built on the observed \ha luminosity functions up to $z = 2.5$ and over a range of limiting fluxes $F(\rm H\alpha)_{\rm lim}$ from 0.5 to $3\times10^{-16}\, \mathrm{erg\,s^{-1}\,cm^{-2}}$, to provide predictions for the \Euclid\ and WFIRST missions. 

The intrinsic emission line luminosities are also attenuated for dust by applying the extinction law of \cite{Calzetti_2000}. The differential attenuation between nebular emission lines and the stellar continuum is parameterised through a redshift-dependent attenuation factor, following the approach of \citet{Pannella_2015}, as $E(B-V)_{\rm gas}$ = $f\,\, E(B-V)_{\rm star}$, where $f = 1.3$, $2.0$, and $1.5$ at $z = 0$, 1, and 2, respectively. A piecewise linear interpolation is adopted between these redshifts, while $f$ is kept constant at 1.5 for $z\ge 2$. Attenuation due to Galactic extinction is not included in the simulated SEDs.

Starting from this MAMBO mock catalogue, we selected two samples (one for the EWS and one for the EDS) to be simulated according to the following criteria:
\begin{itemize}

\item[$\bullet$]  redshift $z\in [0,4]$, encompassing the range where the emission lines of interest for this study can be detected with \Euclid, based on the wavelength coverage of the NISP-S grisms; 

\item[$\bullet$] SF classification, in the $UVJ$ diagram (SF-$UVJ$), since no emission lines above the NISP-S detection limit are expected in quiescent galaxies;

\item[$\bullet$] \HE\ magnitude below the \Euclid\ survey limits: $\HE \leq 24$ for the EWS, and $\HE \leq 26$ for the EDS;

\item[$\bullet$] at least one emission line with input flux above $2\times 10^{-17}\,\mathrm{erg\, s^{-1}\, cm^{-2}}$ within the NISP-S wavelength range, to limit undetected sources and computational cost\footnote{This threshold is ten and three times fainter than the EWS and EDS detection limits, respectively, and does not affect individual-detection analyses while enabling stacking analyses (see \citetalias{EP-Quai}) down to the faintest \Euclid\ spectroscopic fluxes.};

\item[$\bullet$] no AGN \citep[see][for a version of the MAMBO catalogue that includes AGN]{Lopez-Lopez_2024}.
\end{itemize}
We obtained a sample of \num{255743} galaxies for the EWS and \num{366927}  for the EDS. These two MAMBO-based samples of galaxies, along with their corresponding  SEDs (hereafter also referred as true spectra), were used to  construct realistic  spectroscopic simulations, as detailed in Sect. \ref{sc:FastSpec Simul}.

Constructing the three scaling relations presented in this paper also requires two key photometrically derived  quantities: stellar mass and nebular colour excess. In real data, these  will be estimated from \Euclid\ photometry and ancillary data, ideally  fixing the redshift to the spectroscopic value through ML techniques  \citep{EP-Enia} or SED fitting. Assessing the accuracy of these  estimates would require dedicated photometric simulations, which are beyond the scope of this work. We therefore adopt their true MAMBO values, hereafter $M_{*,\rm true}$ and  $E(B-V)_{\rm true,gas}$, to focus on the spectroscopic recovery of the  scaling relations without introducing additional uncertainties  from photometric estimates. 

\subsection{\label{sc:FastSpec Simul}\fssp\ bypass spectroscopic simulations}
We simulated the \Euclid\ spectra using \fssp\ (de la Torre et al., in prep.), a Python-based tool designed to model the instrumental and environmental effects affecting NISP spectra while bypassing the computationally intensive pixel-level simulations. In the standard \Euclid\ simulation workflow, pixel-level photometric (VIS and NISP-P) and dispersed (NISP-S) images are first generated by the SIM processing function \citep[PF;][]{EP-Serrano}. The photometric products are then processed by the VIS \citep{Q1-TP002} and NIR PFs \citep{Q1-TP003}, whose outputs are used by the MER PF to produce the source catalogue \citep{Q1-TP004}. The dispersed frames and MER catalogue subsequently serve as input to the SIR PF for spectroscopic data reduction, calibration, and processing up to the extraction and combination of the final 1D spectra \citep{Q1-TP006}. As illustrated in the flowchart in Appendix~\ref{apdx:FastSpec_flow}, \fssp\ bypasses this entire chain by directly generating `euclidised' 2D and 1D spectra from an input galaxy catalogue containing source positions, multi-band photometry, morphological parameters describing the surface-brightness profiles, and the corresponding incident spectra.

\fssp\ is significantly faster than the official \Euclid\ pixel-level spectroscopic simulator \citep[see][hereafter \citetalias{Gabarra-EP31}]{Zoubian_2014,Gabarra-EP31}, while retaining key NISP instrumental characteristics, including the point-spread function (PSF), spectral dispersion, and dark-current noise. These are parameterised using median values measured across the NISP focal plane during ground tests \citep{Costille_2019, Maciaszek22}. The astrophysical background is modelled by including environmental noise from zodiacal and stray light.
The simulations are nevertheless simplified with respect to real observations. In particular, they do not include contamination from overlapping spectra, cosmic rays, persistence effects, or spectral orders other than the first (i.e. neither zero-th nor higher orders). 

These simplifications may lead to an overestimation of the expected performance, primarily through an underestimate of spectral contamination. Some results presented here should therefore be regarded as dependent on the adopted simulations, although the analysis methodology (Sect.~\ref{sc:Analysis} and Appendices~\ref{apdx:SPE_flux}--\ref{apdx:Reliability}) can in principle also be applied to real data.
Mutual spectral contamination is expected to reduce the redshift-recovery success rate by about $25\%$ for \ha emitters \citep[cf.,][]{EP-Passalacqua}. However, given the large \Euclid\ survey area, this reduction in sample size should have little impact on the scaling relations presented in Sect.~\ref{sc:ScalingRel}. Despite these limitations, the simulations provide a valuable benchmark for assessing the scientific potential of \Euclid\ spectroscopy and guiding the analysis of future observations. 

\section{\label{sc:Analysis} Data analysis} 
\subsection{\label{sc:SPE redshift and flux} Redshift and flux measurements from SPE pipeline}
The simulated spectra are processed with the standard \Euclid\ pipeline for redshift and flux measurements, namely the spectroscopic processing function \citep[SPE PF;][]{Q1-TP007}. The main results from the SPE PF output catalogue\footnote{For a full description of the SPE PF catalogue columns, see \url{http://st-dm.pages.euclid-sgs.uk/data-product-doc/dm10/spedpd/dpcards/spe_spepfoutputcatalog.html}.} adopted in the following analysis are summarised below.

\subsubsection{Redshifts}\label{sc:redshifts} 
The spectroscopic redshifts are derived by the SPE PF using a modified version of the algorithm for massive automated $z$ evaluation and determination (AMAZED; \citealt{Schmitt_2019}). The fitting procedure returns a redshift probability density function (PDF) and retains up to five candidate redshift solutions, corresponding to the most significant peaks of the PDF. For each solution, a probability score (hereafter $P_{\mathrm{z}}$) is assigned, defined as the integral of the PDF under the corresponding peak within $\pm 3\,\sigma$, and ranging from 0 to 1, from lowest to highest confidence. In this work, we adopt only the best-fitting redshift solution (hereafter $z_{\mathrm{meas}}$).
The SPE PF redshift measurement is designed to maximise the number of \ha detections at cosmic noon, in order to fulfil the requirements of the \Euclid primary science goal of GC \citep{EuclidSkyOverview}. To this end, redshifts in the EWS are estimated using the so-called ‘\ha prior’, whereby single emission line detections are predominantly interpreted as \ha. While this approach enhances the success rate, it also increases the risk of contamination by interlopers at lower or higher redshifts, due to misidentification of [\ion{S}{iii}]$\lambda$9531 or [\ion{O}{ii}] for example, within the redshift interval where \ha falls within the EWS NISP grism range ($z\in[0.84, 1.88]$). 
In contrast, the prior is not applied in the EDS, where the fraction of galaxies with \ha as the only detected emission line is significantly reduced thanks to the broader wavelength coverage and the deeper line-flux sensitivity.

For the simulated data used in this work, the SPE PF also provides a redshift reliability metric ($R_z$). This parameter, ranging from 0 to 1, is a machine learning (ML) based estimate of the robustness of the redshift measurement, with larger values indicating higher reliability  \citep{Jamal_2018}. As reported by \citet{Q1-TP007}, $R_z$ has been extensively tested on simulations and shown to outperform the $P_z$ score in evaluating the robustness of measured redshifts. Its computation, however, was not feasible for the Q1 data release, since training requires a sufficiently large sample of galaxies with secure spectroscopic redshifts. This will only become possible for real data when such a sample will be available in the EDS, possibly starting from DR2. In this paper, we adopt the redshift reliability metric rather than $P_z$  (used, for instance, in~\citetalias{EP-Quai}) in order to assess its performance on our simulations for sample cleaning (see Appendix~\ref{apdx:Reliability}).

\subsubsection{Emission line fluxes}\label{sc:fluxes}
The SPE PF implements two complementary methods for measuring line fluxes. The first, direct integration (DI), estimates the total flux by integrating the signal around the line peak after continuum subtraction. 
The second method, Gaussian fitting (GF), models blended lines with multiple Gaussian components on top of a constant continuum, and is specifically designed to fit the \ha blended with [\ion{N}{ii}]$\lambda6548$ and [\ion{N}{ii}]$\lambda6584$ at the NISP spectral resolution (hereafter \ha--[\ion{N}{ii}]). This method relies on a Levenberg–Marquardt algorithm and provides, for each emission line, flux with relative  uncertainty, integrated S/N, continuum level, full width at half maximum (FWHM), equivalent width (EW), and centroid wavelength. Since our analysis requires the individual deblended fluxes of all components of the \ha–[\ion{N}{ii}] complex, we rely exclusively on the GF measurements for the MAMBO EWS and EDS simulated samples.
The emission lines fitted include: [\ion{O}{ii}], H\,$\epsilon$, H\,$\delta$, H\,$\gamma$, \hb, [\ion{O}{iii}]$\lambda$5008, [\ion{N}{ii}]$\lambda$6548, \ha, [\ion{N}{ii}]$\lambda$6584, [\ion{S}{ii}]$\lambda$6718, [\ion{S}{iii}]$\lambda$9068, [\ion{S}{iii}]$\lambda$9531, Pa\,$\delta$, Pa\,$\gamma$, and Pa\,$\beta$.
In particular, for the forecasts presented in the next sections, we define reliable flux measurements as those with $F(\mathrm{line})/F(\mathrm{line})_{\mathrm{err}} \geq 3$, corresponding to relative uncertainties below $33\%$ on the Gaussian-fit flux. While this criterion ensures acceptable fits, the $F(\mathrm{line})/F(\mathrm{line})_{\rm err}$ ratio differs from the integrated $\mathrm{S/N}(\mathrm{line})$, reported in the SPE PF catalogue, which characterises line detectability. We therefore apply an additional threshold of $\mathrm{S/N}(\mathrm{line}) \geq 3.5$ to reduce contamination without discarding an excessive number of detections. The determination of this S/N threshold is discussed in Appendix~\ref{apdx:SPE_flux}.

Flux-loss corrections are required for the emission-line fluxes derived from the \fssp\ simulations used in this work, since the one-dimensional (1D) spectra are extracted using a simplified procedure based on a fixed summation aperture of five pixels. This approach differs from that adopted by the SIR pipeline \citep[][]{Q1-TP006} for Q1, which employs a variable aperture ranging from 5 to 31 pixels depending on galaxy size. It also differs more markedly from the procedure adopted for the DR1 data release, which includes optimal extraction (Scodeggio et al., in prep.). 
While the use of a fixed aperture does not significantly affect the emission-line fluxes or stellar continuum of galaxies at $z>0.8$, whose small effective radii ($R_{\rm eff}<1''$) do not require apertures larger than five pixels, it can result in greater flux losses than in the SIR pipeline products for the largest sources, predominantly at lower redshift (cf. also \citetalias{Gabarra-EP31}).

These effects are quantified in detail for the main emission lines in the companion paper EC:C. We therefore adopt the multiplicative flux corrections reported in their Table~1: 1.324, 1.380, 1.203, 1.257, and 1.792 for \ha, \hb, [\ion{O}{ii}], [\ion{O}{iii}], and Pa\,$\beta$, respectively\footnote{We note that the EC:C aperture-correction factors for the Paschen and Balmer lines implicitly account, on average, for both aperture losses and underlying stellar absorption, as they are derived by comparing the measured emission-line fluxes with the intrinsic MAMBO values, which do not include stellar absorption.}. The impact is strongest at low redshift, where the larger average galaxy sizes lead to greater flux losses, particularly for near-infrared rest-frame lines such as those of the Paschen series. Figure~\ref{fig:flux} in Appendix \ref{apdx:SPE_flux_zRel} shows that, after applying these corrections, the measured and input emission-line fluxes are in good overall agreement.

\subsection{\label{sc:SPE_redshift} Redshift accuracy}
\begin{figure}[!htbp]
\centering
\includegraphics[width=0.99\linewidth]{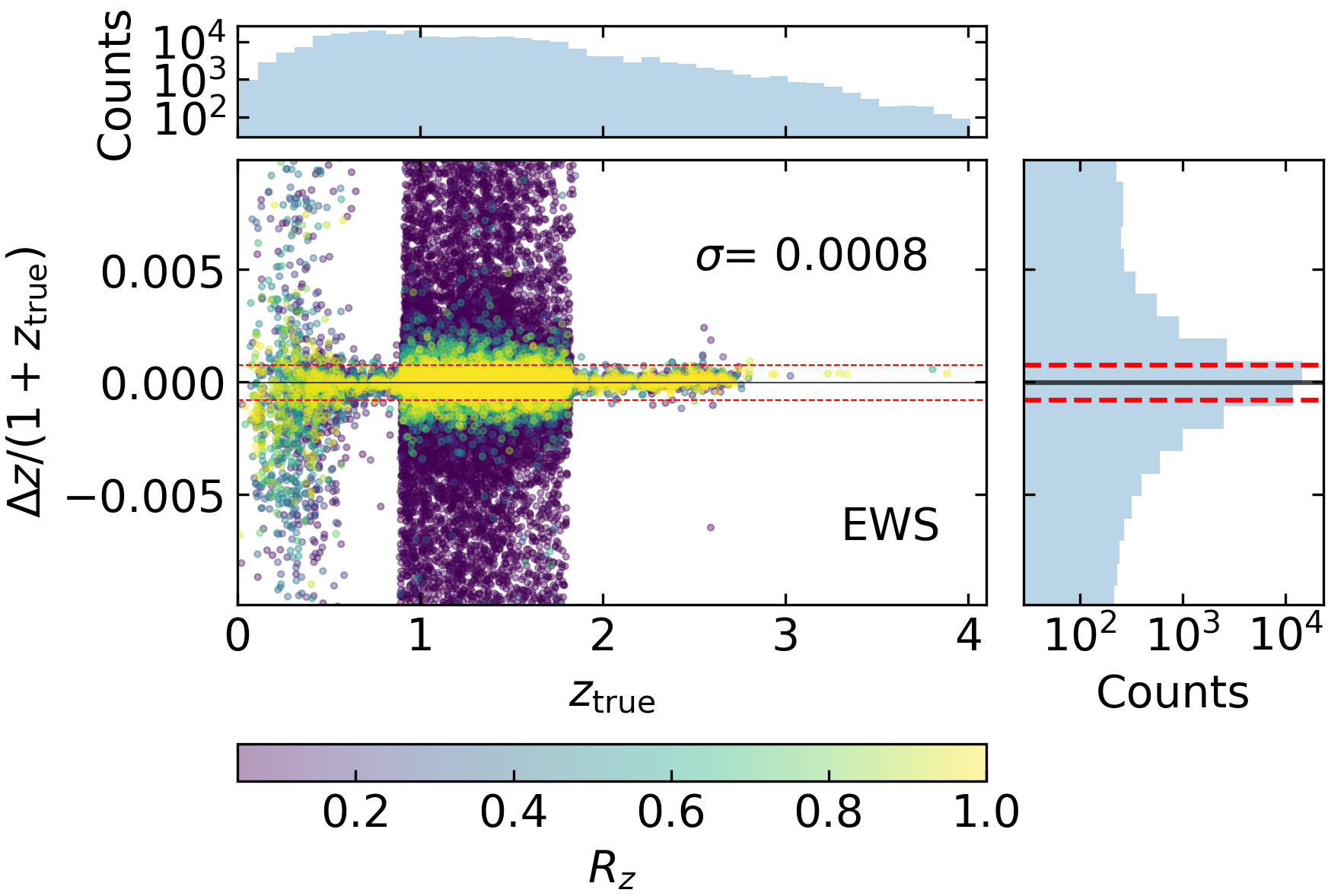}\\
\includegraphics[width=0.99\linewidth]{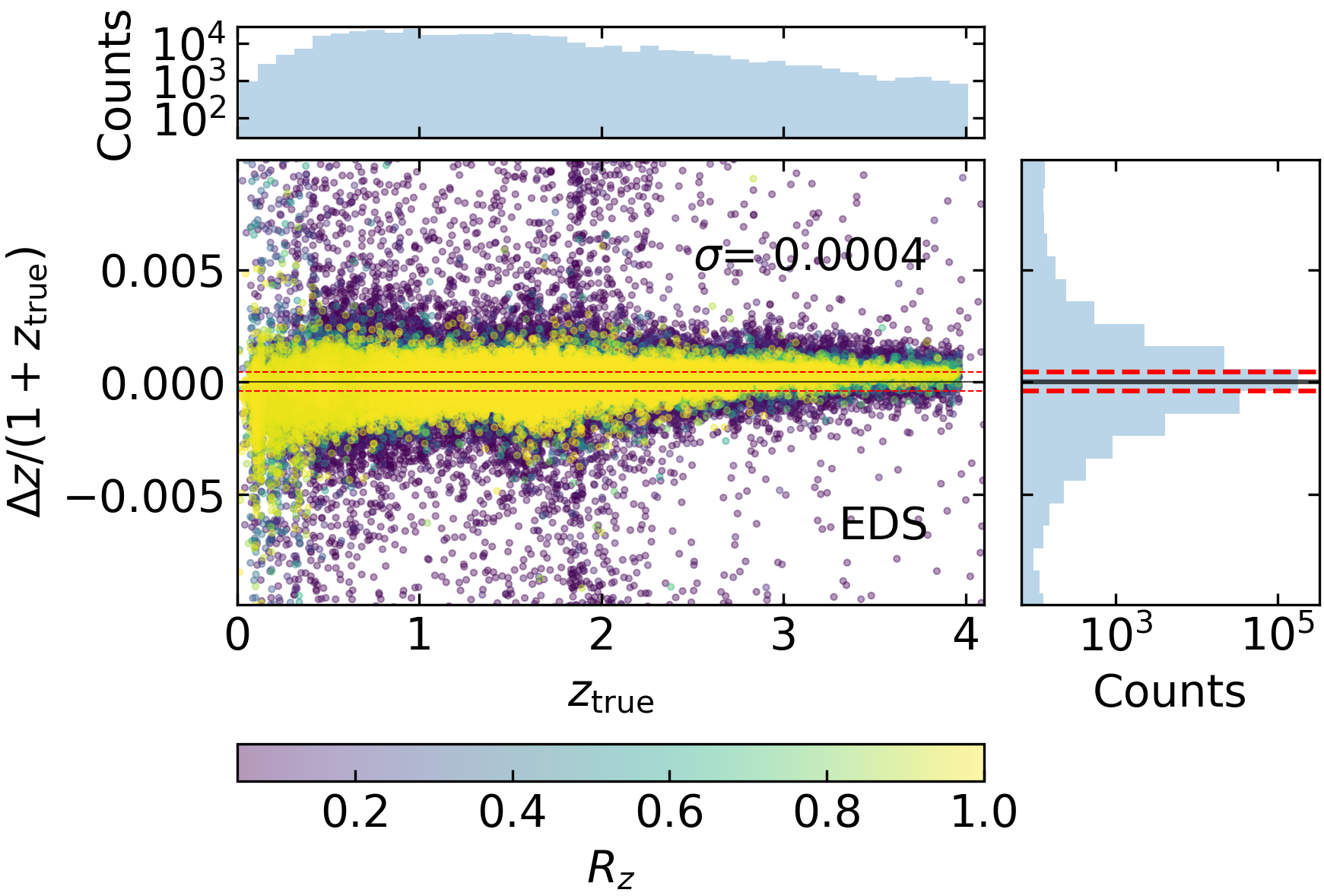}
\caption{Assessment of the accuracy of the best-fit measured redshifts, $z_{\rm meas}$, through comparison with the true input redshifts, $z_{\rm true}$, from the MAMBO simulated data for the EWS (\emph{top}) and EDS (\emph{bottom}). The data are colour-coded by the redshift reliability parameter, $R_z$. The side histograms show the distributions of $z_{\rm true}$ ($x$-axis) and $\Delta z/(1+ z_{\rm true})$ ($y$-axis). The redshift accuracy, $\sigma_z$, as defined in the text, is indicated by the dashed red lines.}
\label{fig:zaccuracy}
\end{figure}

We use our simulations to assess the accuracy of the measured redshifts adopted in the following analysis. Figure~\ref{fig:zaccuracy} shows the redshift accuracy for our simulated samples in the EWS (top panel) and EDS (bottom panel), expressed as
$\Delta z/(1+ z_{\rm true}) \equiv (z_{\rm meas} - z_{\rm true})/(1+ z_{\rm true})$,
as a function of $z_{\rm true}$ and colour-coded by $R_z$, where yellow and light-green points correspond to the most reliable redshifts. For this analysis, we apply no quality cuts on $R_z$. However, we mask catastrophic outliers by excluding objects with $|\Delta z|/(1+ z_{\rm true}) > 0.01$, broadly following the approach adopted for the real Q1 data by \citet{Q1-TP007}. This cut is particularly important for the EWS, as it limits the bias induced by the \ha prior used in the redshift determination, clearly visible in the top panel of Fig.~\ref{fig:zaccuracy}. In this regime, a large number of objects spanning a wide redshift range are misidentified as \ha at $0.84 \leq z_{\rm true} \leq 1.88$ and are associated with low reliability values ($R_{\rm z} \lesssim 0.1$–$0.2$).

The redshift accuracy, $\sigma_z$, is computed as the standard deviation of $\Delta z/(1+ z_{\rm true})$ after applying a $3\,\sigma$ clipping procedure. This yields $\sigma_z = 0.0008$ for the EWS and $\sigma_z = 0.0004$ for the EDS. Throughout the remainder of this paper, a redshift is considered to be correctly recovered if it lies within $3\,\sigma_z$ of the true value, that is,
\begin{align}
\Delta z &\leq 0.002(1+z_{\rm true})\, , 
&& \text{for the EWS}\, , \label{eq:deltaz_EWS} \\
\Delta z &\leq 0.001(1+z_{\rm true})\, ,
&& \text{for the EDS}\, . \label{eq:deltaz_EDS}
\end{align}

\section{Scaling relations from \Euclid line-emitters}\label{sc:ScalingRel}
\begin{figure}[h!]
   \includegraphics[width=\linewidth]{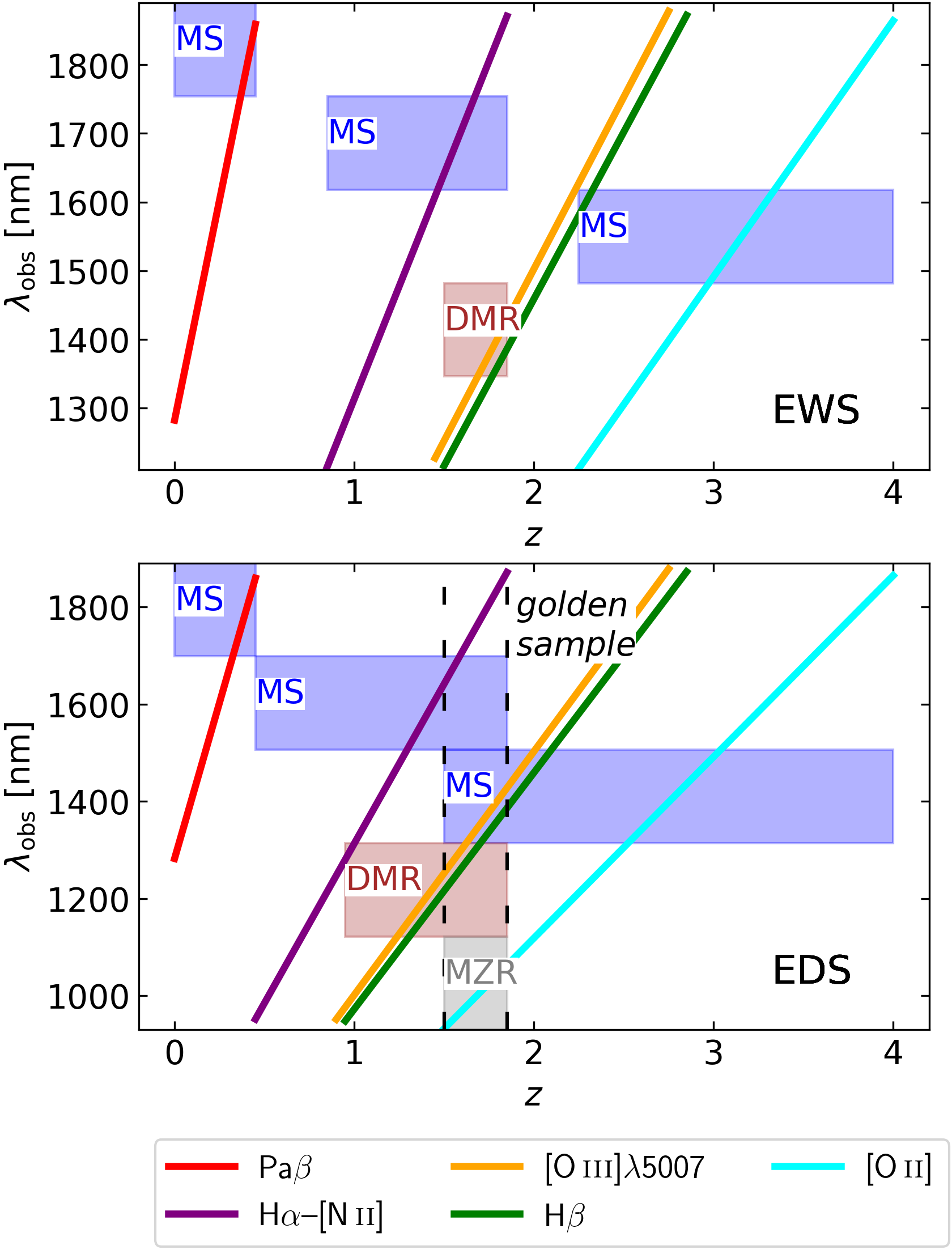}
\caption
{Redshift ranges ($x$-axis) over which the scaling relations analysed in this paper can be derived from emission-line detections (observed wavelengths on the $y$-axes) in the \Euclid\ NISP grism(s), for the EWS (\emph{top}) and EDS (\emph{bottom}). 
The MS (blue stripes) is traced using Pa\,$\beta$, \ha, or [\ion{O}{ii}] as SFR indicators, covering the whole range $z\in [0, 4]$ in the EDS, but only specific intervals in the EWS. 
The DMR (red stripes) is measured in the redshift windows where both \ha and \hb are observed. 
The MZR (grey stripes) is accessible only at $1.49 \leq z \leq 1.88$ in the EDS, where the main optical lines from [\ion{O}{ii}] to the blended \ha--[\ion{N}{ii}]$\lambda$6548,6584 complex are detected. 
The overlap of these ranges defines the golden sample defined by the black dashed lines.}
\label{fig:wl_vs_z}
\end{figure}

The redshift intervals over which the MS, DMR, and MZR can be investigated with \Euclid\ spectroscopy are set by the wavelength coverage of the NISP-S grisms and by the availability of the emission lines required to trace SFR, dust attenuation, and gas-phase metallicity. Figure~\ref{fig:wl_vs_z} summarises these intervals for the RGS coverage in the EWS and the combined RGS+BGS coverage in the EDS. From lower to higher redshift, the relevant emission features are Pa\,$\beta$, the \ha--[\ion{N}{ii}] complex, [\ion{O}{iii}]$\lambda5007$, \hb, and [\ion{O}{ii}].

The MS can be traced over nearly the entire redshift range $z\in[0,4]$ using different emission lines as SFR indicators. In the EWS, Pa\,$\beta$, \ha, and [\ion{O}{ii}] probe $z\leq0.47$, $z\in [0.84, 1.88]$, and $z \in [2.24, 4]$, respectively. The broader EDS wavelength coverage extends the \ha range down to $z=0.42$ and provides simultaneous access to \ha and [\ion{O}{ii}] for $z\in [1.49, 1.88]$.
The DMR requires the simultaneous detection of \ha and \hb to measure the Balmer decrement and can therefore be investigated at $z\in [1.49, 1.88]$ in the EWS and over the wider interval $z\in[0.9, 1.88]$ in the EDS.
Finally, a robust determination of the MZR requires the simultaneous detection of multiple strong emission lines to mitigate metallicity degeneracies \citep{Curti_2020}. The full set of lines from [\ion{O}{ii}] to \ha--[\ion{N}{ii}] is accessible only over $1.49\leq z\leq1.88$ in the EDS. Galaxies detected above the EDS flux limit in this range constitute our ‘golden sample’, providing the line coverage needed to characterise gas-phase metallicity, dust attenuation, and SFR, and hence all three scaling relations considered in this work.

\subsection{\label{sc:MS} Evolution of the star-forming main sequence}
\begin{figure}[h!]    
\includegraphics[width=0.99\linewidth]{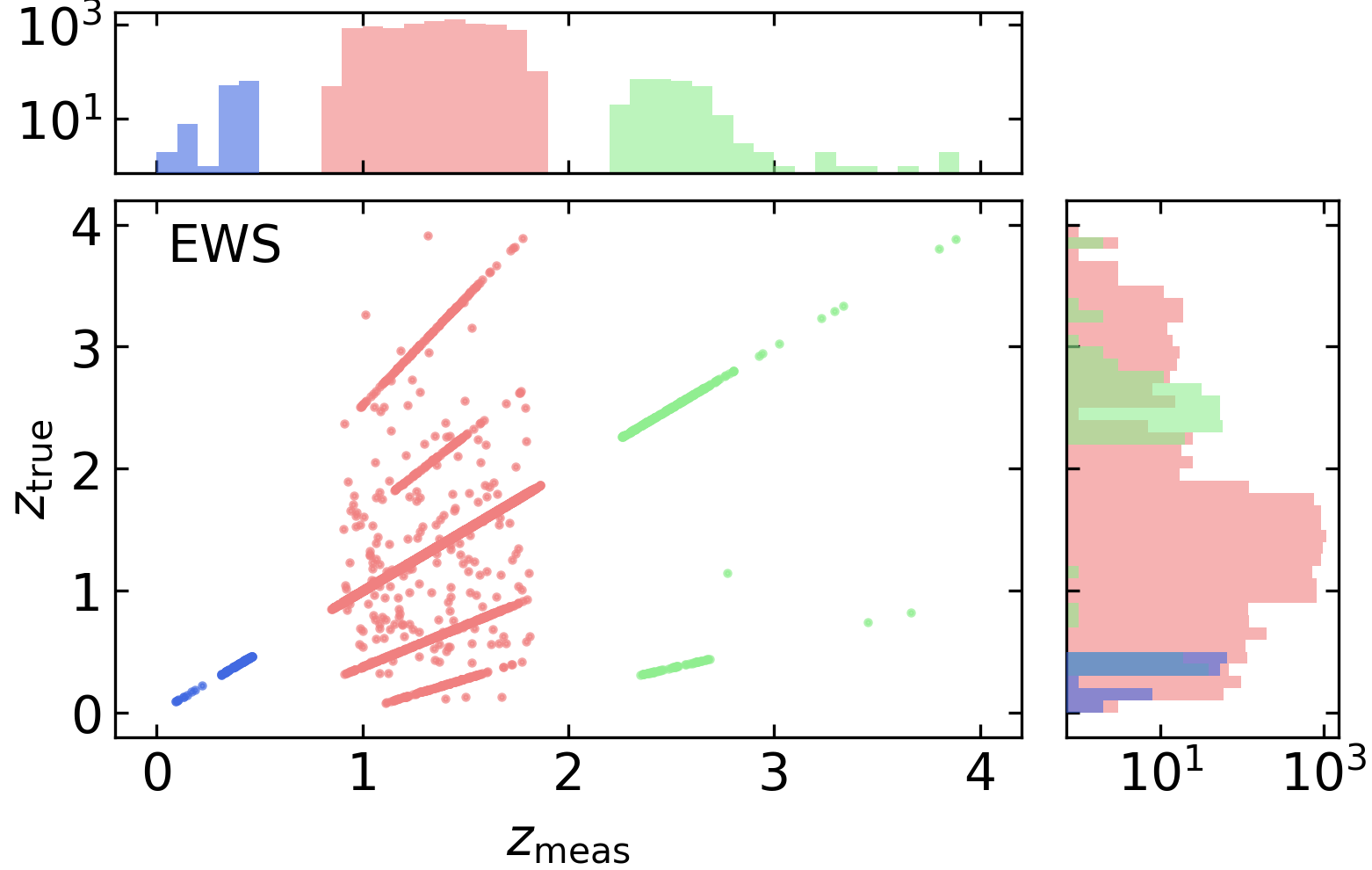}\\
\includegraphics[width=0.99\linewidth]{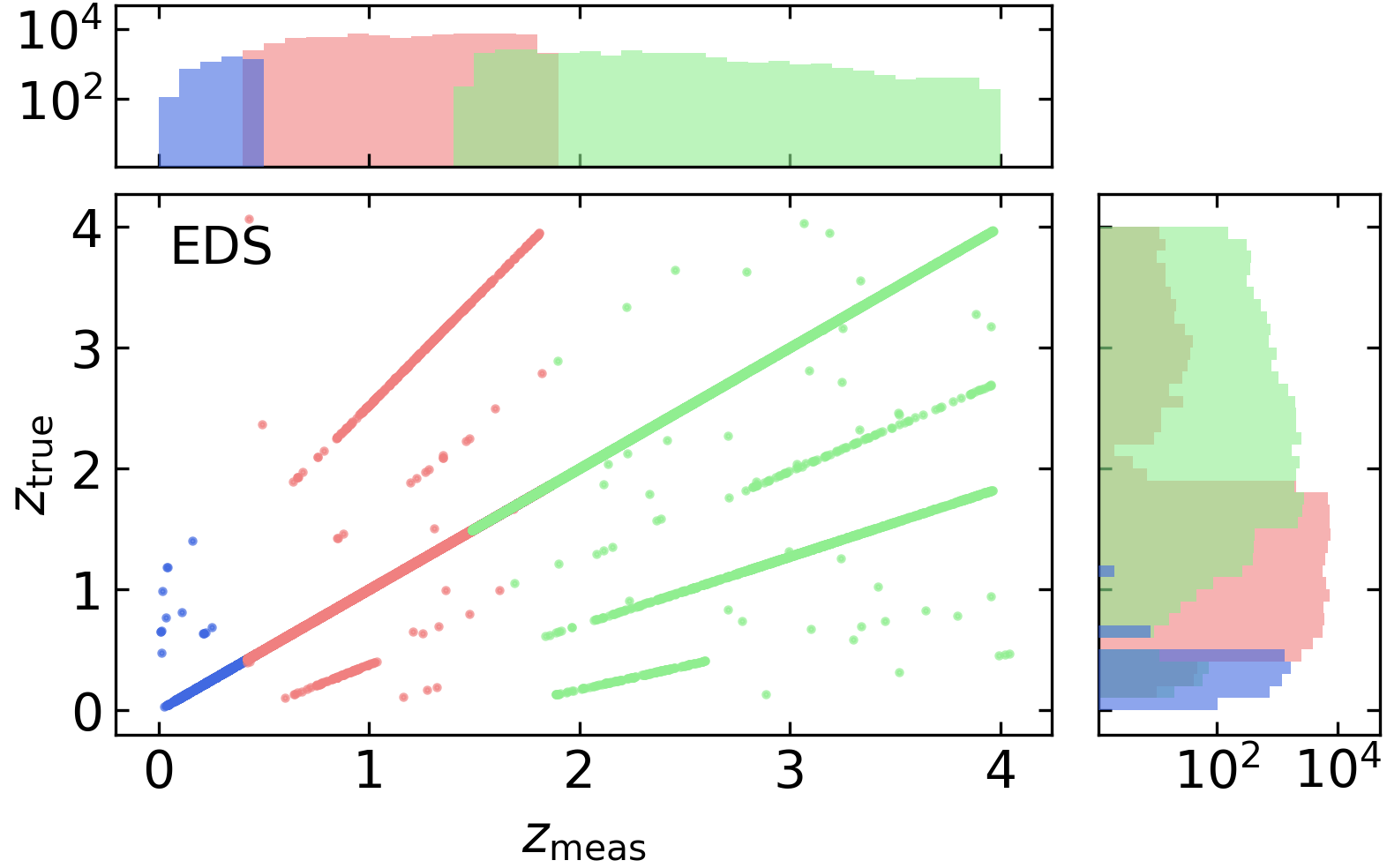}
\caption{Comparison between the measured ($z_{\rm meas}$) and the true input ($z_{\rm true}$) redshifts for the three selected samples of Pa\,$\beta$ (blue), \ha (pink), and [\ion{O}{ii}] (green) emitters in the EWS (\emph{top}) and in the EDS (\emph{bottom}). The top and right histograms show the distributions of the measured and true redshifts, respectively, for the three samples, adopting the same colour coding.}
\label{fig:MS_z_in_out_EWS_EDS}
\end{figure}

To derive the MS of star-forming galaxies, we select Pa\,$\beta$, \ha, and [\ion{O}{ii}] emitters from the simulated star-forming galaxy population (see Sect.~\ref{sc:MAMBO}) based on the following measured quantities:
\begin{itemize}
\item[$\bullet$] measured redshift in the range $z_{\rm meas} \in [z_{\rm min,line},\, z_{\rm max,line}]$, over which the relevant emission line falls within the NISP wavelength coverage  in the EWS and EDS (Fig.~\ref{fig:wl_vs_z}); 

\item [$\bullet$]  line flux relative error and S/N above the thresholds defined in Sect.~\ref{sc:fluxes}, corresponding to well-constrained flux measurements and robust line detections;

\item[$\bullet$] redshift reliability above a given threshold, $R_{\rm z} \geq R_{z, \rm min}$, with $R_{z, \rm min} = 0.2$ for the EWS and $R_{z, \rm min} = 0.15$ for the EDS. These thresholds are chosen and optimised on the simulated data as described in Appendix~\ref{apdx:Reliability}, so to maximise the success rate and minimise contamination across the three emission lines simultaneously.

\end{itemize}
Figure~\ref{fig:MS_z_in_out_EWS_EDS} shows the comparison between input ($z_{\rm true}$) and measured ($z_{\rm meas}$) redshifts for the selected samples of Pa\,$\beta$, \ha, and [\ion{O}{ii}] emitters in the EWS and EDS. By construction, the adopted reliability thresholds result in a relatively low level of residual contamination in the selected samples.
In the EWS, the \ha and [\ion{O}{ii}] samples are the most affected by contamination ($\leq 20\%$, see Appendix~\ref{apdx:Reliability} and EC:C). Most contaminants arise from single-line misidentifications of lower-redshift interlopers, predominantly [\ion{S}{iii}]$\lambda$9531 for both samples, followed by Pa\,$\beta$ in the case of the \ha sample. By contrast, for this sample, the adopted $R_{z, \rm min}$ efficiently suppresses spurious line detections due to noise, reducing their contribution to the total contamination to 2\%. 
In the EDS the highest residual contamination is found in the [\ion{O}{ii}] sample ($\leq 10\%$), mainly due to misidentified \ha emitters, while the fraction of incorrect redshifts caused by spurious detections is negligible.

Depending on the required level of sample purity for specific studies, contamination can be further reduced, at the expense of the success rate, by adopting more stringent reliability thresholds, e.g.  $R_{z, \rm min}=0.9$, as illustrated in Fig.~\ref{fig:reliability} of Appendix \ref{apdx:Reliability}. In this work, we seek the best compromise between minimising contamination and maximising the success rate, to enhance the effective sampling of the $M_*$–SFR relation with \Euclid. Anyway, we verified that adopting progressively higher values of the reliability threshold, up to its maximum, reduces the sample size but does not significantly affect the slope and scatter of the star-forming MS across the full redshift range explored.

\subsubsection{Star formation rates}\label{sc:SFR_and_M}
For the selected samples, the measured Pa\,$\beta$, \ha, and [\ion{O}{ii}] emission-line fluxes are corrected for aperture effects using the correction factors provided by EC:C (see Appendix~\ref{apdx:SPE_flux} for the flux accuracy after aperture correction), and then converted into luminosities using the measured spectroscopic redshift, $z_{\mathrm{meas}}$.
 
We convert the aperture-corrected \ha and Pa\,$\beta$ luminosities into SFRs using the \citet{Kennicutt_1998} calibration, assuming Case B recombination at $T_{\rm e} = 10^{4}\,{\rm K}$ and  $n_{\rm e} =  10^{2}\,{\rm cm}^{-3}$, with an intrinsic line flux ratio $F(\ha)/ F(\rm Pa\, \beta) = 17.6$ \citep{Osterbrock_Ferland_2006}. The corresponding relations are then rescaled by a factor of 1.58 to account for the difference between the \citet{Salpeter_1955} initial mass function (IMF), originally assumed, and the \citet{Chabrier_2003} IMF adopted in MAMBO as

\begin{equation}\label{L2SFR_Ha}
\mathrm{SFR}(\mathrm{H\alpha})\ = \frac{4.98 \times 10^{-42}\ L(\rm H\alpha)}{\mathrm{erg\ s^{-1}}} \ M_\odot\, \mathrm{yr^{-1}}\, ,
\end{equation}

\begin{equation}\label{L2SFR_Pab}
\mathrm{SFR} (\mathrm{ Pa\beta})\ = \frac{ 8.77 \times 10^{-41}\ L(\mathrm{Pa\beta)}}{ \mathrm{erg\ s^{-1}}}\ M_{\odot}\,\mathrm{yr^{-1}}\, .
\end{equation}
For [\ion{O}{ii}], we adopt the \citet{Kewley_2004} conversion, derived from the \citet{Kennicutt_1998} calibration assuming an average $F(\ha)/F([\ion{O}{ii}])$ ratio of $1.2 \pm 0.3$, and also rescaled to the \citet{Chabrier_2003} IMF,

\begin{equation}\label{L2SFR_O2}
\mathrm {SFR}([\ion{O}{ii}])= \frac{4.15 \times 10^{-42}\ L([\ion{O}{ii}])}{ \mathrm{erg\ s^{-1}}}\ M_\odot\, \mathrm{yr^{-1}}\, .
\end{equation}
The luminosities $L(\ha)$, $L(\rm Pa\,\beta)$, and $L([\ion{O}{ii}])$ in Eqs.~(\ref{L2SFR_Ha})--(\ref{L2SFR_O2}) are assumed to be already corrected for dust attenuation, using the \citet{Calzetti_2000} law.

\begin{figure*}[h!]
    \centering \includegraphics[width=\linewidth]{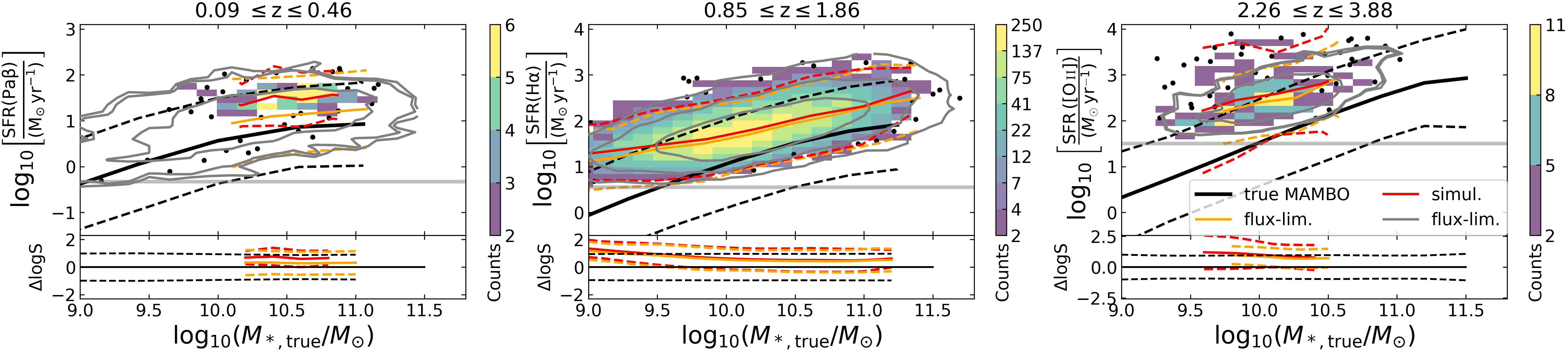}
    \caption{The MS since $z=4$ in the EWS. The SFR derived as described in Sect.~\ref{sc:SFR_and_M} are plotted as a function of the MAMBO stellar masses ($M_{*, \rm true}$) for the three samples of Pa\,$\beta$ (\emph{left}), \ha (\emph{middle}), and [\ion{O}{ii}]  (\emph{right}). The data are colour-coded by number counts (bin size 0.15$\times$0.15 dex) over the full 3.14 deg$^{2}$ MAMBO lightcone area (see side bars). Bins  containing only one galaxy are shown by individual black dots. 
    The black solid and dashed curves are the running median and $\pm 2.5$ sMAD, respectively, computed for the full sample of SF galaxies above the EWS \HE-band detection limit ($\HE \leq 24$) in the MAMBO catalogue. The orange and red solid and dashed curves show the corresponding quantities for the line-flux-limited sample, and the simulated sample. In the bottom panels, the offset between the orange and red median relations compared to the one from the \HE-band-limited sample, $\Delta \mathrm{logS}$ (see definition in Sect.~\ref{sc:results}), 
    illustrates the biases due to the survey line-flux sensitivity limits.}.
    \label{fig:ms_ews}
\end{figure*}

\begin{figure*}[h!]
    \centering \includegraphics[width=\linewidth]{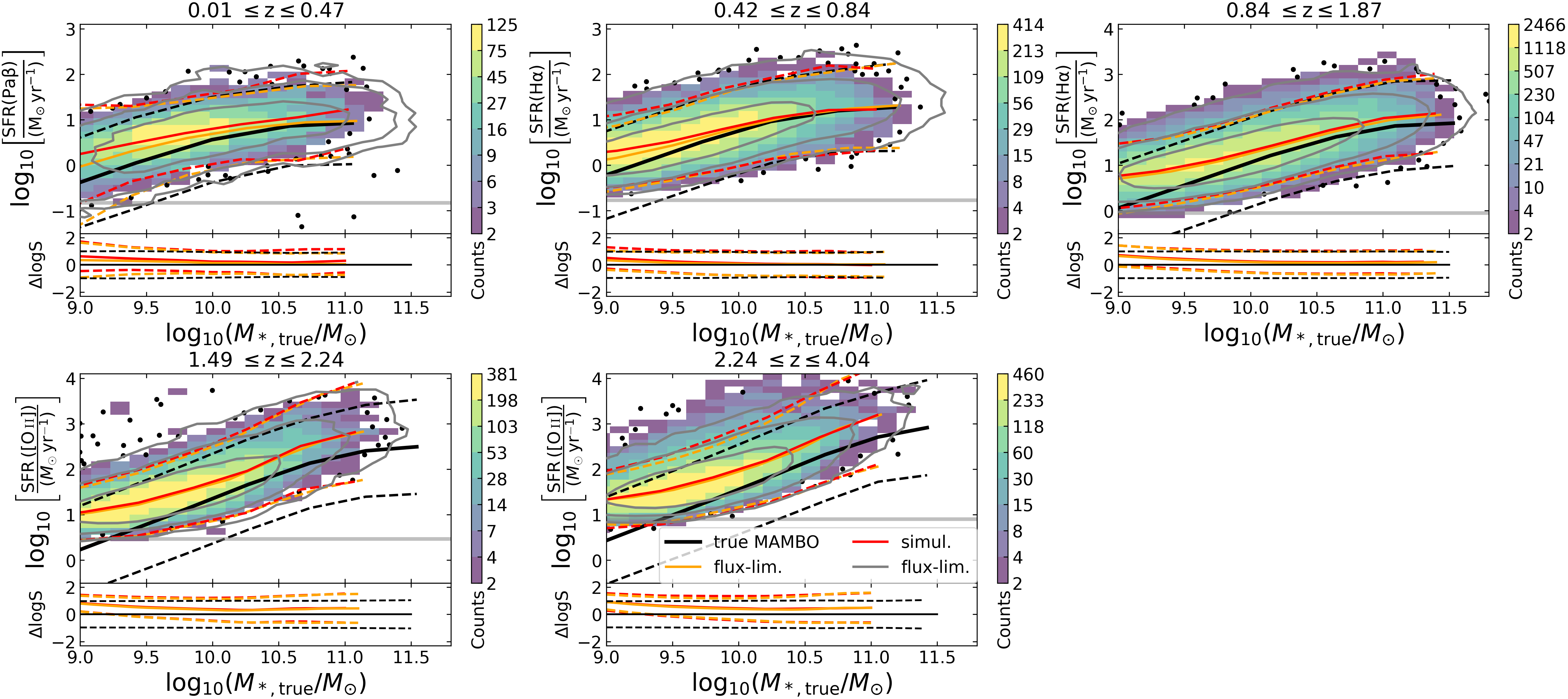}
    \caption{Similarly to Fig.~\ref{fig:ms_ews}, the figure shows the MS since $z=4$ in the EDS. The two additional panels compared to the EWS, display the MS in the redshift ranges where \ha and [\ion{O}{ii}] are detected by the BGS.}
    \label{fig:ms_eds}
\end{figure*}

\subsubsection{Results}\label{sc:results}
 The three panels of Fig.~\ref{fig:ms_ews} show the MS traced by Pa\,$\beta$, \ha, and [\ion{O}{ii}] over the redshift ranges where these lines are detectable with the RGS. Figure~\ref{fig:ms_eds} extends this analysis to the EDS, with two additional panels covering the redshift intervals accessible with the BGS alone, $z\in [0.42, 0.84]$ for \ha and $z\in [1.49, 2.24]$ for [\ion{O}{ii}].
In each panel, the distribution of simulated galaxies in the $M_{*,\rm true}$–$\mathrm{SFR(line)}$ plane, colour-coded by number counts, is shown together with the grey contours of the true, flux-limited MAMBO sample, for which $\mathrm{SFR(line)}$ is derived from the intrinsic line fluxes.
The median relation and the $\pm\,2.5\,\sigma$ robust scatter (where $\sigma$ denotes the scaled median absolute deviation, sMAD) for the simulated and the line-flux-limited 
sample are compared with the intrinsic MAMBO MS derived from the true \HE-band-limited ($\HE \leq 24$) star-forming galaxy sample in the same redshift range. 
The true MAMBO MS follows the parameterisation of \citet{Schreiber_2015}, which includes a high-mass bending that evolves with redshift and nearly vanishes by $z = 2$.
The curves in the bottom panels show the median offset of the simulated and line-flux-limited MAMBO samples, respectively, from the true MAMBO MS, computed  at fixed stellar mass as $ \Delta \mathrm{logS}=\operatorname{median}\!\left[\logten\!\left(\mathrm{SFR}({\rm line})_{\rm sample}/\mathrm{SFR}({\rm line})_{\rm true\ MAMBO}\right)\right]$.

From Fig.~\ref{fig:ms_ews}, it is evident that only very limited regions of the MS can be recovered in the EWS at $z < 0.45$ (from Pa\,$\beta$) and at $z > 1.88$ (from [\ion{O}{ii}]). Based on the survey flux sensitivity limits, we would expect to recover a larger number of Pa\,$\beta$ and [\ion{O}{ii}] emitters than are actually detected in the simulations. For the [\ion{O}{ii}] sample, the discrepancy is relatively small and arises mainly from the \ha prior adopted by the EWS for redshift determination. This causes about 25\% of the detected [\ion{O}{ii}] emitters to be misidentified as \ha emitters, thereby reducing the success rate of the recovered sample.

A similar effect is also present for the Pa\,$\beta$ sample, for which about 10\% of the detected emitters are misidentified as \ha. Since this prior is part of the official \Euclid SPE pipeline, the same behaviour is expected in the analysis of real \Euclid\ data.
The much larger deficit of Pa\,$\beta$ emitters with intrinsic flux above the detection limit is, however, primarily caused by the fixed five-pixel extraction aperture adopted in the \fssp simulations  (Sect.~\ref{sc:FastSpec Simul}). Because low-redshift Pa\,$\beta$ emitters appear more extended than the higher-redshift sources, they are observed at a lower effective spectral resolution, which broadens the emission lines and reduces their peak amplitudes. Combined with the larger flux losses introduced by the fixed extraction aperture, these effects make sources close to the detection limit substantially harder to detect, particularly in the shallower EWS. We therefore expect the resulting deficit of recovered Pa\,$\beta$ emitters to be less pronounced in the real \Euclid\ data.

Furthermore, even for \ha emitters, the MS in the EWS is strongly biased towards galaxies with high SFRs across all stellar masses. The offset between the measured and true relations is $\Delta\mathrm{logS}=0.5$ dex at $\logten(M_{*,\rm true}/M_\odot)\in [10.5, 11]$, increasing to 1 dex at the lowest masses.
This bias produces an overall increase in the MS normalisation and a flattening of its slope at the low-mass end. In contrast, the measured $2.5\,\sigma$ scatter around the median relation (0.33 dex) is comparable with the true MAMBO value (0.38 dex). Similar results were reported by \citet{EP-Scharre} for \ha emitters in the EWS, using the \textsc{Gaea} (GAlaxy Evolution and Assembly) semi-analytic model \citep{Delucia_2014,Hirschmann_2016,Fontanot_2020}. 

In the EDS, the deeper line-flux limit results in a noticeably weaker bias. Indeed, Fig.~\ref{fig:ms_eds} shows that, at least at high stellar masses,  ${\logten (M_{*, \rm true}/M_{\odot})} \geq 10.3$, the MS distribution traced by the \Euclid\  
\ha sample is fully recovered at $z\in [0.42, 0.84]$, with $\Delta \mathrm{logS}\leq 0.05$ dex. 
A larger offset of $\Delta \mathrm{logS}= 0.2$ dex is found for the Pa\,$\beta$ sample, and for the \ha sample at $z\in [0.84,1.88]$, although still substantially reduced compared to the EWS. In all other redshift bins the measured median offset is larger, reaching $\Delta\mathrm{logS} =$ 0.35 and 0.45 dex for the [\ion{O}{ii}] samples at $z\in [1.49, 2.24]$ and at $z\in[2.24, 4]$, respectively. The scatter of the recovered relation remains stable around 0.35 dex across all redshifts, in close agreement with the true value.

\subsection{\label{sc:DM}Dust–mass relation from Balmer decrement}
In this section we present the forecast for the DMR in the redshift intervals where the Balmer decrement can be measured in the EWS and EDS (cf. Fig.~\ref{fig:wl_vs_z}).  The DMR is shown in terms of the measured colour excess of nebular regions, $E(B-V)_{\rm gas, meas}$, as a function of the MAMBO true stellar masses, $M_{*,\rm true}$. 

The colour excess (in mag units) is derived from the Balmer decrement as:

\begin{equation}
E(B-V)_{\rm gas, meas} = \frac{-2.5 }{k_{{\rm H}\beta}-k_{{\rm H}\alpha}}\,\logten \left(\frac{2.87}{F({\rm H}\alpha)_{\rm meas}/F({\rm H}\beta)_{\rm meas}} \right)\, ,
  \label{eq:EBV_from_BD} 
\end{equation}
where $F({\rm H}\alpha)_{\rm meas}/F({\rm H}\beta)_{\rm meas}$ is the measured \ha to \hb line flux ratio already corrected for aperture effects (cf. Sect.~\ref{sc:fluxes}, and Appendix ~\ref{apdx:SPE_flux}), and 2.87 is the intrinsic line flux ratio assuming Case B recombination at a temperature of $T$=$10^4\, \rm K$ and electron density $n_{\rm e} = 10^2\,{\rm cm}^{-3}$ \citep{Osterbrock_Ferland_2006}.
The \citet{Calzetti_2000} extinction law is used to derive the reddening curve $k_{\lambda}$ as a function of wavelength, with $k_{\rm H\alpha}$ and $k_{\rm H\beta}$ indicating its values evaluated at the \ha and \hb wavelengths, respectively. 

Sample selection in the EWS and EDS requires that both \ha and \hb lines are robustly detected and that their fluxes are accurately measured (see Sect.~\ref{sc:fluxes}). We note that no redshift reliability threshold is required for the DMR samples, since the selection criteria used (implying the detection of at least two lines) naturally  yield sources with reliable ($R_{\rm z} \geq 0.7$) and accurate redshifts (see the discussion in  Appendix~\ref{apdx:Reliability}). 

In Fig.~\ref{fig:EBV_in_out}, we present a direct comparison between the nebular colour excess measured from the Balmer decrement, $E(B-V)_{\rm gas, meas}$, and the true values, $E(B-V)_{\rm gas, true}$, for the sample in the EDS. The points are colour-coded by the S/N of the \hb line, which is more challenging to detect than \ha because of its stronger attenuation by dust and underlying stellar absorption.
Due to measurement uncertainties, some galaxies have $F(\mathrm{H}\alpha)_{\rm meas}/F(\mathrm{H}\beta)_{\rm meas}<2.87$, which would formally imply $E(B-V)_{\rm gas,meas}<0$. 
Since negative colour excesses are not physically meaningful, these sources, representing 23\% of the sample, are treated as left-censored measurements using their corresponding $3\,\sigma$ upper limits, shown as downward arrows in Fig.~\ref{fig:EBV_in_out}. 
We find good overall agreement between the measured and true colour excess, with the running median (red solid curve) closely following the one-to-one relation. The running median and percentiles are derived using the non-parametric Reverse Kaplan--Meier estimator, which accounts for left-censored data when reconstructing the underlying distribution \citep{Feigelson_1985}. The median bias, $\Delta E(B-V)_{\rm gas}$, defined as the median difference between measured and true nebular colour excess, remains below 0.03 mag across the explored range, while the corresponding scatter, traced by the 16th and 84th percentiles, ranges from 0.1 to 0.3 mag (red dashed curves).
The errors in the recovered colour excess, estimated by propagating those of the measured \ha and \hb fluxes, are typically $\sigma_{E(B-V)_{\rm gas,meas}}=0.17\,{\rm mag}$, indicating that the measurement bias is small compared with the statistical uncertainty.
Similar results are obtained for this comparison in the EWS, although the statistics are more limited over the MAMBO 3.14 deg$^2$ area, owing to the shallower flux sensitivity limit and the narrower redshift range in which both \ha and \hb can be simultaneously detected.

In Fig.~\ref{fig:DustMass-EWS-EDS}, we compare the DMR derived from the simulations 
with that obtained from the true MAMBO \HE-band-limited and line-flux-limited samples for the EWS and EDS. Upper limits on $E(B-V)_{\rm gas, meas}$ are not shown in this figure, but are included as left-censored data in the derivation of the running median, 16th, and 84th percentiles, as for Fig.~\ref{fig:EBV_in_out}. 
The simulation results closely resemble those of the line-flux-limited samples, in terms of both the running median and scatter, as indicated by the agreement between the red and orange curves in the right panels. However, relative to the median DMR of the true MAMBO \HE-band-limited samples, the \Euclid\ samples are biased towards less obscured galaxies at all stellar masses. This bias is primarily driven by the difficulty of detecting the faint \hb line in heavily obscured sources and is particularly severe in the EWS, where only a few tens of galaxies per square degree have reliable Balmer decrement measurements (see EC:C), mostly at intermediate stellar masses 
and with low dust attenuation. The deeper EDS flux limit provides a more complete sampling of the true MAMBO DMR, extending the coverage to a wider range of stellar masses and nebular colour excess values and yielding approximately a thousand galaxies per square degree.

On the other hand, the median DMR derived from the true MAMBO \HE-band-limited sample appears steeper than the DMR relation reported in the literature by \citet[][blue dashed curve in Fig.~\ref{fig:DustMass-EWS-EDS}]{Garn_Best_2010}, which instead lies in the region of the plane populated by our detected simulated galaxies. This difference arises because the relation between stellar mass and colour excess derived from the Balmer decrement is not explicitly imposed on the MAMBO sample (see Sect.~\ref{sc:MAMBO}). The steeper relation suggests that the nebular dust attenuation may be overestimated in MAMBO. 
Overall, these results indicate that the DMR may be recoverable in the EDS, at least for star-forming galaxies in the less obscured region of the $E(B-V)_{\rm gas}$–$M_*$ plane, close to the \citet{Garn_Best_2010} relation.

\begin{figure}[!htbp]
\centering
\includegraphics[width=0.85\linewidth]{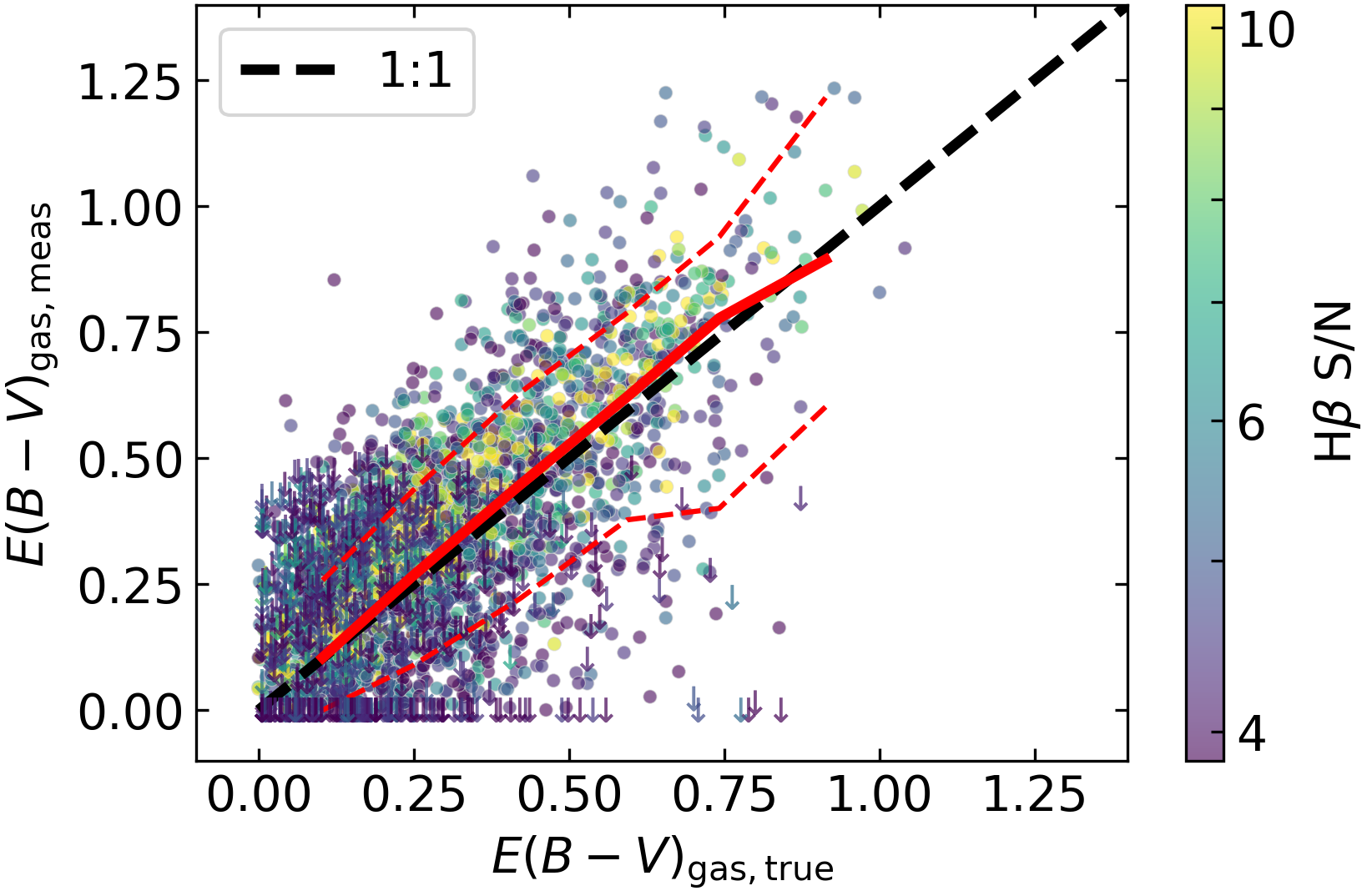}
\caption{The filled dots show the comparison between true colour excess from MAMBO, $E(B-V)_{\rm gas, true}$, and measured values from simulated spectra, $E(B-V)_{\rm gas, meas}$. Downwards arrows are $3\,\sigma$ upper limits on galaxies for which the measurement Balmer decrement is below the theoretical value. Symbols are colour-coded by \hb S/N.  Red solid and dashed lines show the running median and the 16th and 84th percentiles of the distribution (see text for details).} 
\label{fig:EBV_in_out}
\end{figure}

\begin{figure}[h!]
\includegraphics[width=\linewidth]{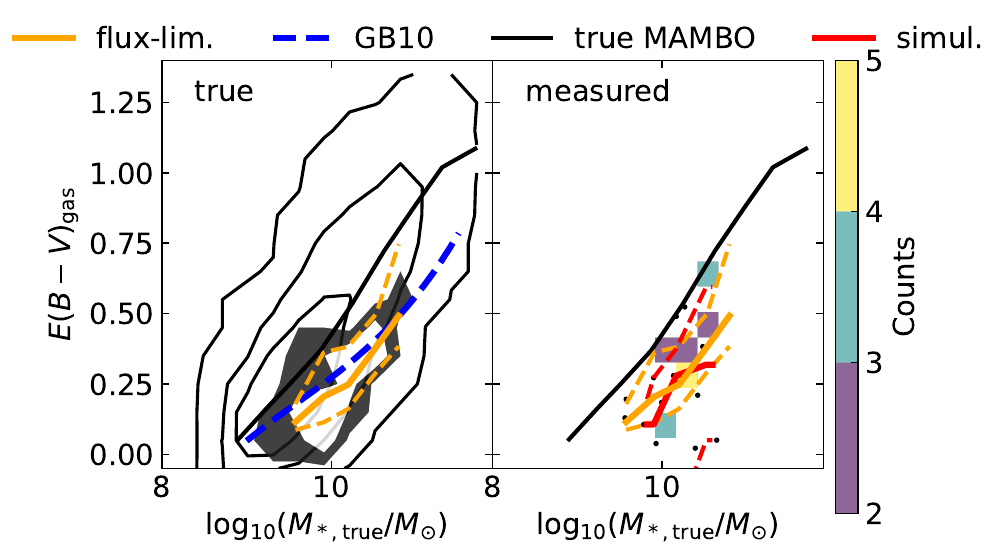}\\
\includegraphics[width=\linewidth]{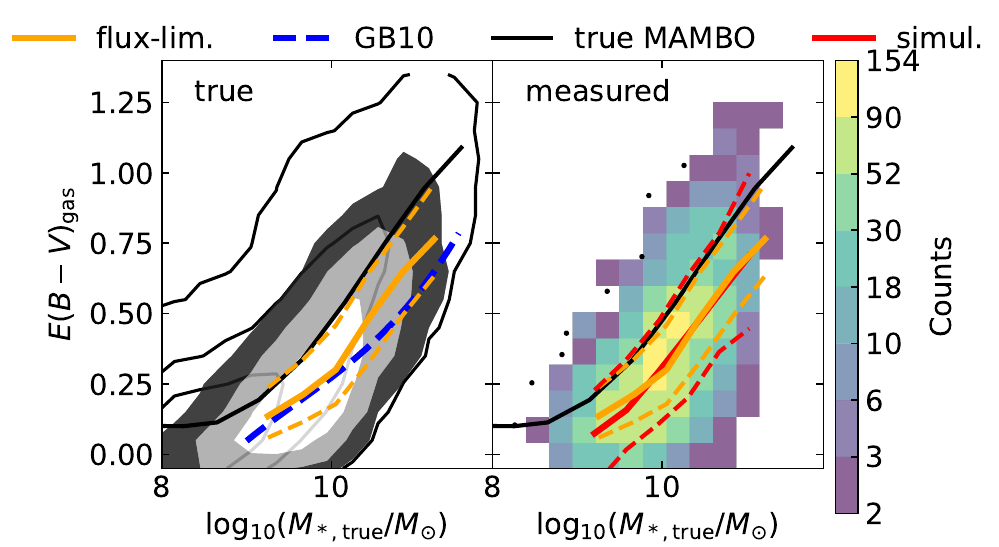}
\caption{Forecasts for the DMR in the EWS at $z\in [1.49, 1.88]$ ({\it top}) and in the EDS at $z\in [0.9, 1.88]$ ({\it bottom}).
\emph{Left panels}: Relation between $M_{*, {\rm true}}$ and $E(B-V)_{\rm gas,true}$ for MAMBO SF galaxies below the \HE-band survey limits, shown by black open contours ($1\,\sigma$, $2\,\sigma$, and $3\,\sigma$), and for the line-flux-limited samples, shown by grey filled contours ($1\,\sigma$ and $2\,\sigma$ for the EWS, and up to $3\,\sigma$ for the EDS). The black, orange solid/dashed, and blue curves show, respectively, the running median DMR of the \HE-band-limited sample, the running median and 16th and 84th percentiles of the line-flux-limited sample, and the \citet{Garn_Best_2010} relation.
\emph{Right panels}: Relation between $M_{*, {\rm true}}$ and $E(B-V)_{\rm gas,meas}$ for the simulated galaxies, colour-coded by number counts (bin size 0.3 dex $\times$ 0.1 mag). The red solid and dashed curves show the running median and corresponding 16th and 84th percentiles, respectively. Bins containing a single galaxy are shown as black dots.}
\label{fig:DustMass-EWS-EDS}
\end{figure}

\subsection{The mass–metallicity relation for a golden sample}\label{sc:golden_sample} 
As introduced in Sect.~\ref{sc:ScalingRel}, a golden sample for galaxy evolution science with \Euclid\ spectroscopy consists of EDS galaxies over $z\in [1.49, 1.88]$ with all key emission lines (from [\ion{O}{ii}] to \ha--[\ion{N}{ii}]) above the flux detection limit. This sample provides an ideal laboratory to assess the capability of \Euclid\ spectroscopy to recover the gas-phase metallicity and the MZR at cosmic noon, enabling the use of multiple strong-line metallicity diagnostics. Moreover, these galaxies can be placed simultaneously on the three scaling relations considered in this work: the MS, traced by \ha and [\ion{O}{ii}], the DMR, and the MZR. The simultaneous detection of \ha and \hb also provides the dust-attenuation corrections needed for SFR estimates and for metallicity diagnostics involving lines widely separated in wavelength. 

The diagnostics most commonly used in both the local and high-redshift Universe are $R_2=F([\ion{O}{ii}])/F(\rm H\beta)$, $R_3=F([\ion{O}{iii}]\lambda5007)/F(\rm H\beta)$, $R_{23}=[F([\ion{O}{ii}])+F([\ion{O}{iii}]\lambda4959,5007)]/F(\rm H\beta)$, and ${\rm O3N2}=R_3/N_2$, where $N_2=F([\ion{N}{ii}]\lambda6584)/F(\rm H\alpha)$. These are typically combined with $O_{32}=F([\ion{O}{iii}]\lambda5007)/F([\ion{O}{ii}])$, which is primarily sensitive to the ionisation parameter, $U$, and helps break the degeneracy between $U$ and metallicity \citep{Dopita_2000, Kewley_Dopita_2002, Shapley_2015, Strom_2017, Kashino_2019b}.
In the MAMBO simulated galaxy spectra, the emission-line fluxes and their relative ratios are, by construction, tied to the gas-phase metallicity through calibrations based on observations and photoionisation models 
\citep[following the \egg\ convention, cf.][]{Osterbrock_Ferland_2006, Maiolino_2008, Shapley_2015, Curti_2017}. 

We select the golden sample based on the measured redshifts, requiring robust detection and accurate flux measurement for all  emission lines.  
As for DMR, no reliability threshold is needed for this sample, as the multiple line detections ensure a highly pure sample (free from contamination) by construction. 
Several hundred galaxies per deg$^2$ meet these criteria (see Table~4 in EC:C). To ensure accurate nebular dust-attenuation corrections for the lines used in the diagnostics, we exclude sources with Balmer decrement measurements below the intrinsic value of 2.87, which further reduces the sample (see Sect.~\ref{sc:DM}).
The gas-phase metallicity of each galaxy is determined by minimising a $\chi^2$ statistic, identifying the metallicity value that best reproduces the observed set of line ratios for the adopted diagnostics \citep[see Eq.~1 in][]{Curti_2020}.  Here, we use $\rm O_{32}$, $\rm R_3$, and $\rm O_3N_2$, which are monotonic with metallicity in MAMBO over the range of interest, $8.1 \leq 12+\logten({\rm O/H}) \leq 8.7$, as shown in Figs.~\ref{fig:R3}--\ref{fig:O3N2} in Appendix~\ref{apdx:lineratios}.  

Figure~\ref{fig:compareZ} compares the true and measured gas-phase metallicities for the detected galaxies, colour-coded by \hb S/N, as \hb is typically the faintest of the lines used in the diagnostics.
Overall, individual metallicities are recovered with a typical precision of about $0.1$ dex over most of the explored range, based on the $\Delta\chi^2=1$ confidence intervals, with the largest deviations occurring at the low-metallicity end. 
The recovered values show a modest systematic bias of up to $0.15$ dex in this regime, which decreases with increasing metallicity and becomes nearly negligible for $12+\logten(\mathrm{O/H})_{\rm true}\geq8.45$. This may result from low-metallicity galaxies tending to have lower stellar masses and fainter emission-line fluxes, leading to larger measurement uncertainties. 

The MZR of the golden sample is shown in Fig.~\ref{fig:MZ_relation}, adopting the same colour coding as in Fig.~\ref{fig:DustMass-EWS-EDS} for the true MAMBO \HE-band-limited sample, the line-flux-limited sample, and the simulated data.
Compared to the \HE-band-limited sample, the golden sample is biased toward galaxies with high SFRs, moderate dust attenuation, and intermediate stellar masses, $\logten (M_*/ M_{\odot}) \in [9.5,10.5]$, primarily due to the requirement that all rest-frame optical emission lines be detected with ${\rm S/N(line)} \geq 3.5$.  Within the framework of a universal FMR \citep{Curti_2020}, this selection favours galaxies with lower metallicity at fixed stellar mass, effectively sampling the upper part of the MS and the lower part of the MZR.
The MZR recovered from the simulated sample (red solid curve) agrees well with the true MAMBO line-flux-limited relation (orange solid curve) in the most populated stellar-mass bin, $\logten(M_*/M_{\odot})\in[10,10.5]$, but shows a $\sim2$ times larger scatter at all stellar masses. As already seen in Fig.~\ref{fig:compareZ}, the median metallicity is overestimated at lower masses.

In conclusion, despite the biases discussed above, the golden sample will enable detailed studies of the interplay between star formation, dust, gas inflows, and chemical enrichment at cosmic noon for intermediate-mass, moderately obscured galaxies.

\begin{figure}[!htbp]
\centering
\includegraphics[width=0.85\linewidth]{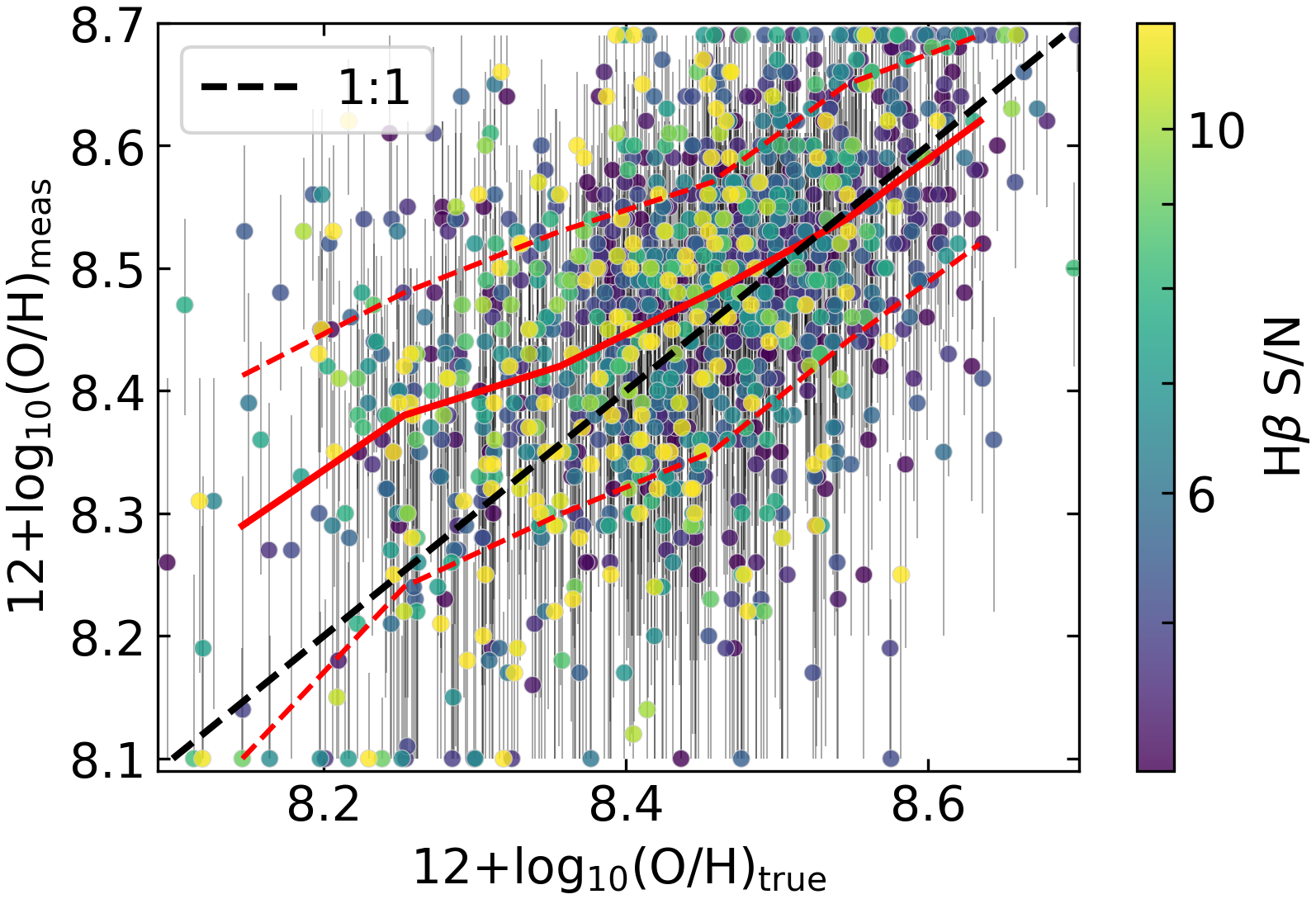}\\
\caption{Comparison between the true and measured gas-phase metallicities. 
Symbols are colour-coded by \hb S/N and plotted from lowest to highest S/N. The black dashed line indicates the 1:1 relation, while the red solid and dashed curves show the running median and corresponding sMAD of the distribution, respectively, computed in bins of $12+\logten(\rm O/H)_{\rm true}$.}\label{fig:compareZ}
\end{figure}

\begin{figure}[!htbp]
\includegraphics[width=0.99\linewidth]{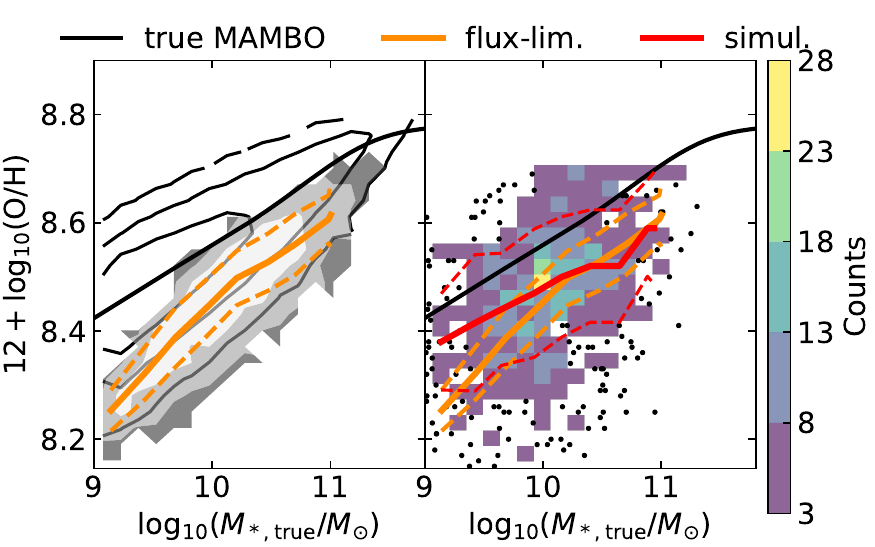}
\caption{\emph{Left}: MZR for the true MAMBO $\HE \leq 26$ sample of SF galaxies at $1.49 \leq z_{\rm true} \leq 1.88$ (black open contours), and for the line-flux-limited sub-sample obtained after applying the flux sensitivity cut (grey-scale filled contours). The black solid curve shows the median MZR of the \HE-band-limited sample, while the orange solid and dashed curves indicate the running median and sMAD of the line-flux-limited distribution.
\emph{Right}: MZR derived from simulations, colour-coded by object number counts (bin size of 0.15 $\times$ 0.025 dex). The red solid and dashed curves indicate the running median and its sMAD, respectively. The black dots indicate galaxies in bins containing fewer than three objects.} \label{fig:MZ_relation}
\end{figure}

\section{\label{sc:Conclusions} Summary and conclusions}
This work forecasts the capability of \Euclid spectroscopic data to trace three key scaling relations in star-forming galaxies, linking stellar mass to star formation rate, dust attenuation, and gas-phase metallicity: the star-forming galaxy main sequence, the dust--mass relation, and the mass--metallicity relation, which we refer throughout the paper as MS, DMR, and MZR. The analysis is based on realistic \fssp\ simulations of the MAMBO mock catalogue over 3.14 deg$^2$, processed with the official SPE PF pipelines used for \Euclid\ data to measure spectroscopic redshifts and emission-line properties.

We assess the redshift accuracy over $z\in[0,4]$, finding $\Delta z\leq0.002(1+z_{\rm true})$ for the EWS and $\Delta z\leq0.001(1+z_{\rm true})$ for the EDS, and define selection criteria for reliable line measurements, requiring relative flux uncertainties $\leq33\%$ and $\rm S/N(line)\geq3.5$. The recovery of the three scaling relations and the biases introduced by the NISP flux sensitivity limits are then examined, with the main results summarised below.

{\bf Star-forming main sequence -- MS}. We use Pa\,$\beta$, \ha, and [\ion{O}{ii}] as SFR tracers to construct the MS over $z \in [0,4]$, within the redshift ranges accessible to NISP in the EWS and EDS. Since most galaxies have redshifts based on a single emission-line detection, we adopt optimised selection criteria to limit contamination while preserving the redshift success rate and sample size, requiring redshift reliability $R_{\rm z}\geq0.2$ in the EWS and $\geq 0.15$ in the EDS. These thresholds keep contamination below 15--20\% and 10\%, respectively.
In the EWS, the shallow flux limit substantially biases the recovered MS, preferentially selecting the least obscured among the most actively star-forming galaxies at intermediate masses. Even the most populated  \ha sample, is biased towards the upper MS at all stellar masses, producing a higher normalisation and an artificially flattened low-mass end relative to the true \HE-band-limited MAMBO relation. Outside the \ha detection range, the \ha prior further reduces the redshift success rate of the Pa\,$\beta$ and [\ion{O}{ii}] samples, with 25\% and 10\% of sources, respectively, misidentified as \ha emitters. Pa\,$\beta$ detections are additionally reduced by the fixed extraction aperture in our simulations, an effect that will not affect real data.
The deeper sensitivity limit of the EDS allows a less biased recovery of the MS. \ha detections provide an almost complete sampling of the true MAMBO MS at $\logten(M_*/M_\odot)\geq10.3$ in the range $z \in [0.42,0.84]$, and show only a mild bias of 0.2 dex at $z\in [0.84,1.88]$. The Pa\,$\beta$ sample at $z\leq0.47$ shows a similar bias of 0.2 dex. At the highest redshifts, the bias increases  because the intrinsically fainter [\ion{O}{ii}] line preferentially selects higher-SFR galaxies at a fixed line-flux sensitivity.

{\bf Dust–mass relation - DMR.} The DMR can be traced through the Balmer decrement in the redshift ranges $z\in[1.49,1.88]$ in the EWS and $z\in[0.9,1.88]$ in the EDS, using galaxies with reliable \ha and \hb measurements. 
The measured nebular colour excess agrees well with the true MAMBO values across the explored $E(B-V)_{\rm gas,true}$ range, with negligible median bias and a scatter of 0.1--0.3 mag. Galaxies with a Balmer decrement below the intrinsic theoretical value, yielding negative $E(B-V)_{\rm gas,meas}$, are treated as $3\,\sigma$ upper limits and account for 23\% of the sample in both surveys. The main limitation is the faint \hb line: the shallow EWS flux limit restricts DMR studies to a small sample of bright, least-obscured galaxies, whereas the deeper EDS provides a more representative sampling across stellar mass and dust attenuation.
In the EDS, the DMR traced by detected galaxies is offset towards lower $E(B-V)_{\rm gas, true}$ at all stellar masses relative to the true MAMBO DMR derived from the \HE-band-limited sample of star-forming galaxies. Its shallower slope, however, is closer to the observed relation reported by \citet{Garn_Best_2010}, suggesting that MAMBO may overestimate dust attenuation. If real galaxies at cosmic noon follow a relation similar to \citet{Garn_Best_2010}, the EDS should therefore recover the DMR over $z\in [0.9, 1.88]$.

{\bf Mass–metallicity relation - MZR.}
 We define the golden sample as EDS galaxies at $z\in [1.49,1.88]$ with reliable measurements of the key emission lines entering the main strong-line metallicity diagnostics, including [\ion{O}{ii}], \hb, [\ion{O}{iii}], \ha, and [\ion{N}{ii}].
Using the $O_{32}$, $R_3$, and O$_3$N$_2$ diagnostics, we show that gas-phase metallicities can be recovered with typical uncertainties of $0.1$ dex for galaxies with $12+\logten(\mathrm{O/H})_{\rm true}\geq8.45$, corresponding to $\logten(M_*/M_{\odot})\geq10$ in the MZR.  Lower metallicities tend to be overestimated by up to $0.15$ dex, possibly due to larger uncertainties affecting faint emission lines in low-mass, low-metallicity galaxies.  The recovered MZR therefore agrees well with the true MAMBO line-flux-limited relation in the most populated intermediate-mass bin, $\logten(M_*/M_{\odot})\in[10,10.5]$, although with 2 times larger scatter, while metallicities are overestimated at lower stellar masses.
Comparison with the true MAMBO $\HE\leq26$ sample highlights the main selection bias affecting the FMR and MZR in the \Euclid\ golden sample:  the EDS line-flux limit preferentially selects the brightest, highest-SFR galaxies, sampling the upper MS and lower FMR and MZR envelopes. Overall, the golden sample provides a valuable probe of galaxy evolution at cosmic noon in intermediate-mass, moderately obscured galaxies, linking star formation, dust, gas inflows, and chemical enrichment.

The forecasts and methodologies developed here provide benchmarks for forthcoming \Euclid\ data, highlighting both the strengths and limitations of the spectroscopic survey. Advanced stacking techniques (\citetalias{EP-Quai}) will help overcome individual detection limits and probe fainter populations, while comparison with survey data will be essential to validate and refine these forecasts for studies of galaxy evolution.
% Add the acknowledgement using the achnowledgements environment.
% Do not use \acknowledgement{....} as this affects the formatting
% of the references.

\begin{acknowledgements}
\AckEC
 The authors acknowledge support from the ELSA project. "ELSA: Euclid Legacy Science Advanced analysis tools" (Grant Agreement no. 101135203) is funded by the European Union. Views and opinions expressed are however those of the author(s) only and do not necessarily reflect those of the European Union or Innovate UK. Neither the European Union nor the granting authority can be held responsible for them. UK participation is funded through the UK Horizon guarantee scheme under Innovate UK grant 10093177.
\end{acknowledgements}
%
% Here comes the reference list, generated via bibtex from
% your bibfile my.bib and Euclid.bib. Please make sure that
% the same paper is not referenced twice, one from your my.bib
% file, and once from Euclid.bib.
%
%\let\oldclearpage\clearpage
%\let\oldnewpage\newpage
%\clearpage
%\clearpage
%\renewcommand{\clearpage}{} % Remove the page break 
%\renewcommand{\newpage}{}   % Remove column/page interruption 
%\bibliography{my, Euclid}
\bibliography{bib_new,Euclid_v1.37}
%\let\oldclearpage\clearpage
%\let\oldnewpage\newpage

% Now you can add appendices.
% In this example, the appendices are in one column mode.
% If that is not requires, comment out \onecolumn
% Note that appendices in A\&A come {\it after\/} the references.
\begin{appendix}

%\onecolumn %If you don't want single column for the Appendix, please
%comment this out
\section{\label{apdx:SPE_flux_zRel} Data analysis methodology}  
\subsection{\label{apdx:FastSpec_flow}\fssp bypass code flow chart}
Figure~\ref{apdx:FastSpec_flow} shows the flowchart of the \fssp\ code, designed to bypass the standard \Euclid\ spectroscopic pipeline chain for simulated data described in Sect.~\ref{sc:FastSpec Simul}. The latter includes the three SIM PF pixel-level simulators (VIS, NISP-P, and NISP-S), followed by the VIS and NIR PFs for photometric data reduction, the MER PF for catalogue creation, and the SIR PF for spectroscopic data reduction and calibration, culminating in the production of the final 2D and 1D spectra.

\begin{figure}[!h]
\centering
    \includegraphics[width=\linewidth]{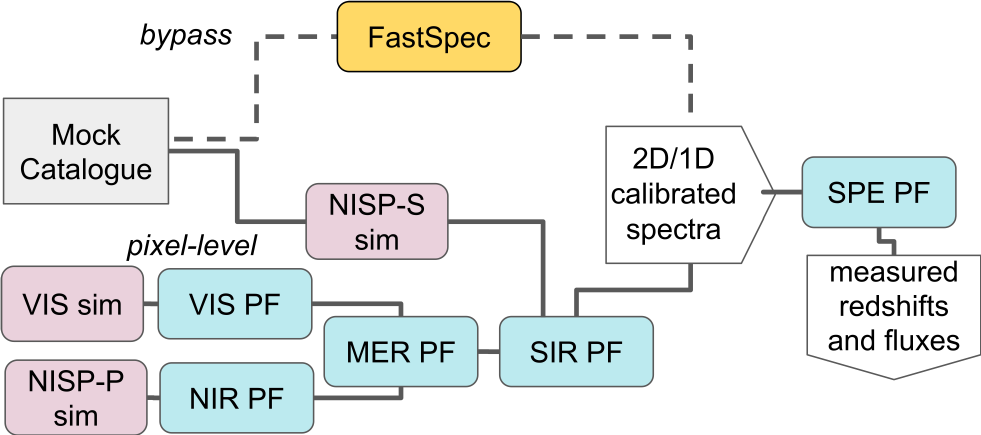}
    \caption{Flowchart illustrating the input and output products of the \fssp bypass code in comparison with the standard \Euclid spectroscopic pipeline chain for pixel-level simulations. The mock catalogue and associated spectra are used as input to generate euclidised 1D spectra, then fed to the SPE PF for redshift and line flux measurements.}
    \label{fig:fastspec}
\end{figure}

\subsection{\label{apdx:SPE_flux}Emission line fluxes and S/N threshold}
In this Appendix, we test the emission-line fluxes and integrated S/N measured from our simulated spectra using the SPE PF pipeline (see Sect.~\ref{sc:Analysis}), with the aim of establishing a S/N threshold for robust line detections, and assessing the accuracy of the flux measurements in our data. In the following, we show the results for the \ha emission line, detected in the RGS over the redshift range $0.84 \leq z \leq 1.88$ (see Fig.~\ref{fig:wl_vs_z}). However, we verified that similar results hold for all the other emission lines considered in this paper.  

\begin{figure}[htbp!]
    \includegraphics[width=0.99\linewidth]{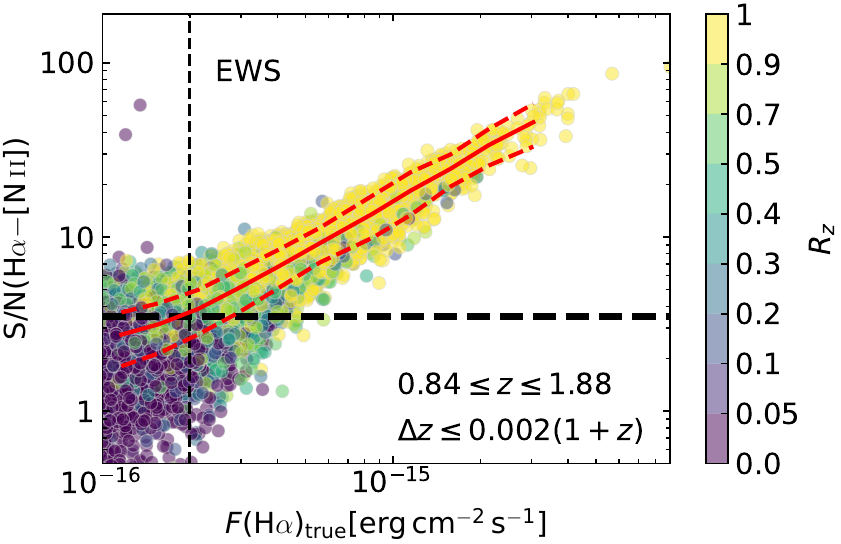}\\
    \includegraphics[width=0.99\linewidth]{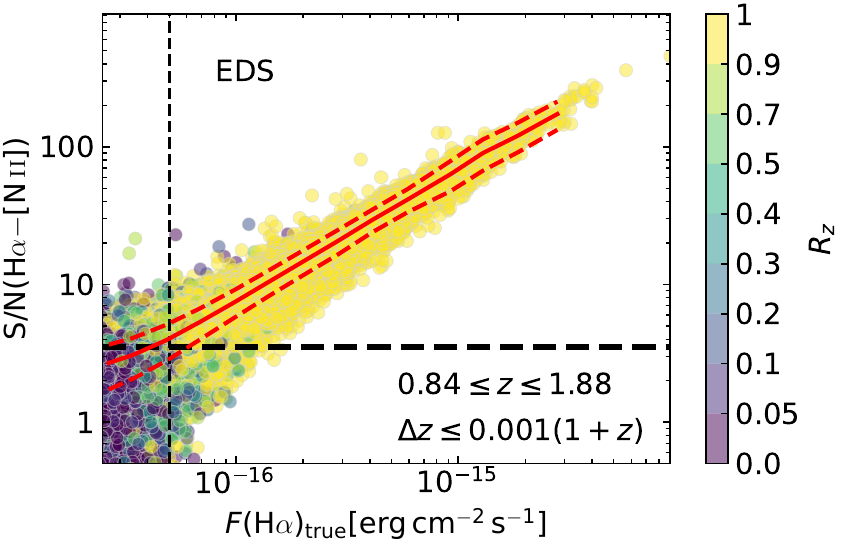}\\

    \caption{The SPE S/N ratio on the \ha-[\ion{N}{ii}] complex is shown as a function of the input \ha line flux, ${F(\rm H\alpha)_{\rm true}}$, for the EWS (\emph{top}) and EDS (\emph{bottom}). The red solid and dashed lines show the running median and sMAD of the distribution, while the black dashed horizontal lines mark the locus with S/N(\ha-[\ion{N}{ii}])\,=\,3.5. In the four panels the data points are colour-coded in $R_z$, and the survey requirement for the $3.5\,\sigma$ line flux limit (see the text) are also shown by thin dashed vertical lines. }
    \label{fig:snr}
\end{figure}

\begin{figure}[ht!] 
\begin{center}
\includegraphics[width=\linewidth]{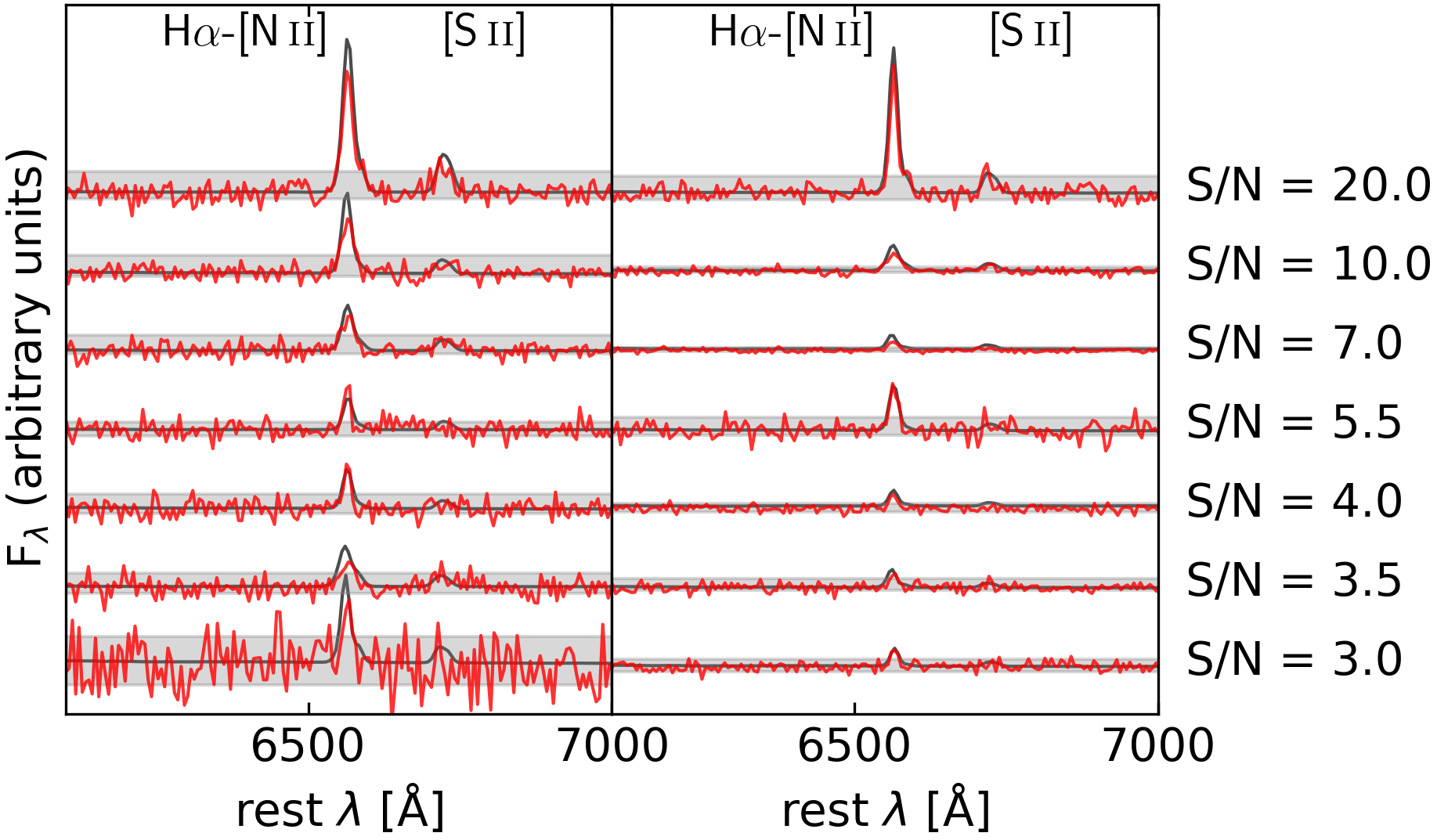}
\caption{A collection of simulated spectra of galaxies at $z= 1.5$, converted to rest-frame, cropped and centered on the region of the  \ha-[\ion{N}{ii}], for the EWS (\emph{left}) and EDS (\emph{right}). The \fssp output spectra are shown in red, and the input MAMBO spectra, degraded to the expected \Euclid resolution, in black. In both panels the S/N on the \ha-[\ion{N}{ii}] complex increases from the bottom to the top, as labelled on the right. The shaded grey area shows the root mean square measured on the continuum.}
\label{fig:spec_examples}
\end{center}
\end{figure}

In Fig.~\ref{fig:snr}, we examine the variation of the SPE integrated S/N measured on the \ha--[\ion{N}{ii}] complex as a function of the input \ha line flux, $F({\rm H\alpha})_{\rm true}$. From these figures, it can be seen that in our simulated EWS an average ${\rm S/N}({\rm H}\alpha\text{--}[\ion{N}{ii}]) = 3.5$ is measured for galaxies with $F(\mathrm{ H\alpha})_{\rm true} = F_{\rm lim}$, the line flux sensitivity limit.
A slightly higher value, ${\rm S/N}({\rm H}\alpha\text{--}[\ion{N}{ii}]) = 4$, is found at the line flux sensitivity limit in the EDS. This is consistent with expectations based on the $3.5\,\sigma$ line-flux sensitivity limits adopted in the simulations (see Sect.~\ref{sc:Intro}), further confirming the reliability of the SPE PF flux and S/N measurements \citep{Q1-TP007}. Based on these results, together with a careful visual inspection of spectra spanning a range of S/N, we adopt a threshold of ${\rm S/N(line)} \geq 3.5$ to identify reliable line detections. Examples of simulated spectra for galaxies at $1.4 \leq z \leq 1.6$, in which the \ha–[\ion{N}{ii}] complex is detected with increasing ${\rm S/N}({\rm H}\alpha$–[\ion{N}{ii}]) in the EWS and EDS, are shown in Fig.~\ref{fig:spec_examples}.

\begin{figure}[htbp!]
\centering
    \includegraphics[width=0.99\linewidth]{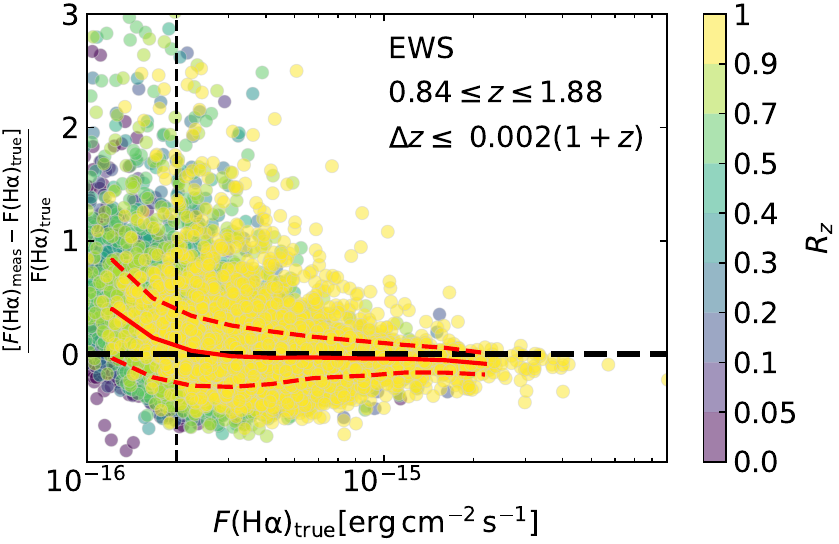}\\

    \includegraphics[width=0.99\linewidth]{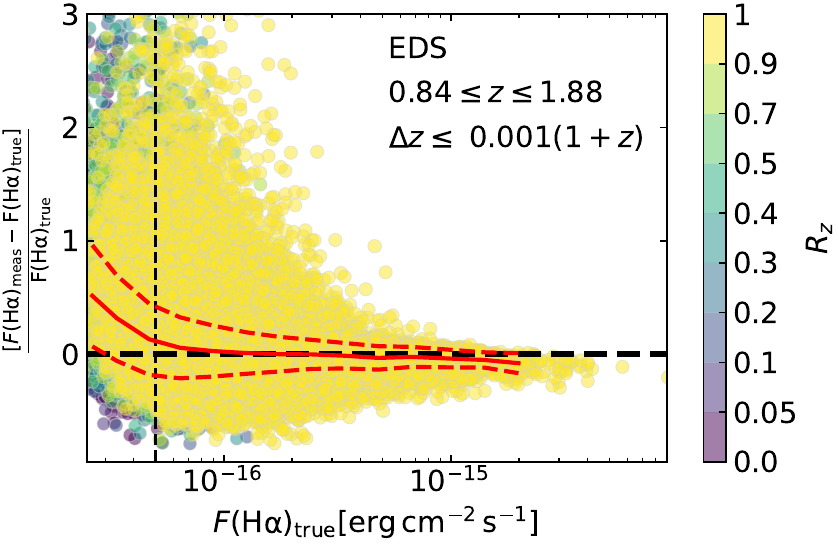}
    \caption{ The comparison between the input and the measured \ha line fluxes after applying the aperture corrections calculated by EC:C, for the EWS (\emph{top}) and EDS (\emph{bottom}). The red solid and dashed lines show the running median and sMAD of the distribution. The dashed horizontal and vertical lines are a visual reference to indicate flux differences equal to zero, and the survey flux senstivity limit.}
    \label{fig:flux}
\end{figure}

 In Fig.~\ref{fig:flux} we evaluate the flux accuracy for \ha emitters, expressed as $F(\mathrm {H\alpha})_{\rm acc} =\left[ F(\mathrm {H\alpha})_{\rm meas}-F(\mathrm {H\alpha})_{\rm true}]\right/F(\mathrm {H\alpha})_{\rm true}$ (relative difference between measured and true input flux), as a function of the input flux.
The sample shown in this figure is selected according to the following criteria: (i) correctly identified redshifts in the RGS, namely satisfying Eq.~(\ref{eq:deltaz_EWS}) or Eq.~(\ref{eq:deltaz_EDS}) in Sect.~\ref{sc:SPE_redshift}, in the EWS and EDS respectively, (ii) robust flux measurements, consistent with $F(\mathrm{H}\alpha)_{\rm meas} / F(\mathrm{H}\alpha)_{\rm meas,err} \geq 3$, and (iii) secure line detections, with $\mathrm{S/N(line)} \geq 3.5$.
The measured fluxes are corrected for aperture effects using the correction factors derived by EC:C, and shown in  their Table~1.  We find good agreement between the aperture-corrected measured fluxes and the input values, with the median flux accuracy, $F({\rm H}\alpha)_{\rm acc}$, centred around zero and a dispersion, quantified by the scaled median absolute deviation (sMAD), increasing from 0.1 for the brightest sources to 0.3 near the survey flux limits (indicated by the vertical dashed lines). Below the survey flux limit, a positive bias in the median flux accuracy is observed. This is due to the combined effect of the simulated instrument characteristics and the selection criteria, which strongly suppress the number of galaxies with measured fluxes below the survey limit, thus leading to an average overestimation of the recovered flux. 

For the flux accuracy of the other emission lines used in this work, we refer the reader to Fig.~1 of the EC:C companion paper, where the aperture-correction factors are estimated. The only differences compared to Fig.~\ref{fig:flux} are that the flux accuracy is shown as a function of measured rather than corrected flux, and the correction factors are not applied.

\begin{figure}[!htbp]
\includegraphics[width=\linewidth]{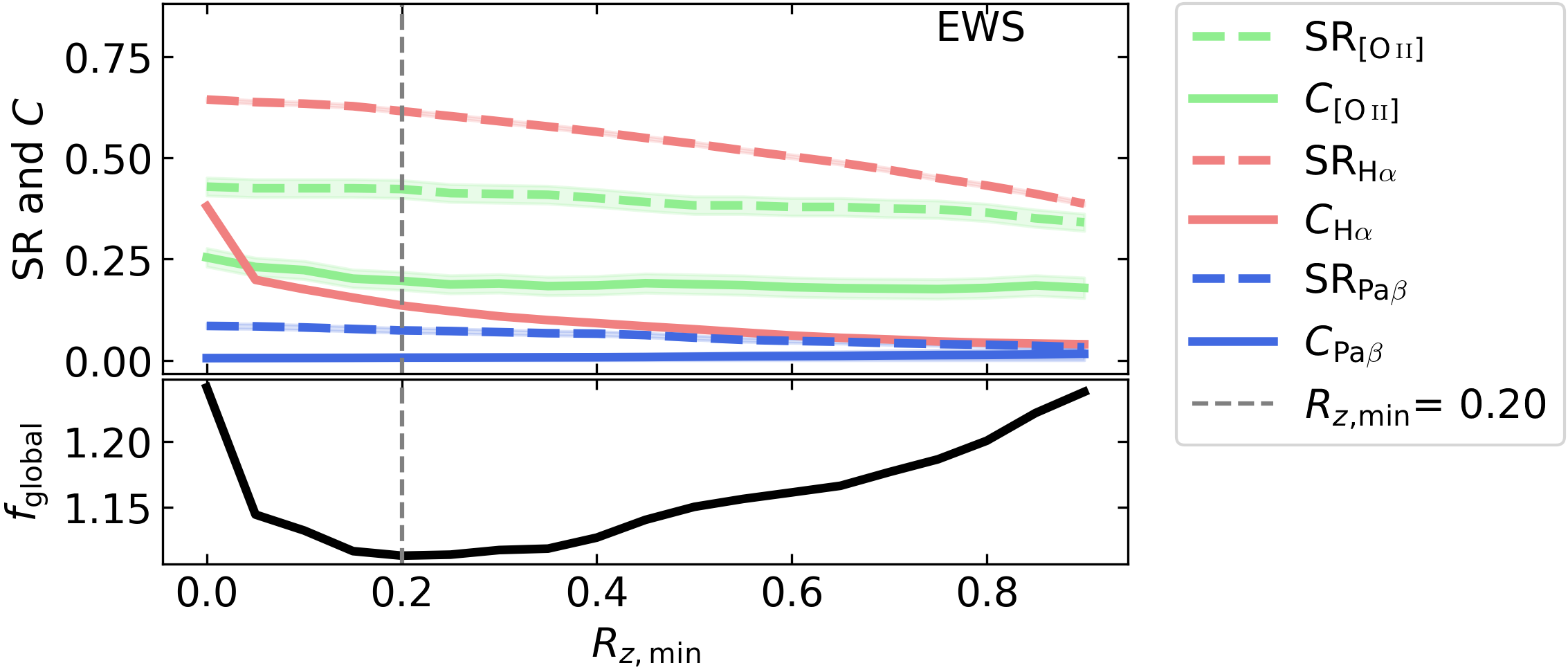}\\
\includegraphics[width=\linewidth]{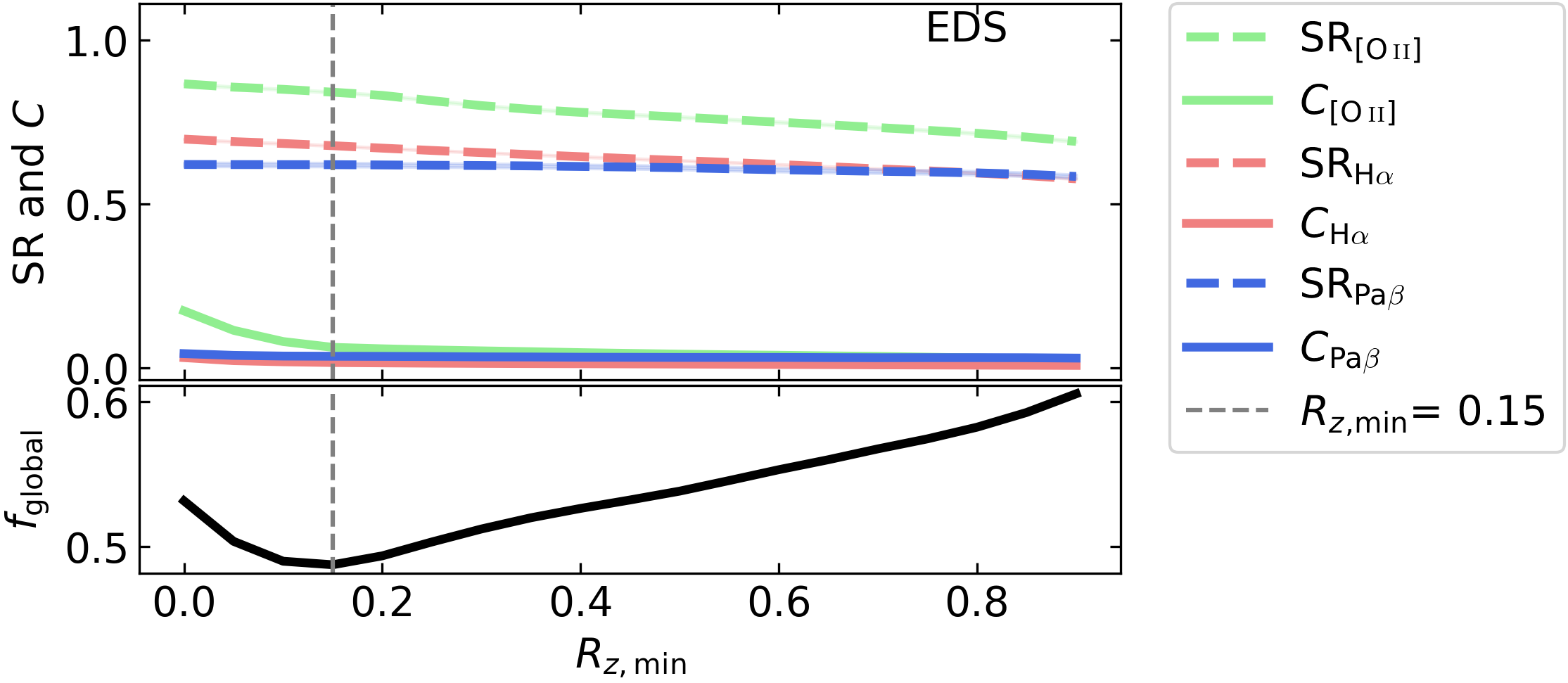}
\caption{Success rate (SR) and contamination ($C$) as a function of $R_{z, \rm min}$ for the EWS (\emph{top}) and EDS (\emph{bottom}), shown for [\ion{O}{ii}] (green), \ha (red), and Pa\,$\beta$ (blue) emitters. Shaded regions indicate binomial standard errors. The lower panels show the global objective functions $f_{\rm global}$ used to optimise the threshold selection, while vertical grey dashed lines mark the adopted optimal values for the reliability thresholds, $R_{z, \rm min}$ values.}
    \label{fig:reliability}
\end{figure}

\subsection{\label{apdx:Reliability} Redshift reliability threshold for sample selection}
The sample of galaxies with measured redshifts includes contaminants at both lower and higher redshifts (see Fig.~\ref{fig:zaccuracy}). By construction, contamination peaks in the redshift range where \ha is expected in the EWS, owing to the \ha prior adopted in the redshift estimation. Sample cleaning is therefore particularly important in this redshift interval. In addition to the $P_z$, the $R_z$ parameter provides a key diagnostic for identifying and rejecting contaminants.
This parameter exploits the features of the redshift probability density function (PDF), through a machine-learning approach designed to reproduce the classification typically performed during expert visual inspection. 
It is worth to note that the $R_z$ evaluation requires a training sample for effective application. It will therefore only be available in \Euclid data once a sufficiently large subsample with known spectroscopic redshifts becomes available in the EDS, starting from Data Release 2 \citep{Q1-TP007}. The choice of the reliability threshold, $R_{z, \rm min}$, may depend on the scientific goals. In particular, the value suggested to derive purity and completeness for the GC probe, set at $R_{z, \rm min}=0.9$, may be too conservative for galaxy evolution studies.

We investigated the impact of the adopted redshift reliability threshold on galaxy scaling relations derived from emission-line properties. Based on extensive analyses as a function of both measured and intrinsic input parameters (including redshift, line flux, and stellar mass), we conclude that the choice of $R_{z, \rm min}$ is particularly critical for samples in which most redshift measurements rely on the detection of a single emission line.
In this work, this primarily concerns the samples used to construct the star-forming MS, where the SFR is derived from the Pa\,$\beta$, \ha, and [\ion{O}{ii}] emission lines at low, intermediate, and high redshifts, respectively (see Sect.~\ref{sc:MS}). In fact, these lines represent the brightest emission features used for redshift determination within their respective observable redshift ranges. 
Accordingly, we focus here on the identification of the redshift reliability threshold adopted for the selection of the Pa\,$\beta$, \ha, and [\ion{O}{ii}] emitter samples.

To this end, we consider two key quantities for each emission line and for each survey (EWS and EDS), namely the success rate and the contamination, defined in  Eq.\,(\ref{eq:SR}) and Eq.\,(\ref{eq:C}), respectively. The success rate ${\rm SR}_{\rm line}$ quantifies the number of line emitters with correctly measured redshifts  
relative to the true number of line emitters 
above the survey detection limit. 
Conversely, the contamination $C_{\rm line}$ measures the fraction of false positives among the identified line emitters with measured redshifts   
and emission-line fluxes robustly detected,
expressed as the complement of the fraction of true positives in Eq.\,(\ref{eq:C}). 
Both quantities are evaluated as a function of a variable redshift reliability threshold, $R_{\rm z}\geq R_{z, \rm min}$, with $0\leq R_{z, \rm min}\leq 0.9$, as

\begin{equation}\label{eq:SR}
\mathrm {SR_{\rm line}}(R_{z, \rm min}) = N_{\rm corr}(R_{z, \rm min})/N_{\rm true}\, ,
\end{equation}
and 
\begin{equation}\label{eq:C}
C_{\rm line} (R_{z, \rm min})= 1-N_{\rm corr}(R_{z, \rm min})/N_{\rm meas}(R_{z, \rm min})\, ,
\end{equation}
where $N_{\rm true}$ is the number of sources with $z_{\rm true}$ in the redshift interval where the line can be revealed by NISP, and with true line flux above the survey 3.5\,$\sigma$ sensitivity limit ($F_{\rm line,true}\geq F_{\rm lim}$). $N_{\rm meas}$ is the number of sources with the line detected with $\mathrm{S/N(line)} \geq 3.5$, and $N_{\rm corr}$, is the subsample of $N_{\rm meas}$ with correctly measured redshifts, satisfying Eq.\,(\ref{eq:deltaz_EWS}) in the EWS or  Eq.\,(\ref{eq:deltaz_EDS}) in the EDS. 

Our goal is to identify, for each survey, the value of $R_{z, \rm min}$ that simultaneously maximises ${\rm SR_{\rm line}}$ and minimises $C_{\rm line}$ for the three emission lines.
To this end, we define a global objective function $f_{\rm global}(R_{z, \rm min})$, to be minimised, which balances the competing trends of ${\rm SR_{\rm line}}$ and $C_{\rm line}$ as a function of $R_{z, \rm min}$ across all three emission lines,

\begin{equation}
f_{\text{global}}(R_{z, \rm min}) = \sum_{\text{lines}} \Bigl\{ \alpha\ [1 - \text{SR}_{\text{line}}(R_{z, \rm min})] + (1-\alpha)\ C_{\text{line}}(R_{z, \rm min}) \Bigr\}\ .
\label{eq:obj_funct}
\end{equation}

\noindent In Eq.~(\ref{eq:obj_funct}), ${\rm SR}_{\text{line}}(R_{z, \rm min})$ and $C_{\text{line}}(R_{z, \rm min})$ denote the success rate and the contamination for each emission line, respectively. The weights $\alpha$ and $(1-\alpha)$ quantify the relative importance assigned to ${\rm SR}_{\text{line}}$ and $C_{\text{line}}$; in this work, we adopt $\alpha = 0.5$, thereby assigning equal weight to success rate and contamination.

In the upper panels of Fig.\,\ref{fig:reliability}, we show the trends of ${\rm SR_{\rm line}}$ (dashed lines) and $C_{\rm line}$ (solid lines) as functions of $R_{z, \rm min}$, with different colours indicating the three line-emitter samples. For each line, both quantities either decrease or remain approximately constant (within the uncertainties) with increasing $R_{z, \rm min}$. This behaviour is expected: increasing $R_{z, \rm min}$ progressively reduces the numerator of ${\rm SR_{\rm line}}$, while its denominator remains fixed. Likewise, $C_{\rm line}$ is expected to approach zero as $R_{z, \rm min}$ increases, since the sample of objects with correctly measured redshifts increasingly overlaps with that of all sources with measured redshifts.\footnote{The $C_{\rm line}$ curve for the Pa\,$\beta$ sample in the EWS should be regarded as effectively flat. The slight increase at high $R_{z, \rm min}$ is driven by the progressive reduction of the sample size (only one source with an incorrect redshift remains at $R_{z, \rm min} = 0.9$) which is reflected in the associated uncertainties.} The lower panels of Fig.\,\ref{fig:reliability} show the global objective function $f_{\rm global}(R_{z, \rm min})$. The optimal redshift reliability thresholds adopted for the selection of the line-emitter samples are identified at the minima of $f_{\rm global}$, corresponding to $R_{z, \rm min}=0.2$ for the EWS, and $R_{z, \rm min}=0.15$ for the EDS.  These thresholds are used throughout this paper for samples based on single emission-line detections, which are employed in the derivation of the MS of SF galaxies (Sect.~\ref{sc:MS}). Sources with multiple detected emission lines, by contrast, typically exhibit a well-constrained, single-peaked redshift probability distribution and are associated with high values of $R_z$. As a result, the choice of $R_{z, \rm min}$ has a negligible impact on these samples. This applies to the samples used to study the DMR and the MZR, which, by construction, include galaxies with at least two detected emission lines.

\section{Effects of photometric stellar mass and dust attenuation on the recovery of the MS}\label{apdx:Lephare}

\begin{figure*}[h!]    
\includegraphics[width=\linewidth]{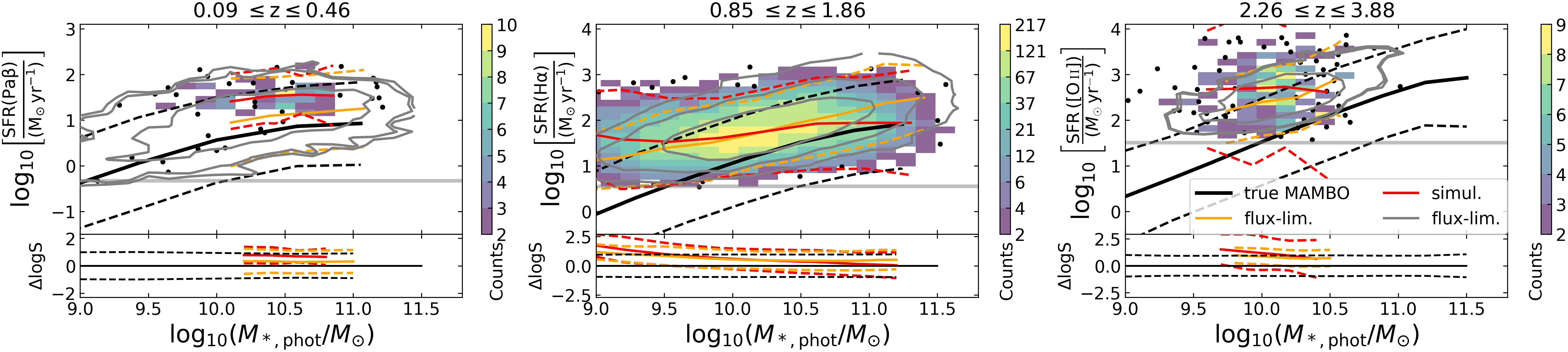}
\caption{Forecasts for the star-forming MS in the EWS, analogous to Fig.~\ref{fig:ms_ews}, but obtained using photometrically derived stellar masses ($M_{*,\mathrm{phot}}$) and dust-corrected SFRs based on photometry, as described in the text.}
\label{fig:MS_phot_EWS}
\end{figure*}

\begin{figure*}[h!]
\includegraphics[width=\linewidth]{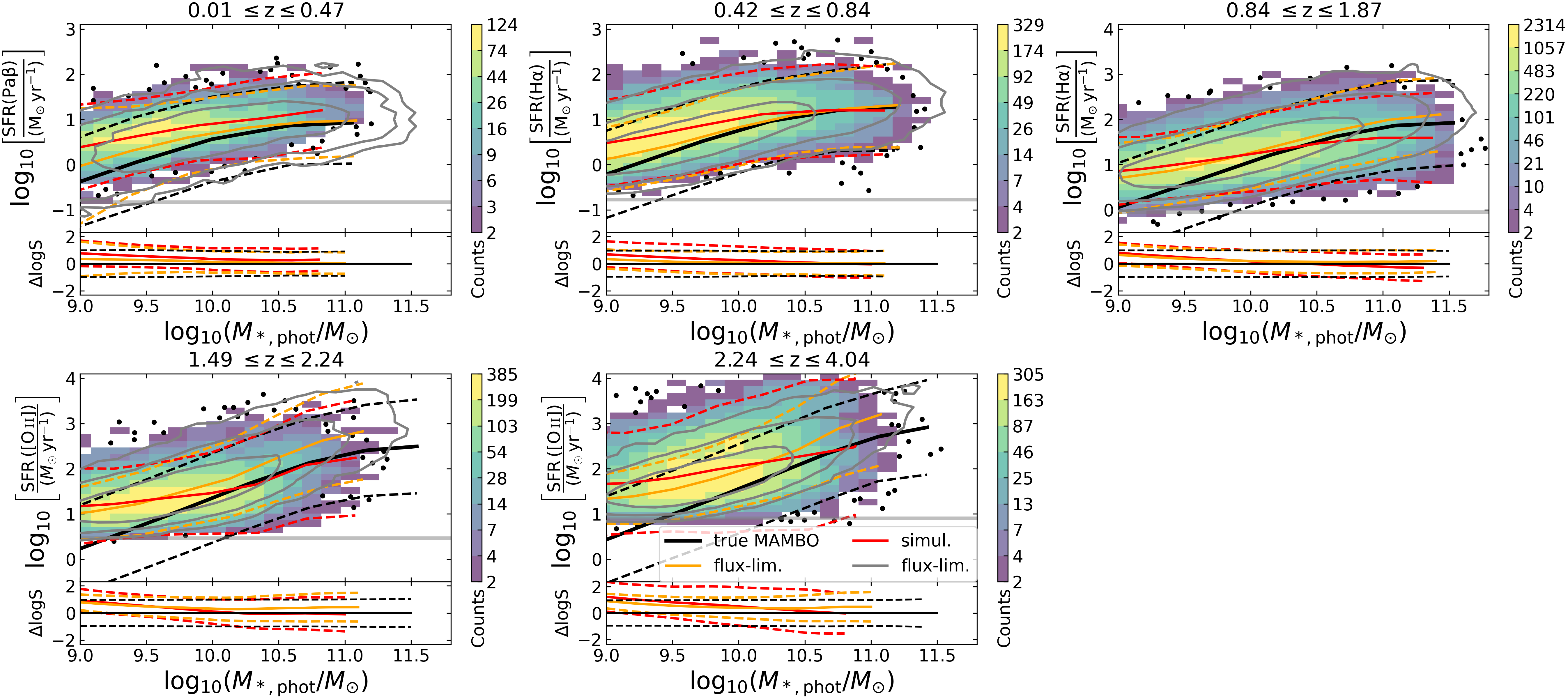}
\caption{Same as Fig.~\ref{fig:MS_phot_EWS}, but for the EDS, to be compared with Fig.~\ref{fig:ms_eds}.}
\label{fig:MS_phot_EDS}
\end{figure*}

The \Euclid\ survey includes near-IR imaging, and benefits from ancillary ground-based optical photometry across the entire survey area. The largest share of ancillary photometric data comes from the Ultraviolet Near-Infrared Optical Northern Survey \citep[UNIONS;][]{Gwyn_2025}, the Dark Energy Survey \citep[DES;][]{Abbott_2016}, and the forthcoming Legacy Survey of Space and Time \citep[LSST;][]{LSST_ScienceBook_2009}. 
The PHZ PF primarily uses these data to derive photometric redshifts, but also to estimate galaxy stellar masses ($M_*$), dust attenuation for the stellar continuum, in terms of the colour excess $E(B-V)_{\rm star, phot}$, and other physical properties.

As mentioned in Sect.~\ref{sc:ScalingRel}, \Euclid\ spectroscopy will only allow us to quantify the dust attenuation in galaxy nebular regions, via the Balmer decrement, within the limited redshift ranges $z\in[1.49,1.88]$ and $z\in [0.9, 1.88]$ in the EWS and EDS, respectively. For galaxies at different redshifts, and in fields where mid-IR/far-IR/radio observations are not available to derive the dust-obscured SFR, one must rely on quantities inferred from the multi-band \Euclid\ and ground-based photometry.

For analyses based on spectroscopic redshifts, an improvement over the direct use of the official PHZ PF catalogues would consist in re-running the SED-fitting (or ML) procedure used to derive stellar masses and dust attenuation while fixing the redshifts to the spectroscopic values.  
Although a full assessment of photometrically derived quantities is beyond the scope of this paper, in this Appendix we perform an exploratory and simplified exercise to evaluate the impact of photometrically derived stellar masses and dust attenuation on the results presented in Sect.~\ref{sc:MS} for the star-forming MS and its redshift evolution. To this end, we simulated the fluxes and their associated uncertainties by perturbing the MAMBO true photometry (assuming Gaussian noise) in both the \Euclid\ and the ground-based ancillary bands, according to the magnitude depths expected for real data. We then performed a SED-fitting analysis with {\it LePhare} \citep{Arnouts_1999, Ilbert_2006}, fixing the redshifts to the values measured by the SPE PF.

The photometrically derived stellar masses ($M_{*,\rm phot}$) are recovered with an accuracy of 0.15 to 0.2 dex from the SED-fitting analysis, with slightly improved performance in the EDS. In contrast, the dust attenuation, expressed as the colour excess $E(B-V)_{\rm star, phot}$, exhibits larger uncertainties in both the EWS and the EDS. When analysed as a function of stellar mass, a mild systematic trend towards underestimating the colour excess emerges at the high-mass end. The median of the offset with respect to the true value, defined as $\Delta E(B-V) = E(B-V)_{\rm star, true} - E(B-V)_{\rm star, phot}$ varies from $-0.15$ to $0.2$ across the $M_{*,\rm true}$ range. The sMAD of $\Delta E(B-V)$ ranges from 0.25 to 0.3. 

The star-forming MS for the EWS and EDS derived using these photometric quantities is shown in Figs.~\ref{fig:MS_phot_EWS} and \ref{fig:MS_phot_EDS}, respectively. The SFRs derived from the Pa\,$\beta$, \ha, and [\ion{O}{ii}] emission lines are corrected for dust attenuation using the nebular, photometrically derived colour excess $E(B-V)_{\rm gas, phot}$, obtained from $E(B-V)_{\rm star, phot}$ by assuming the same redshift-dependent differential attenuation between stellar and nebular emission as adopted in the MAMBO simulations \citep[][see Sect.~\ref{sc:SFR_and_M}]{Calzetti_1994}. The comparison between Figs.~\ref{fig:MS_phot_EWS} and ~\ref{fig:MS_phot_EDS} with Figs.~\ref{fig:ms_ews} and \ref{fig:ms_eds} shows that the photometrically derived $M_{*,\rm phot}$ and $E(B-V)_{\rm gas, phot}$ do not significantly affect the recovery of the average MS trend at any redshift. However, in some redshift bins (e.g. the \ha sample in the RGS at $0.84\leq z\leq 1.88$ in both the EWS and EDS), the MS derived from photometric quantities presents a shallower slope at the high-mass end relative to the relation based on $M_{*, \rm true}$ and $ E(B-V)_{\rm gas, true}$. This behaviour is consistent with the observed systematic underestimation of the photometric colour excess, which leads to weaker dust corrections and consequently lower SFRs at the high-mass end. Although modest, this effect should be taken into account in studies aimed at constraining the high-mass slope of the MS and assessing its possible bending.

\section{Strong-line ratios used to derive the Mass-metallicity relation}\label{apdx:lineratios}
\begin{figure}[!htbp]
\includegraphics[width=\linewidth]{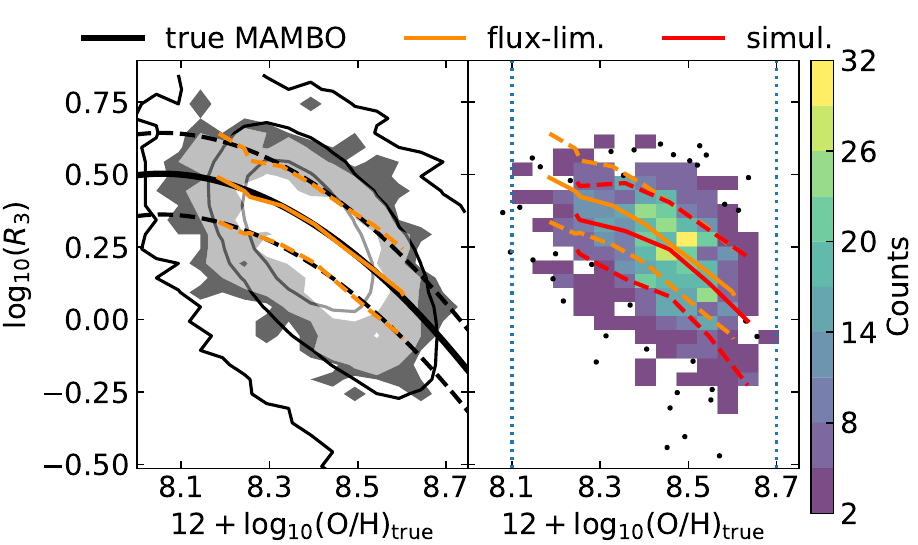}\\
\caption{$R_3$ strong-line diagnostic. The left panel shows the input MAMBO line ratio for the full \HE$\leq26$ SF galaxy sample (black contours) and the line-flux-limited sample (grey filled contours). The right panel show the corresponding simulated distribution for the golden sample. Black solid and dashed curves indicate the \egg\ median relation and scatter adopted in MAMBO; orange and red curves show the running median and sMAD of the true line-flux-limited and simulated samples, respectively.}\label{fig:R3}
\end{figure}
\begin{figure}[!htbp]
\includegraphics[width=\linewidth]{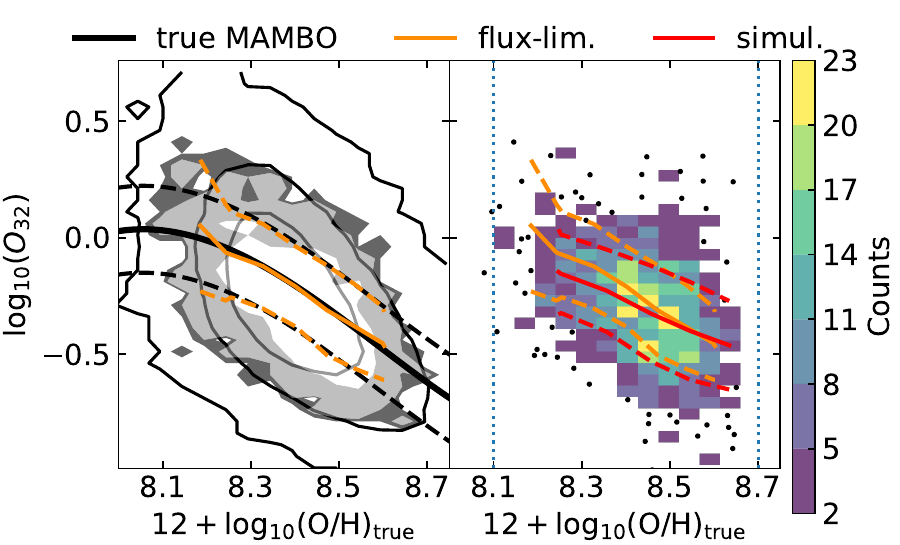}\\
\caption{Same as Fig.~\ref{fig:R3} for $O_{32}$.}\label{fig:O32}
\end{figure}
\begin{figure}[!htbp]
\includegraphics[width=\linewidth]{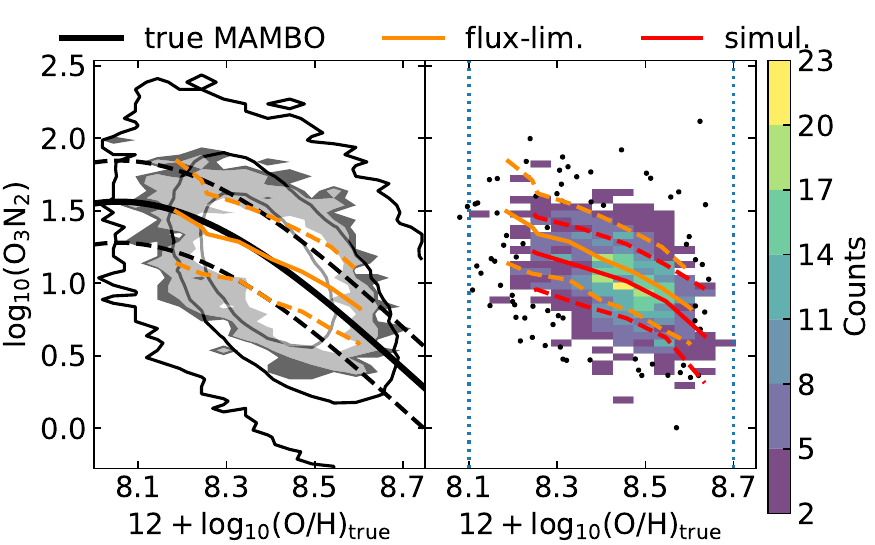}\\
\caption{Same as Fig.~\ref{fig:R3} and \ref{fig:O32} for $O_3N_2$.}\label{fig:O3N2}
\end{figure}

Figures~\ref{fig:R3}--\ref{fig:O3N2} compare the $R_3$, $O_{32}$, and $O_3N_2$ true line ratios in the \HE-band-limited and line-flux-limited MAMBO parent samples over $z\in [1.49, 1.88]$ with those measured from the simulated data for the golden sample, as a function of the true gas-phase metallicity, $12+\logten (\mathrm {O/H})_{\rm true}$.
About 60\% of the galaxies in the flux-limited sample meet all our selection criteria for robust line measurements, and reliable Balmer decrement in the simulated sample. Sources with intrinsically faint \hb and consequently high $R_3$ and $O_3N_2$ are most affected by the selection. The lower S/N of these faint lines produces noisier spectra, particularly in low-mass, low-metallicity galaxies, and reduces the completeness of the recovered sample. At $12+\logten(\mathrm{O/H})_{\rm true}<8.4$, this results in slightly shallower median relations in Figs.~\ref{fig:R3} and \ref{fig:O3N2} than those of the line-flux-limited sample (red and orange curves, respectively).

\clearpage

\end{appendix}

\end{document}

%% file: author_list.tex
%%%% Version Wednesday 16th of September 2026 04:06:09 PM UT *PRELIMINARY*
%%%% Assumes the new A&A style file from Oct 2025 or later
%%%% Please do not edit the author list -- contact ECEB Bureau for changes
\newcommand{\orcid}[1]{} %% if already defined in aa.cls: comment, or use renewcommand			   
\author{Euclid Collaboration: C.~Mancini\orcid{0000-0002-4297-0561}\thanks{\email{chiara.mancini@inaf.it}}\inst{\ref{aff1}}
\and L.~Gabarra\orcid{0000-0002-8486-8856}\inst{\ref{aff2}}
\and S.~Quai\orcid{0000-0002-0449-8163}\inst{\ref{aff3},\ref{aff4}}
\and L.~Pozzetti\orcid{0000-0001-7085-0412}\inst{\ref{aff4}}
\and P.~Cassata\orcid{0000-0002-6716-4400}\inst{\ref{aff5},\ref{aff6}}
\and M.~Talia\orcid{0000-0003-4352-2063}\inst{\ref{aff3},\ref{aff4}}
\and M.~Scodeggio\inst{\ref{aff1}}
\and M.~Bolzonella\orcid{0000-0003-3278-4607}\inst{\ref{aff4}}
\and M.~Moresco\orcid{0000-0002-7616-7136}\inst{\ref{aff3},\ref{aff4}}
\and D.~Vergani\orcid{0000-0003-0898-2216}\inst{\ref{aff4}}
\and E.~Maiorano\orcid{0000-0003-2593-4355}\inst{\ref{aff4}}
\and V.~Le~Brun\orcid{0000-0002-5027-1939}\inst{\ref{aff7}}
\and B.~R.~Granett\orcid{0000-0003-2694-9284}\inst{\ref{aff8}}
\and S.~Bisogni\orcid{0000-0003-3746-4565}\inst{\ref{aff1}}
\and A.~Gargiulo\orcid{0000-0002-3351-1216}\inst{\ref{aff1}}
\and G.~Vietri\orcid{0000-0001-9155-8875}\inst{\ref{aff1}}
\and M.~Siudek\orcid{0000-0002-2949-2155}\inst{\ref{aff9},\ref{aff10},\ref{aff11}}
\and R.~Ragusa\inst{\ref{aff12}}
\and G.~Herzog\orcid{0000-0001-5909-4054}\inst{\ref{aff1}}
\and G.~De~Lucia\orcid{0000-0002-6220-9104}\inst{\ref{aff13}}
\and L.~Scharr\'e\orcid{0000-0003-2551-4430}\inst{\ref{aff14}}
\and M.~Fumana\orcid{0000-0001-6787-5950}\inst{\ref{aff1}}
\and M.~Hirschmann\orcid{0000-0002-3301-3321}\inst{\ref{aff14}}
\and V.~Allevato\orcid{0000-0001-7232-5152}\inst{\ref{aff12}}
\and L.~Bisigello\orcid{0000-0003-0492-4924}\inst{\ref{aff6}}
\and A.~Enia\orcid{0000-0002-0200-2857}\inst{\ref{aff15},\ref{aff4}}
\and A.~Feltre\orcid{0000-0001-6865-2871}\inst{\ref{aff16}}
\and G.~Rodighiero\orcid{0000-0002-9415-2296}\inst{\ref{aff5},\ref{aff6}}
\and G.~Zamorani\orcid{0000-0002-2318-301X}\inst{\ref{aff4}}
\and P.~Corcho-Caballero\orcid{0000-0001-6327-7080}\inst{\ref{aff17}}
\and X.~Lopez~Lopez\orcid{0009-0008-5194-5908}\inst{\ref{aff4}}
\and S.~Andreon\orcid{0000-0002-2041-8784}\inst{\ref{aff8}}
\and N.~Auricchio\orcid{0000-0003-4444-8651}\inst{\ref{aff4}}
\and C.~Baccigalupi\orcid{0000-0002-8211-1630}\inst{\ref{aff18},\ref{aff13},\ref{aff19},\ref{aff20}}
\and M.~Baldi\orcid{0000-0003-4145-1943}\inst{\ref{aff21},\ref{aff4},\ref{aff22}}
\and A.~Balestra\orcid{0000-0002-6967-261X}\inst{\ref{aff6}}
\and S.~Bardelli\orcid{0000-0002-8900-0298}\inst{\ref{aff4}}
\and P.~Battaglia\orcid{0000-0002-7337-5909}\inst{\ref{aff4}}
\and A.~Biviano\orcid{0000-0002-0857-0732}\inst{\ref{aff13},\ref{aff18}}
\and E.~Branchini\orcid{0000-0002-0808-6908}\inst{\ref{aff23},\ref{aff24},\ref{aff8}}
\and M.~Brescia\orcid{0000-0001-9506-5680}\inst{\ref{aff25},\ref{aff12}}
\and S.~Camera\orcid{0000-0003-3399-3574}\inst{\ref{aff26},\ref{aff27},\ref{aff28}}
\and V.~Capobianco\orcid{0000-0002-3309-7692}\inst{\ref{aff28}}
\and C.~Carbone\orcid{0000-0003-0125-3563}\inst{\ref{aff1}}
\and J.~Carretero\orcid{0000-0002-3130-0204}\inst{\ref{aff29},\ref{aff30}}
\and M.~Castellano\orcid{0000-0001-9875-8263}\inst{\ref{aff31}}
\and G.~Castignani\orcid{0000-0001-6831-0687}\inst{\ref{aff4}}
\and S.~Cavuoti\orcid{0000-0002-3787-4196}\inst{\ref{aff12},\ref{aff32}}
\and P.-Y.~Chabaud\inst{\ref{aff7}}
\and A.~Cimatti\inst{\ref{aff33}}
\and C.~Colodro-Conde\inst{\ref{aff9}}
\and G.~Congedo\orcid{0000-0003-2508-0046}\inst{\ref{aff34}}
\and L.~Conversi\orcid{0000-0002-6710-8476}\inst{\ref{aff35},\ref{aff36}}
\and Y.~Copin\orcid{0000-0002-5317-7518}\inst{\ref{aff37}}
\and F.~Courbin\orcid{0000-0003-0758-6510}\inst{\ref{aff38},\ref{aff39},\ref{aff40}}
\and H.~M.~Courtois\orcid{0000-0003-0509-1776}\inst{\ref{aff41}}
\and H.~Degaudenzi\orcid{0000-0002-5887-6799}\inst{\ref{aff42}}
\and S.~de~la~Torre\inst{\ref{aff7}}
\and H.~Dole\orcid{0000-0002-9767-3839}\inst{\ref{aff43}}
\and F.~Dubath\orcid{0000-0002-6533-2810}\inst{\ref{aff42}}
\and C.~A.~J.~Duncan\orcid{0009-0003-3573-0791}\inst{\ref{aff34}}
\and X.~Dupac\inst{\ref{aff36}}
\and S.~Escoffier\orcid{0000-0002-2847-7498}\inst{\ref{aff44}}
\and M.~Farina\orcid{0000-0002-3089-7846}\inst{\ref{aff45}}
\and R.~Farinelli\inst{\ref{aff4}}
\and F.~Faustini\orcid{0000-0001-6274-5145}\inst{\ref{aff31}}
\and S.~Ferriol\inst{\ref{aff37}}
\and F.~Finelli\orcid{0000-0002-6694-3269}\inst{\ref{aff4},\ref{aff46}}
\and S.~Fotopoulou\orcid{0000-0002-9686-254X}\inst{\ref{aff47}}
\and N.~Fourmanoit\orcid{0009-0005-6816-6925}\inst{\ref{aff44}}
\and M.~Frailis\orcid{0000-0002-7400-2135}\inst{\ref{aff13}}
\and E.~Franceschi\orcid{0000-0002-0585-6591}\inst{\ref{aff4}}
\and S.~Galeotta\orcid{0000-0002-3748-5115}\inst{\ref{aff13}}
\and K.~George\orcid{0000-0002-1734-8455}\inst{\ref{aff48}}
\and B.~Gillis\orcid{0000-0002-4478-1270}\inst{\ref{aff34}}
\and C.~Giocoli\orcid{0000-0002-9590-7961}\inst{\ref{aff4},\ref{aff22}}
\and J.~Gracia-Carpio\orcid{0000-0003-4689-3134}\inst{\ref{aff49}}
\and A.~Grazian\orcid{0000-0002-5688-0663}\inst{\ref{aff6}}
\and F.~Grupp\inst{\ref{aff49},\ref{aff50}}
\and S.~V.~H.~Haugan\orcid{0000-0001-9648-7260}\inst{\ref{aff51}}
\and S.~Hemmati\orcid{0000-0003-2226-5395}\inst{\ref{aff52}}
\and W.~Holmes\orcid{0009-0007-8554-4646}\inst{\ref{aff53}}
\and F.~Hormuth\inst{\ref{aff54}}
\and A.~Hornstrup\orcid{0000-0002-3363-0936}\inst{\ref{aff55},\ref{aff56}}
\and K.~Jahnke\orcid{0000-0003-3804-2137}\inst{\ref{aff57}}
\and M.~Jhabvala\inst{\ref{aff58}}
\and B.~Joachimi\orcid{0000-0001-7494-1303}\inst{\ref{aff59}}
\and S.~Kermiche\orcid{0000-0002-0302-5735}\inst{\ref{aff44}}
\and A.~Kiessling\orcid{0000-0002-2590-1273}\inst{\ref{aff53}}
\and B.~Kubik\orcid{0009-0006-5823-4880}\inst{\ref{aff37}}
\and M.~K\"ummel\orcid{0000-0003-2791-2117}\inst{\ref{aff50}}
\and M.~Kunz\orcid{0000-0002-3052-7394}\inst{\ref{aff60}}
\and H.~Kurki-Suonio\orcid{0000-0002-4618-3063}\inst{\ref{aff61},\ref{aff62}}
\and A.~M.~C.~Le~Brun\orcid{0000-0002-0936-4594}\inst{\ref{aff63}}
\and S.~Ligori\orcid{0000-0003-4172-4606}\inst{\ref{aff28}}
\and P.~B.~Lilje\orcid{0000-0003-4324-7794}\inst{\ref{aff51}}
\and V.~Lindholm\orcid{0000-0003-2317-5471}\inst{\ref{aff61},\ref{aff62}}
\and I.~Lloro\orcid{0000-0001-5966-1434}\inst{\ref{aff64}}
\and M.~Magliocchetti\orcid{0000-0001-9158-4838}\inst{\ref{aff45}}
\and G.~Mainetti\orcid{0000-0003-2384-2377}\inst{\ref{aff65}}
\and O.~Mansutti\orcid{0000-0001-5758-4658}\inst{\ref{aff13}}
\and O.~Marggraf\orcid{0000-0001-7242-3852}\inst{\ref{aff66}}
\and M.~Martinelli\orcid{0000-0002-6943-7732}\inst{\ref{aff31},\ref{aff67}}
\and N.~Martinet\orcid{0000-0003-2786-7790}\inst{\ref{aff7}}
\and F.~Marulli\orcid{0000-0002-8850-0303}\inst{\ref{aff3},\ref{aff4},\ref{aff22}}
\and R.~J.~Massey\orcid{0000-0002-6085-3780}\inst{\ref{aff68}}
\and N.~Mauri\orcid{0000-0001-8196-1548}\inst{\ref{aff33},\ref{aff22}}
\and E.~Medinaceli\orcid{0000-0002-4040-7783}\inst{\ref{aff4}}
\and S.~Mei\orcid{0000-0002-2849-559X}\inst{\ref{aff69},\ref{aff70}}
\and M.~Meneghetti\orcid{0000-0003-1225-7084}\inst{\ref{aff4},\ref{aff22}}
\and E.~Merlin\orcid{0000-0001-6870-8900}\inst{\ref{aff6}}
\and G.~Meylan\orcid{0000-0001-6503-0209}\inst{\ref{aff71}}
\and P.~Monaco\orcid{0000-0003-2083-7564}\inst{\ref{aff72},\ref{aff13},\ref{aff19},\ref{aff18}}
\and A.~Mora\orcid{0000-0002-1922-8529}\inst{\ref{aff73}}
\and C.~Moretti\orcid{0000-0003-3314-8936}\inst{\ref{aff13},\ref{aff18},\ref{aff19}}
\and L.~Moscardini\orcid{0000-0002-3473-6716}\inst{\ref{aff3},\ref{aff4},\ref{aff22}}
\and C.~Neissner\orcid{0000-0001-8524-4968}\inst{\ref{aff74},\ref{aff30}}
\and S.-M.~Niemi\orcid{0009-0005-0247-0086}\inst{\ref{aff75}}
\and J.~W.~Nightingale\orcid{0000-0002-8987-7401}\inst{\ref{aff76}}
\and C.~Padilla\orcid{0000-0001-7951-0166}\inst{\ref{aff74}}
\and S.~Paltani\orcid{0000-0002-8108-9179}\inst{\ref{aff42}}
\and F.~Pasian\orcid{0000-0002-4869-3227}\inst{\ref{aff13}}
\and V.~Pettorino\orcid{0000-0002-4203-9320}\inst{\ref{aff75}}
\and A.~Pezzotta\orcid{0000-0003-0726-2268}\inst{\ref{aff8}}
\and S.~Pires\orcid{0000-0002-0249-2104}\inst{\ref{aff77}}
\and G.~Polenta\orcid{0000-0003-4067-9196}\inst{\ref{aff78}}
\and M.~Poncet\inst{\ref{aff79}}
\and L.~A.~Popa\inst{\ref{aff80}}
\and F.~Raison\orcid{0000-0002-7819-6918}\inst{\ref{aff49}}
\and R.~Rebolo\orcid{0000-0003-3767-7085}\inst{\ref{aff9},\ref{aff81},\ref{aff10}}
\and A.~Renzi\orcid{0000-0001-9856-1970}\inst{\ref{aff5},\ref{aff82},\ref{aff4}}
\and J.~Rhodes\orcid{0000-0002-4485-8549}\inst{\ref{aff53}}
\and G.~Riccio\inst{\ref{aff12}}
\and I.~Risso\orcid{0000-0003-2525-7761}\inst{\ref{aff23},\ref{aff24},\ref{aff8}}
\and F.~Rizzo\orcid{0000-0002-9407-585X}\inst{\ref{aff13}}
\and E.~Romelli\orcid{0000-0003-3069-9222}\inst{\ref{aff13}}
\and M.~Roncarelli\orcid{0000-0001-9587-7822}\inst{\ref{aff4}}
\and B.~Rusholme\orcid{0000-0001-7648-4142}\inst{\ref{aff52}}
\and R.~Saglia\orcid{0000-0003-0378-7032}\inst{\ref{aff50},\ref{aff49}}
\and Z.~Sakr\orcid{0000-0002-4823-3757}\inst{\ref{aff83},\ref{aff84},\ref{aff85}}
\and D.~Sapone\orcid{0000-0001-7089-4503}\inst{\ref{aff86}}
\and B.~Sartoris\orcid{0000-0003-1337-5269}\inst{\ref{aff49},\ref{aff13}}
\and P.~Schneider\orcid{0000-0001-8561-2679}\inst{\ref{aff66}}
\and T.~Schrabback\orcid{0000-0002-6987-7834}\inst{\ref{aff87}}
\and A.~Secroun\orcid{0000-0003-0505-3710}\inst{\ref{aff44}}
\and E.~Sihvola\orcid{0000-0003-1804-7715}\inst{\ref{aff88}}
\and P.~Simon\inst{\ref{aff66}}
\and C.~Sirignano\orcid{0000-0002-0995-7146}\inst{\ref{aff5},\ref{aff82}}
\and G.~Sirri\orcid{0000-0003-2626-2853}\inst{\ref{aff22}}
\and L.~Stanco\orcid{0000-0002-9706-5104}\inst{\ref{aff82}}
\and J.-L.~Starck\orcid{0000-0003-2177-7794}\inst{\ref{aff77}}
\and P.~Tallada-Cresp\'{i}\orcid{0000-0002-1336-8328}\inst{\ref{aff29},\ref{aff30}}
\and A.~N.~Taylor\inst{\ref{aff34}}
\and H.~I.~Teplitz\orcid{0000-0002-7064-5424}\inst{\ref{aff89}}
\and I.~Tereno\orcid{0000-0002-4537-6218}\inst{\ref{aff90},\ref{aff91}}
\and N.~Tessore\orcid{0000-0002-9696-7931}\inst{\ref{aff92}}
\and S.~Toft\orcid{0000-0003-3631-7176}\inst{\ref{aff93},\ref{aff94}}
\and R.~Toledo-Moreo\orcid{0000-0002-2997-4859}\inst{\ref{aff95},\ref{aff96}}
\and F.~Torradeflot\orcid{0000-0003-1160-1517}\inst{\ref{aff30},\ref{aff29}}
\and I.~Tutusaus\orcid{0000-0002-3199-0399}\inst{\ref{aff11},\ref{aff97},\ref{aff84}}
\and J.~Valiviita\orcid{0000-0001-6225-3693}\inst{\ref{aff61},\ref{aff62}}
\and T.~Vassallo\orcid{0000-0001-6512-6358}\inst{\ref{aff13},\ref{aff48}}
\and A.~Veropalumbo\orcid{0000-0003-2387-1194}\inst{\ref{aff8},\ref{aff24},\ref{aff23}}
\and D.~Vibert\orcid{0009-0008-0607-631X}\inst{\ref{aff7}}
\and Y.~Wang\orcid{0000-0002-4749-2984}\inst{\ref{aff52}}
\and J.~Weller\orcid{0000-0002-8282-2010}\inst{\ref{aff50},\ref{aff49}}
\and A.~Zacchei\orcid{0000-0003-0396-1192}\inst{\ref{aff13},\ref{aff18}}
\and F.~M.~Zerbi\orcid{0000-0002-9996-973X}\inst{\ref{aff8}}
\and E.~Zucca\orcid{0000-0002-5845-8132}\inst{\ref{aff4}}
\and M.~Ballardini\orcid{0000-0003-4481-3559}\inst{\ref{aff98},\ref{aff99},\ref{aff4}}
\and P.~Bergamini\orcid{0000-0003-1383-9414}\inst{\ref{aff4}}
\and S.~Borgani\orcid{0000-0001-6151-6439}\inst{\ref{aff72},\ref{aff18},\ref{aff13},\ref{aff19},\ref{aff100}}
\and A.~Boucaud\orcid{0000-0001-7387-2633}\inst{\ref{aff69}}
\and E.~Bozzo\orcid{0000-0002-8201-1525}\inst{\ref{aff42}}
\and C.~Burigana\orcid{0000-0002-3005-5796}\inst{\ref{aff101},\ref{aff46}}
\and R.~Cabanac\orcid{0000-0001-6679-2600}\inst{\ref{aff84}}
\and F.~Caro\orcid{0009-0003-1053-0507}\inst{\ref{aff31}}
\and T.~Castro\orcid{0000-0002-6292-3228}\inst{\ref{aff102},\ref{aff13},\ref{aff18}}
\and G.~Daste\inst{\ref{aff7}}
\and J.~A.~Escartin~Vigo\inst{\ref{aff49}}
\and J.~Garc\'ia-Bellido\orcid{0000-0002-9370-8360}\inst{\ref{aff83}}
\and T.~Gasparetto\orcid{0000-0002-7913-4866}\inst{\ref{aff31}}
\and J.~Macias-Perez\orcid{0000-0002-5385-2763}\inst{\ref{aff103}}
\and R.~Maoli\orcid{0000-0002-6065-3025}\inst{\ref{aff104},\ref{aff31}}
\and R.~B.~Metcalf\orcid{0000-0003-3167-2574}\inst{\ref{aff3},\ref{aff4}}
\and A.~A.~Nucita\inst{\ref{aff105},\ref{aff106},\ref{aff107}}
\and M.~P\"ontinen\orcid{0000-0001-5442-2530}\inst{\ref{aff61}}
\and E.~Sarpa\orcid{0000-0002-1256-655X}\inst{\ref{aff13}}
\and V.~Scottez\orcid{0009-0008-3864-940X}\inst{\ref{aff108},\ref{aff109}}
\and M.~Sereno\orcid{0000-0003-0302-0325}\inst{\ref{aff4},\ref{aff22}}
\and M.~Tenti\orcid{0000-0002-4254-5901}\inst{\ref{aff22}}
\and M.~Tucci\inst{\ref{aff42}}
\and M.~Viel\orcid{0000-0002-2642-5707}\inst{\ref{aff18},\ref{aff13},\ref{aff20},\ref{aff19},\ref{aff100}}
\and M.~Wiesmann\orcid{0009-0000-8199-5860}\inst{\ref{aff51}}
\and J.~A.~Acevedo~Barroso\orcid{0000-0002-9654-1711}\inst{\ref{aff53}}
\and Y.~Akrami\orcid{0000-0002-2407-7956}\inst{\ref{aff83},\ref{aff110}}
\and I.~T.~Andika\orcid{0000-0001-6102-9526}\inst{\ref{aff50}}
\and G.~Angora\orcid{0000-0002-0316-6562}\inst{\ref{aff12},\ref{aff98}}
\and S.~Anselmi\orcid{0000-0002-3579-9583}\inst{\ref{aff82},\ref{aff5},\ref{aff111}}
\and M.~Archidiacono\orcid{0000-0003-4952-9012}\inst{\ref{aff112},\ref{aff113}}
\and G.~Aric\`o\orcid{0000-0002-2802-2928}\inst{\ref{aff22}}
\and F.~Atrio-Barandela\orcid{0000-0002-2130-2513}\inst{\ref{aff114}}
\and M.~Baes\orcid{0000-0002-3930-2757}\inst{\ref{aff115}}
\and L.~Bazzanini\orcid{0000-0003-0727-0137}\inst{\ref{aff98},\ref{aff4}}
\and D.~Bertacca\orcid{0000-0002-2490-7139}\inst{\ref{aff5},\ref{aff6},\ref{aff82}}
\and M.~Bethermin\orcid{0000-0002-3915-2015}\inst{\ref{aff116}}
\and F.~Beutler\orcid{0000-0003-0467-5438}\inst{\ref{aff34}}
\and A.~Blanchard\orcid{0000-0001-8555-9003}\inst{\ref{aff84}}
\and L.~Blot\orcid{0000-0002-9622-7167}\inst{\ref{aff117},\ref{aff63}}
\and M.~Bonici\orcid{0000-0002-8430-126X}\inst{\ref{aff118},\ref{aff1}}
\and M.~L.~Brown\orcid{0000-0002-0370-8077}\inst{\ref{aff119}}
\and S.~Bruton\orcid{0000-0002-6503-5218}\inst{\ref{aff120}}
\and A.~Calabro\orcid{0000-0003-2536-1614}\inst{\ref{aff31}}
\and B.~Camacho~Quevedo\orcid{0000-0002-8789-4232}\inst{\ref{aff18},\ref{aff20},\ref{aff13}}
\and C.~S.~Carvalho\inst{\ref{aff91}}
\and F.~Cogato\orcid{0000-0003-4632-6113}\inst{\ref{aff3},\ref{aff4}}
\and T.~E.~Collett\orcid{0000-0001-5564-3140}\inst{\ref{aff121}}
\and S.~Conseil\orcid{0000-0002-3657-4191}\inst{\ref{aff37}}
\and T.~Contini\orcid{0000-0003-0275-938X}\inst{\ref{aff84}}
\and A.~R.~Cooray\orcid{0000-0002-3892-0190}\inst{\ref{aff122}}
\and B.~Csizi\orcid{0000-0003-3227-6581}\inst{\ref{aff87}}
\and O.~Cucciati\orcid{0000-0002-9336-7551}\inst{\ref{aff4}}
\and H.~Dannerbauer\orcid{0000-0001-7147-3575}\inst{\ref{aff9},\ref{aff10}}
\and S.~Davini\orcid{0000-0003-3269-1718}\inst{\ref{aff24}}
\and T.~de~Boer\orcid{0000-0001-5486-2747}\inst{\ref{aff123}}
\and G.~Desprez\orcid{0000-0001-8325-1742}\inst{\ref{aff17}}
\and A.~D\'iaz-S\'anchez\orcid{0000-0003-0748-4768}\inst{\ref{aff124}}
\and S.~Di~Domizio\orcid{0000-0003-2863-5895}\inst{\ref{aff23},\ref{aff24}}
\and J.~M.~Diego\orcid{0000-0001-9065-3926}\inst{\ref{aff125}}
\and V.~Duret\orcid{0009-0009-0383-4960}\inst{\ref{aff44}}
\and M.~Y.~Elkhashab\orcid{0000-0001-9306-2603}\inst{\ref{aff13},\ref{aff19},\ref{aff72},\ref{aff18}}
\and Y.~Fang\orcid{0000-0002-0334-6950}\inst{\ref{aff50}}
\and A.~Farina\orcid{0009-0000-3420-929X}\inst{\ref{aff8},\ref{aff24}}
\and A.~Finoguenov\orcid{0000-0002-4606-5403}\inst{\ref{aff61}}
\and F.~Fontanot\orcid{0000-0003-4744-0188}\inst{\ref{aff13},\ref{aff18}}
\and A.~Franco\orcid{0000-0002-4761-366X}\inst{\ref{aff105},\ref{aff106},\ref{aff107}}
\and Y.~Fu\orcid{0000-0002-0759-0504}\inst{\ref{aff126},\ref{aff17}}
\and K.~Ganga\orcid{0000-0001-8159-8208}\inst{\ref{aff69}}
\and E.~Gaztanaga\orcid{0000-0001-9632-0815}\inst{\ref{aff11},\ref{aff97},\ref{aff121}}
\and Z.~Ghaffari\orcid{0000-0002-6467-8078}\inst{\ref{aff13},\ref{aff18}}
\and F.~Giacomini\orcid{0000-0002-3129-2814}\inst{\ref{aff22}}
\and F.~Gianotti\orcid{0000-0003-4666-119X}\inst{\ref{aff4}}
\and E.~J.~Gonzalez\orcid{0000-0002-0226-9893}\inst{\ref{aff11},\ref{aff127}}
\and G.~Gozaliasl\orcid{0000-0002-0236-919X}\inst{\ref{aff128},\ref{aff61}}
\and A.~Gruppuso\orcid{0000-0001-9272-5292}\inst{\ref{aff4},\ref{aff22}}
\and M.~Guidi\orcid{0000-0001-9408-1101}\inst{\ref{aff21},\ref{aff4}}
\and C.~M.~Gutierrez\orcid{0000-0001-7854-783X}\inst{\ref{aff9},\ref{aff10}}
\and A.~Hall\orcid{0000-0002-3139-8651}\inst{\ref{aff34}}
\and N.~A.~Hatch\orcid{0000-0001-5600-0534}\inst{\ref{aff129}}
\and H.~Hildebrandt\orcid{0000-0002-9814-3338}\inst{\ref{aff130}}
\and J.~Hjorth\orcid{0000-0002-4571-2306}\inst{\ref{aff131}}
\and J.~J.~E.~Kajava\orcid{0000-0002-3010-8333}\inst{\ref{aff132},\ref{aff133},\ref{aff134}}
\and Y.~Kang\orcid{0009-0000-8588-7250}\inst{\ref{aff87}}
\and V.~Kansal\orcid{0000-0002-4008-6078}\inst{\ref{aff135},\ref{aff136}}
\and D.~Karagiannis\orcid{0000-0002-4927-0816}\inst{\ref{aff98},\ref{aff137}}
\and J.~Kim\orcid{0000-0003-2776-2761}\inst{\ref{aff2}}
\and C.~C.~Kirkpatrick\inst{\ref{aff88}}
\and A.~Kov\'acs\orcid{0000-0002-5825-579X}\inst{\ref{aff138},\ref{aff139}}
\and I.~Kova{\v{c}}i{\'{c}}\orcid{0000-0001-6751-3263}\inst{\ref{aff115}}
\and K.~Koyama\orcid{0000-0001-6727-6915}\inst{\ref{aff121}}
\and S.~Kruk\orcid{0000-0001-8010-8879}\inst{\ref{aff36}}
\and M.~C.~Lam\orcid{0000-0002-9347-2298}\inst{\ref{aff34}}
\and G.~Leroy\orcid{0009-0004-2523-4425}\inst{\ref{aff140},\ref{aff68}}
\and G.~F.~Lesci\orcid{0000-0002-4607-2830}\inst{\ref{aff3},\ref{aff4}}
\and J.~Lesgourgues\orcid{0000-0001-7627-353X}\inst{\ref{aff141}}
\and T.~I.~Liaudat\orcid{0000-0002-9104-314X}\inst{\ref{aff142}}
\and L.~Linke\orcid{0000-0002-2622-8113}\inst{\ref{aff87}}
\and S.~J.~Liu\orcid{0000-0001-7680-2139}\inst{\ref{aff45}}
\and F.~Mannucci\orcid{0000-0002-4803-2381}\inst{\ref{aff16}}
\and F.~R.~Marleau\orcid{0000-0002-1442-2947}\inst{\ref{aff87}}
\and C.~J.~A.~P.~Martins\orcid{0000-0002-4886-9261}\inst{\ref{aff143},\ref{aff144}}
\and M.~Migliaccio\inst{\ref{aff145},\ref{aff146}}
\and M.~Miluzio\inst{\ref{aff36},\ref{aff147}}
\and G.~Morgante\inst{\ref{aff4}}
\and K.~Naidoo\orcid{0000-0002-9182-1802}\inst{\ref{aff121},\ref{aff57}}
\and A.~Navarro-Alsina\orcid{0000-0002-3173-2592}\inst{\ref{aff66}}
\and S.~Nesseris\orcid{0000-0002-0567-0324}\inst{\ref{aff83}}
\and F.~Oppizzi\orcid{0000-0003-3904-8370}\inst{\ref{aff24},\ref{aff82},\ref{aff5}}
\and F.~Pace\orcid{0000-0001-8039-0480}\inst{\ref{aff26},\ref{aff27},\ref{aff28},\ref{aff148}}
\and E.~Palazzi\orcid{0000-0002-8691-7666}\inst{\ref{aff4}}
\and D.~Paoletti\orcid{0000-0003-4761-6147}\inst{\ref{aff4},\ref{aff46}}
\and F.~Passalacqua\orcid{0000-0002-8606-4093}\inst{\ref{aff82}}
\and K.~Paterson\orcid{0000-0001-8340-3486}\inst{\ref{aff57}}
\and L.~Patrizii\inst{\ref{aff22}}
\and C.~Pattison\orcid{0000-0003-3272-2617}\inst{\ref{aff121}}
\and A.~Pisani\orcid{0000-0002-6146-4437}\inst{\ref{aff44}}
\and D.~Potter\orcid{0000-0002-0757-5195}\inst{\ref{aff149}}
\and G.~W.~Pratt\inst{\ref{aff77}}
\and M.~Radovich\orcid{0000-0002-3585-866X}\inst{\ref{aff6}}
\and F.~Ricci\orcid{0000-0001-5742-5980}\inst{\ref{aff150},\ref{aff31}}
\and W.~Roster\orcid{0000-0002-9149-6528}\inst{\ref{aff49}}
\and S.~Sacquegna\orcid{0000-0002-8433-6630}\inst{\ref{aff151}}
\and M.~Sahl\'en\orcid{0000-0003-0973-4804}\inst{\ref{aff152}}
\and D.~B.~Sanders\orcid{0000-0002-1233-9998}\inst{\ref{aff123}}
\and A.~Schneider\orcid{0000-0001-7055-8104}\inst{\ref{aff149}}
\and D.~Sciotti\orcid{0009-0008-4519-2620}\inst{\ref{aff31},\ref{aff67}}
\and E.~Sellentin\orcid{0009-0002-2655-3458}\inst{\ref{aff153},\ref{aff126}}
\and S.~Serjeant\orcid{0000-0002-0517-7943}\inst{\ref{aff154}}
\and F.~Shankar\orcid{0000-0001-8973-5051}\inst{\ref{aff155}}
\and L.~C.~Smith\orcid{0000-0002-3259-2771}\inst{\ref{aff156}}
\and J.~G.~Sorce\orcid{0000-0002-2307-2432}\inst{\ref{aff157},\ref{aff43}}
\and K.~Tanidis\orcid{0000-0001-9843-5130}\inst{\ref{aff158}}
\and C.~Tao\orcid{0000-0001-7961-8177}\inst{\ref{aff44}}
\and F.~Tarsitano\orcid{0000-0002-5919-0238}\inst{\ref{aff159},\ref{aff160},\ref{aff42}}
\and G.~Testera\orcid{0000-0003-2970-766X}\inst{\ref{aff24}}
\and R.~Teyssier\orcid{0000-0001-7689-0933}\inst{\ref{aff161}}
\and S.~Tosi\orcid{0000-0002-7275-9193}\inst{\ref{aff23},\ref{aff8},\ref{aff24}}
\and A.~Troja\orcid{0000-0003-0239-4595}\inst{\ref{aff13}}
\and C.~Uhlemann\orcid{0000-0001-7831-1579}\inst{\ref{aff162},\ref{aff76}}
\and C.~Valieri\inst{\ref{aff22}}
\and A.~Venhola\orcid{0000-0001-6071-4564}\inst{\ref{aff163}}
\and G.~Verza\orcid{0000-0002-1886-8348}\thanks{Deceased}\inst{\ref{aff164},\ref{aff165}}
\and P.~Vielzeuf\orcid{0000-0003-2035-9339}\inst{\ref{aff44}}
\and S.~Vinciguerra\orcid{0009-0005-4018-3184}\inst{\ref{aff7}}
\and M.~von~Wietersheim-Kramsta\orcid{0000-0003-4986-5091}\inst{\ref{aff68},\ref{aff140}}
\and N.~A.~Walton\orcid{0000-0003-3983-8778}\inst{\ref{aff156}}
\and L.~Wang\orcid{0000-0002-6736-9158}\inst{\ref{aff166},\ref{aff17}}
\and A.~H.~Wright\orcid{0000-0001-7363-7932}\inst{\ref{aff130}}
\and H.~W.~Yeung\orcid{0000-0002-4993-9014}\inst{\ref{aff34}}}
										   
%%%% please do not edit the affiliation list -- contact ECEB Bureau for changes
\institute{INAF-IASF Milano, Via Alfonso Corti 12, 20133 Milano, Italy\label{aff1}
\and
Department of Physics, University of Oxford, Keble Road, Oxford OX1 3RH, UK\label{aff2}
\and
Dipartimento di Fisica e Astronomia "Augusto Righi" - Alma Mater Studiorum Universit\`a di Bologna, via Piero Gobetti 93/2, 40129 Bologna, Italy\label{aff3}
\and
INAF-Osservatorio di Astrofisica e Scienza dello Spazio di Bologna, Via Piero Gobetti 93/3, 40129 Bologna, Italy\label{aff4}
\and
Dipartimento di Fisica e Astronomia "G. Galilei", Universit\`a di Padova, Via Marzolo 8, 35131 Padova, Italy\label{aff5}
\and
INAF-Osservatorio Astronomico di Padova, Via dell'Osservatorio 5, 35122 Padova, Italy\label{aff6}
\and
Aix-Marseille Universit\'e, CNRS, CNES, LAM, Marseille, France\label{aff7}
\and
INAF-Osservatorio Astronomico di Brera, Via Brera 28, 20122 Milano, Italy\label{aff8}
\and
Instituto de Astrof\'{\i}sica de Canarias, E-38205 La Laguna, Tenerife, Spain\label{aff9}
\and
Universidad de La Laguna, Dpto. Astrof\'\i sica, E-38206 La Laguna, Tenerife, Spain\label{aff10}
\and
Institute of Space Sciences (ICE, CSIC), Campus UAB, Carrer de Can Magrans, s/n, 08193 Barcelona, Spain\label{aff11}
\and
INAF-Osservatorio Astronomico di Capodimonte, Via Moiariello 16, 80131 Napoli, Italy\label{aff12}
\and
INAF-Osservatorio Astronomico di Trieste, Via G. B. Tiepolo 11, 34143 Trieste, Italy\label{aff13}
\and
Institute of Physics, Laboratory for Galaxy Evolution, Ecole Polytechnique F\'ed\'erale de Lausanne, Observatoire de Sauverny, CH-1290 Versoix, Switzerland\label{aff14}
\and
CEA Saclay, DFR/IRFU, Service d'Astrophysique, Bat. 709, 91191 Gif-sur-Yvette, France\label{aff15}
\and
INAF-Osservatorio Astrofisico di Arcetri, Largo E. Fermi 5, 50125, Firenze, Italy\label{aff16}
\and
Kapteyn Astronomical Institute, University of Groningen, PO Box 800, 9700 AV Groningen, The Netherlands\label{aff17}
\and
IFPU, Institute for Fundamental Physics of the Universe, via Beirut 2, 34151 Trieste, Italy\label{aff18}
\and
INFN, Sezione di Trieste, Via Valerio 2, 34127 Trieste TS, Italy\label{aff19}
\and
SISSA, International School for Advanced Studies, Via Bonomea 265, 34136 Trieste TS, Italy\label{aff20}
\and
Dipartimento di Fisica e Astronomia, Universit\`a di Bologna, Via Gobetti 93/2, 40129 Bologna, Italy\label{aff21}
\and
INFN-Sezione di Bologna, Viale Berti Pichat 6/2, 40127 Bologna, Italy\label{aff22}
\and
Dipartimento di Fisica, Universit\`a di Genova, Via Dodecaneso 33, 16146, Genova, Italy\label{aff23}
\and
INFN-Sezione di Genova, Via Dodecaneso 33, 16146, Genova, Italy\label{aff24}
\and
Department of Physics "E. Pancini", University Federico II, Via Cinthia 6, 80126, Napoli, Italy\label{aff25}
\and
Dipartimento di Fisica, Universit\`a degli Studi di Torino, Via P. Giuria 1, 10125 Torino, Italy\label{aff26}
\and
INFN-Sezione di Torino, Via P. Giuria 1, 10125 Torino, Italy\label{aff27}
\and
INAF-Osservatorio Astrofisico di Torino, Via Osservatorio 20, 10025 Pino Torinese (TO), Italy\label{aff28}
\and
Centro de Investigaciones Energ\'eticas, Medioambientales y Tecnol\'ogicas (CIEMAT), Avenida Complutense 40, 28040 Madrid, Spain\label{aff29}
\and
Port d'Informaci\'{o} Cient\'{i}fica, Campus UAB, C. Albareda s/n, 08193 Bellaterra (Barcelona), Spain\label{aff30}
\and
INAF-Osservatorio Astronomico di Roma, Via Frascati 33, 00078 Monteporzio Catone, Italy\label{aff31}
\and
INFN -- Sezione di Napoli, Via Cinthia 6, 80126, Napoli, Italy\label{aff32}
\and
Dipartimento di Fisica e Astronomia "Augusto Righi" - Alma Mater Studiorum Universit\`a di Bologna, Viale Berti Pichat 6/2, 40127 Bologna, Italy\label{aff33}
\and
Institute for Astronomy, University of Edinburgh, Royal Observatory, Blackford Hill, Edinburgh EH9 3HJ, UK\label{aff34}
\and
European Space Agency/ESRIN, Largo Galileo Galilei 1, 00044 Frascati, Roma, Italy\label{aff35}
\and
ESAC/ESA, Camino Bajo del Castillo, s/n., Urb. Villafranca del Castillo, 28692 Villanueva de la Ca\~nada, Madrid, Spain\label{aff36}
\and
Universit\'e Claude Bernard Lyon 1, CNRS/IN2P3, IP2I Lyon, UMR 5822, Villeurbanne, F-69100, France\label{aff37}
\and
Institut de Ci\`{e}ncies del Cosmos (ICCUB), Universitat de Barcelona (IEEC-UB), Mart\'{i} i Franqu\`{e}s 1, 08028 Barcelona, Spain\label{aff38}
\and
Instituci\'o Catalana de Recerca i Estudis Avan\c{c}ats (ICREA), Passeig de Llu\'{\i}s Companys 23, 08010 Barcelona, Spain\label{aff39}
\and
Institut de Ciencies de l'Espai (IEEC-CSIC), Campus UAB, Carrer de Can Magrans, s/n Cerdanyola del Vall\'es, 08193 Barcelona, Spain\label{aff40}
\and
UCB Lyon 1, CNRS/IN2P3, IUF, IP2I Lyon, 4 rue Enrico Fermi, 69622 Villeurbanne, France\label{aff41}
\and
Department of Astronomy, University of Geneva, ch. d'Ecogia 16, 1290 Versoix, Switzerland\label{aff42}
\and
Universit\'e Paris-Saclay, CNRS, Institut d'astrophysique spatiale, 91405, Orsay, France\label{aff43}
\and
Aix-Marseille Universit\'e, CNRS/IN2P3, CPPM, Marseille, France\label{aff44}
\and
INAF-Istituto di Astrofisica e Planetologia Spaziali, via del Fosso del Cavaliere, 100, 00100 Roma, Italy\label{aff45}
\and
INFN-Bologna, Via Irnerio 46, 40126 Bologna, Italy\label{aff46}
\and
School of Physics, HH Wills Physics Laboratory, University of Bristol, Tyndall Avenue, Bristol, BS8 1TL, UK\label{aff47}
\and
University Observatory, LMU Faculty of Physics, Scheinerstr.~1, 81679 Munich, Germany\label{aff48}
\and
Max Planck Institute for Extraterrestrial Physics, Giessenbachstr. 1, 85748 Garching, Germany\label{aff49}
\and
Universit\"ats-Sternwarte M\"unchen, Fakult\"at f\"ur Physik, Ludwig-Maximilians-Universit\"at M\"unchen, Scheinerstr.~1, 81679 M\"unchen, Germany\label{aff50}
\and
Institute of Theoretical Astrophysics, University of Oslo, P.O. Box 1029 Blindern, 0315 Oslo, Norway\label{aff51}
\and
Caltech/IPAC, 1200 E. California Blvd., Pasadena, CA 91125, USA\label{aff52}
\and
Jet Propulsion Laboratory, California Institute of Technology, 4800 Oak Grove Drive, Pasadena, CA, 91109, USA\label{aff53}
\and
Felix Hormuth Engineering, Goethestr. 17, 69181 Leimen, Germany\label{aff54}
\and
Technical University of Denmark, Elektrovej 327, 2800 Kgs. Lyngby, Denmark\label{aff55}
\and
Cosmic Dawn Center (DAWN), Denmark\label{aff56}
\and
Max-Planck-Institut f\"ur Astronomie, K\"onigstuhl 17, 69117 Heidelberg, Germany\label{aff57}
\and
NASA Goddard Space Flight Center, Greenbelt, MD 20771, USA\label{aff58}
\and
Department of Physics and Astronomy, University College London, Gower Street, London WC1E 6BT, UK\label{aff59}
\and
Universit\'e de Gen\`eve, D\'epartement de Physique Th\'eorique and Centre for Astroparticle Physics, 24 quai Ernest-Ansermet, CH-1211 Gen\`eve 4, Switzerland\label{aff60}
\and
Department of Physics, P.O. Box 64, University of Helsinki, 00014 Helsinki, Finland\label{aff61}
\and
Helsinki Institute of Physics, Gustaf H{\"a}llstr{\"o}min katu 2, University of Helsinki, 00014 Helsinki, Finland\label{aff62}
\and
Laboratoire d'etude de l'Univers et des phenomenes eXtremes, Observatoire de Paris, Universit\'e PSL, Sorbonne Universit\'e, CNRS, 92190 Meudon, France\label{aff63}
\and
SKAO, Jodrell Bank, Lower Withington, Macclesfield SK11 9FT, UK\label{aff64}
\and
Centre de Calcul de l'IN2P3/CNRS, 21 avenue Pierre de Coubertin 69627 Villeurbanne Cedex, France\label{aff65}
\and
Universit\"at Bonn, Argelander-Institut f\"ur Astronomie, Auf dem H\"ugel 71, 53121 Bonn, Germany\label{aff66}
\and
INFN-Sezione di Roma, Piazzale Aldo Moro, 2 - c/o Dipartimento di Fisica, Edificio G. Marconi, 00185 Roma, Italy\label{aff67}
\and
Department of Physics, Institute for Computational Cosmology, Durham University, South Road, Durham, DH1 3LE, UK\label{aff68}
\and
Universit\'e Paris Cit\'e, CNRS, Astroparticule et Cosmologie, 75013 Paris, France\label{aff69}
\and
CNRS-UCB International Research Laboratory, Centre Pierre Bin\'etruy, IRL2007, CPB-IN2P3, Berkeley, USA\label{aff70}
\and
Institute of Physics, Laboratory of Astrophysics, Ecole Polytechnique F\'ed\'erale de Lausanne (EPFL), Observatoire de Sauverny, 1290 Versoix, Switzerland\label{aff71}
\and
Dipartimento di Fisica - Sezione di Astronomia, Universit\`a di Trieste, Via Tiepolo 11, 34131 Trieste, Italy\label{aff72}
\and
Telespazio UK S.L. for European Space Agency (ESA), Camino bajo del Castillo, s/n, Urbanizacion Villafranca del Castillo, Villanueva de la Ca\~nada, 28692 Madrid, Spain\label{aff73}
\and
Institut de F\'{i}sica d'Altes Energies (IFAE), The Barcelona Institute of Science and Technology, Campus UAB, 08193 Bellaterra (Barcelona), Spain\label{aff74}
\and
European Space Agency/ESTEC, Keplerlaan 1, 2201 AZ Noordwijk, The Netherlands\label{aff75}
\and
School of Mathematics, Statistics and Physics, Newcastle University, Herschel Building, Newcastle-upon-Tyne, NE1 7RU, UK\label{aff76}
\and
Universit\'e Paris-Saclay, Universit\'e Paris Cit\'e, CEA, CNRS, AIM, 91191, Gif-sur-Yvette, France\label{aff77}
\and
Space Science Data Center, Italian Space Agency, via del Politecnico snc, 00133 Roma, Italy\label{aff78}
\and
Centre National d'Etudes Spatiales -- Centre spatial de Toulouse, 18 avenue Edouard Belin, 31401 Toulouse Cedex 9, France\label{aff79}
\and
Institute of Space Science, Str. Atomistilor, nr. 409 M\u{a}gurele, Ilfov, 077125, Romania\label{aff80}
\and
Consejo Superior de Investigaciones Cientificas, Calle Serrano 117, 28006 Madrid, Spain\label{aff81}
\and
INFN-Padova, Via Marzolo 8, 35131 Padova, Italy\label{aff82}
\and
Instituto de F\'isica Te\'orica UAM-CSIC, Campus de Cantoblanco, 28049 Madrid, Spain\label{aff83}
\and
Institut de Recherche en Astrophysique et Plan\'etologie (IRAP), Universit\'e de Toulouse, CNRS, UPS, CNES, 14 Av. Edouard Belin, 31400 Toulouse, France\label{aff84}
\and
Universit\'e St Joseph; Faculty of Sciences, Beirut, Lebanon\label{aff85}
\and
Departamento de F\'isica, FCFM, Universidad de Chile, Blanco Encalada 2008, Santiago, Chile\label{aff86}
\and
Universit\"at Innsbruck, Institut f\"ur Astro- und Teilchenphysik, Technikerstr. 25/8, 6020 Innsbruck, Austria\label{aff87}
\and
Department of Physics and Helsinki Institute of Physics, Gustaf H\"allstr\"omin katu 2, University of Helsinki, 00014 Helsinki, Finland\label{aff88}
\and
Infrared Processing and Analysis Center, California Institute of Technology, Pasadena, CA 91125, USA\label{aff89}
\and
Departamento de F\'isica, Faculdade de Ci\^encias, Universidade de Lisboa, Edif\'icio C8, Campo Grande, PT1749-016 Lisboa, Portugal\label{aff90}
\and
Instituto de Astrof\'isica e Ci\^encias do Espa\c{c}o, Faculdade de Ci\^encias, Universidade de Lisboa, Tapada da Ajuda, 1349-018 Lisboa, Portugal\label{aff91}
\and
Mullard Space Science Laboratory, University College London, Holmbury St Mary, Dorking, Surrey RH5 6NT, UK\label{aff92}
\and
Cosmic Dawn Center (DAWN)\label{aff93}
\and
Niels Bohr Institute, University of Copenhagen, Jagtvej 128, 2200 Copenhagen, Denmark\label{aff94}
\and
Universidad Polit\'ecnica de Cartagena, Departamento de Electr\'onica y Tecnolog\'ia de Computadoras,  Plaza del Hospital 1, 30202 Cartagena, Spain\label{aff95}
\and
European University of Technology EUt+, European Union\label{aff96}
\and
Institut d'Estudis Espacials de Catalunya (IEEC),  Edifici RDIT, Campus UPC, 08860 Castelldefels, Barcelona, Spain\label{aff97}
\and
Dipartimento di Fisica e Scienze della Terra, Universit\`a degli Studi di Ferrara, Via Giuseppe Saragat 1, 44122 Ferrara, Italy\label{aff98}
\and
Istituto Nazionale di Fisica Nucleare, Sezione di Ferrara, Via Giuseppe Saragat 1, 44122 Ferrara, Italy\label{aff99}
\and
ICSC - Centro Nazionale di Ricerca in High Performance Computing, Big Data e Quantum Computing, Via Magnanelli 2, Bologna, Italy\label{aff100}
\and
INAF, Istituto di Radioastronomia, Via Piero Gobetti 101, 40129 Bologna, Italy\label{aff101}
\and
Department of Mathematical Physics, Institute of Physics, University of S\~ao Paulo, R. do Mat\~ao 1371, 05508-090, S\~ao Paulo, SP, Brazil\label{aff102}
\and
Univ. Grenoble Alpes, CNRS, Grenoble INP, LPSC-IN2P3, 53, Avenue des Martyrs, 38000, Grenoble, France\label{aff103}
\and
Dipartimento di Fisica, Sapienza Universit\`a di Roma, Piazzale Aldo Moro 2, 00185 Roma, Italy\label{aff104}
\and
Department of Mathematics and Physics E. De Giorgi, University of Salento, Via per Arnesano, CP-I93, 73100, Lecce, Italy\label{aff105}
\and
INFN, Sezione di Lecce, Via per Arnesano, CP-193, 73100, Lecce, Italy\label{aff106}
\and
INAF-Sezione di Lecce, c/o Dipartimento Matematica e Fisica, Via per Arnesano, 73100, Lecce, Italy\label{aff107}
\and
Institut d'Astrophysique de Paris, 98bis Boulevard Arago, 75014, Paris, France\label{aff108}
\and
ICL, Junia, Universit\'e Catholique de Lille, LITL, 59000 Lille, France\label{aff109}
\and
CERCA/ISO, Department of Physics, Case Western Reserve University, 10900 Euclid Avenue, Cleveland, OH 44106, USA\label{aff110}
\and
Laboratoire Univers et Th\'eorie, Observatoire de Paris, Universit\'e PSL, Universit\'e Paris Cit\'e, CNRS, 92190 Meudon, France\label{aff111}
\and
Dipartimento di Fisica "Aldo Pontremoli", Universit\`a degli Studi di Milano, Via Celoria 16, 20133 Milano, Italy\label{aff112}
\and
INFN-Sezione di Milano, Via Celoria 16, 20133 Milano, Italy\label{aff113}
\and
Departamento de F{\'\i}sica Fundamental. Universidad de Salamanca. Plaza de la Merced s/n. 37008 Salamanca, Spain\label{aff114}
\and
Universiteit Gent, Department of Physics and Astronomy, Proeftuinstraat 86 N3, 9000 Ghent, Belgium
\label{aff115}
\and
Universit\'e de Strasbourg, CNRS, Observatoire astronomique de Strasbourg, UMR 7550, 67000 Strasbourg, France\label{aff116}
\and
Center for Data-Driven Discovery, Kavli IPMU (WPI), UTIAS, The University of Tokyo, Kashiwa, Chiba 277-8583, Japan\label{aff117}
\and
Waterloo Centre for Astrophysics, University of Waterloo, Waterloo, Ontario N2L 3G1, Canada\label{aff118}
\and
Jodrell Bank Centre for Astrophysics, Department of Physics and Astronomy, University of Manchester, Oxford Road, Manchester M13 9PL, UK\label{aff119}
\and
California Institute of Technology, 1200 E California Blvd, Pasadena, CA 91125, USA\label{aff120}
\and
Institute of Cosmology and Gravitation, University of Portsmouth, Portsmouth PO1 3FX, UK\label{aff121}
\and
Department of Physics \& Astronomy, University of California Irvine, Irvine CA 92697, USA\label{aff122}
\and
Institute for Astronomy, University of Hawaii, 2680 Woodlawn Drive, Honolulu, HI 96822, USA\label{aff123}
\and
Departamento F\'isica Aplicada, Universidad Polit\'ecnica de Cartagena, Campus Muralla del Mar, 30202 Cartagena, Murcia, Spain\label{aff124}
\and
Instituto de F\'isica de Cantabria, Edificio Juan Jord\'a, Avenida de los Castros, 39005 Santander, Spain\label{aff125}
\and
Leiden Observatory, Leiden University, Einsteinweg 55, 2333 CC Leiden, The Netherlands\label{aff126}
\and
Instituto de Astronomia Teorica y Experimental (IATE-CONICET), Laprida 854, X5000BGR, C\'ordoba, Argentina\label{aff127}
\and
Department of Computer Science, Aalto University, PO Box 15400, Espoo, FI-00 076, Finland\label{aff128}
\and
School of Physics and Astronomy, University of Nottingham, University Park, Nottingham NG7 2RD, UK\label{aff129}
\and
Ruhr University Bochum, Faculty of Physics and Astronomy, Astronomical Institute (AIRUB), German Centre for Cosmological Lensing (GCCL), 44780 Bochum, Germany\label{aff130}
\and
DARK, Niels Bohr Institute, University of Copenhagen, Jagtvej 155, 2200 Copenhagen, Denmark\label{aff131}
\and
Department of Physics and Astronomy, Vesilinnantie 5, University of Turku, 20014 Turku, Finland\label{aff132}
\and
Finnish Centre for Astronomy with ESO (FINCA), Quantum, Vesilinnantie 5, University of Turku, 20014 Turku, Finland\label{aff133}
\and
Serco for European Space Agency (ESA), Camino bajo del Castillo, s/n, Urbanizacion Villafranca del Castillo, Villanueva de la Ca\~nada, 28692 Madrid, Spain\label{aff134}
\and
ARC Centre of Excellence for Dark Matter Particle Physics, Melbourne, Australia\label{aff135}
\and
Centre for Astrophysics \& Supercomputing, Swinburne University of Technology,  Hawthorn, Victoria 3122, Australia\label{aff136}
\and
Department of Physics and Astronomy, University of the Western Cape, Bellville, Cape Town, 7535, South Africa\label{aff137}
\and
MTA-CSFK Lend\"ulet Large-Scale Structure Research Group, Konkoly-Thege Mikl\'os \'ut 15-17, H-1121 Budapest, Hungary\label{aff138}
\and
Konkoly Observatory, HUN-REN CSFK, MTA Centre of Excellence, Budapest, Konkoly Thege Mikl\'os {\'u}t 15-17. H-1121, Hungary\label{aff139}
\and
Department of Physics, Centre for Extragalactic Astronomy, Durham University, South Road, Durham, DH1 3LE, UK\label{aff140}
\and
Institute for Theoretical Particle Physics and Cosmology (TTK), RWTH Aachen University, 52056 Aachen, Germany\label{aff141}
\and
IRFU, CEA, Universit\'e Paris-Saclay 91191 Gif-sur-Yvette Cedex, France\label{aff142}
\and
Centro de Astrof\'{\i}sica da Universidade do Porto, Rua das Estrelas, 4150-762 Porto, Portugal\label{aff143}
\and
Instituto de Astrof\'isica e Ci\^encias do Espa\c{c}o, Universidade do Porto, CAUP, Rua das Estrelas, PT4150-762 Porto, Portugal\label{aff144}
\and
Dipartimento di Fisica, Universit\`a di Roma Tor Vergata, Via della Ricerca Scientifica 1, Roma, Italy\label{aff145}
\and
INFN, Sezione di Roma 2, Via della Ricerca Scientifica 1, Roma, Italy\label{aff146}
\and
HE Space for European Space Agency (ESA), Camino bajo del Castillo, s/n, Urbanizacion Villafranca del Castillo, Villanueva de la Ca\~nada, 28692 Madrid, Spain\label{aff147}
\and
Instituto de Astrof\'isica e Ci\^encias do Espa\c{c}o, Faculdade de Ci\^encias, Universidade de Lisboa, Campo Grande, 1749-016 Lisboa, Portugal\label{aff148}
\and
Department of Astrophysics, University of Zurich, Winterthurerstrasse 190, 8057 Zurich, Switzerland\label{aff149}
\and
Department of Mathematics and Physics, Roma Tre University, Via della Vasca Navale 84, 00146 Rome, Italy\label{aff150}
\and
INAF - Osservatorio Astronomico d'Abruzzo, Via Maggini, 64100, Teramo, Italy\label{aff151}
\and
Theoretical astrophysics, Department of Physics and Astronomy, Uppsala University, Box 516, 751 37 Uppsala, Sweden\label{aff152}
\and
Mathematical Institute, University of Leiden, Einsteinweg 55, 2333 CA Leiden, The Netherlands\label{aff153}
\and
School of Physical Sciences, The Open University, Milton Keynes, MK7 6AA, UK\label{aff154}
\and
School of Physics \& Astronomy, University of Southampton, Highfield Campus, Southampton SO17 1BJ, UK\label{aff155}
\and
Institute of Astronomy, University of Cambridge, Madingley Road, Cambridge CB3 0HA, UK\label{aff156}
\and
Univ. Lille, CNRS, Centrale Lille, UMR 9189 CRIStAL, 59000 Lille, France\label{aff157}
\and
Center for Astrophysics and Cosmology, University of Nova Gorica, Nova Gorica, Slovenia\label{aff158}
\and
Kobayashi-Maskawa Institute for the Origin of Particles and the Universe, Nagoya University, Chikusa-ku, Nagoya, 464-8602, Japan\label{aff159}
\and
Institute for Particle Physics and Astrophysics, Dept. of Physics, ETH Zurich, Wolfgang-Pauli-Strasse 27, 8093 Zurich, Switzerland\label{aff160}
\and
Department of Astrophysical Sciences, Peyton Hall, Princeton University, Princeton, NJ 08544, USA\label{aff161}
\and
Fakult\"at f\"ur Physik, Universit\"at Bielefeld, Postfach 100131, 33501 Bielefeld, Germany\label{aff162}
\and
Space physics and astronomy research unit, University of Oulu, Pentti Kaiteran katu 1, FI-90014 Oulu, Finland\label{aff163}
\and
International Centre for Theoretical Physics (ICTP), Strada Costiera 11, 34151 Trieste, Italy\label{aff164}
\and
Center for Computational Astrophysics, Flatiron Institute, 162 5th Avenue, 10010, New York, NY, USA\label{aff165}
\and
SRON Netherlands Institute for Space Research, Landleven 12, 9747 AD, Groningen, The Netherlands\label{aff166}}